\documentclass{easychair}

\usepackage{xspace} \usepackage{xcolor} \usepackage{graphicx} \usepackage[normalem]{ulem} %
\usepackage{amsmath} \usepackage{amssymb} \usepackage{amsfonts} \usepackage{enumitem} \usepackage{xstring}

\usepackage{cite} \usepackage[numbers]{natbib}

\usepackage{booktabs} \usepackage{multirow} \usepackage{makecell} \usepackage{ragged2e} \usepackage{colortbl} \usepackage{tabu}

\usepackage{caption} \usepackage{subcaption} \usepackage{wrapfig}

\usepackage{listings} \fboxsep=0.8pt 

\usepackage{tikz} \usetikzlibrary{calc} \usetikzlibrary{shapes.geometric} \usetikzlibrary{decorations.pathreplacing} \usetikzlibrary{positioning} \usetikzlibrary{automata} \usetikzlibrary{arrows}

\usepackage[keeplastbox]{flushend} %
\usepackage{pgf} \usepackage{datetime} %

\newcommand{\XSpace}[1]{} \newcommand{\XComment}[1]{} \newcommand{\DefMacro}[2]{\expandafter\newcommand\csname rmk-#1\endcsname{#2}} \newcommand{\UseMacro}[1]{\csname rmk-#1\endcsname}  

\newcommand{\FormattingTask}[1]{} \newcommand{\MyPara}[1]{\noindent\textbf{#1}:}   \newcommand{\reducedstrut}{\vrule width 0pt height .9\ht\strutbox depth .9\dp\strutbox\relax} \newcommand{\InputWithSpace}[1]{\bgroup\def\arraystretch{1.1}\input{#1}\egroup} \newcommand{\Code}[1]{{\ifmmode{\mathtt{#1}}\else$\mathtt{#1}$\fi}} \newcommand{\CodeIn}[1]{{\ifmmode{\mathtt{#1}}\else$\mathtt{#1}$\fi}} \newcommand{\CoqIn}[1]{\lstinline[language=Coq,basicstyle=\normalsize\ttfamily]{#1}} \newcommand{\CoqInSmall}[1]{\lstinline[language=Coq,basicstyle=\small\ttfamily]{#1}} \newcommand{\ColorBack}[1]{%
  \begingroup   \setlength{\fboxsep}{0pt}%
  \colorbox{purple!20}{\reducedstrut#1\/}%
  \endgroup }

\newcommand{\specialcell}[2][c]{%
  \begin{tabular}[#1]{@{}c@{}}#2\end{tabular}} \newcolumntype{R}[1]{>{\RaggedLeft\arraybackslash}p{#1}} \newcolumntype{L}[1]{>{\RaggedRight\arraybackslash}p{#1}}

\newcommand{\numtoword}[1]{%
\IfStrEqCase{#1}{{0}{zero}{1}{one}{2}{two}{3}{three}{4}{four}{5}{five}{6}{six}{7}{seven}{8}{eight}{9}{nine}{10}{ten}}[#1]}

\definecolor{shadecolor}{gray}{1.00} \definecolor{darkgray}{gray}{0.30} \definecolor{violet}{rgb}{0.56, 0.0, 1.0} \definecolor{forestgreen}{rgb}{0.13, 0.55, 0.13}

\lstdefinelanguage{Coq} { mathescape=true,						 texcl=false, morekeywords=[1]{   Add,   All,   Arguments,   Axiom,   Bind,   Canonical,   Check,   Close,   CoFixpoint,   CoInductive,   Coercion,   Contextual,   Corollary,   Defined,   Definition,   Delimit,   End,   Example,   Export,   Fact,   Fixpoint,   Goal,   Graph,   Hint,   Hypotheses,   Hypothesis,   Implicit,   Implicits,   Import,   Inductive,   Lemma,   Let,   Local,   Locate,   Ltac,   Maximal   Module,   Morphism,   Next,   Notation,   Obligation,   Open,   Parameter,   Parameters,   Prenex,   Print,   Printing,   Program,   Projections,   Proof,   Proposition,   Qed,   Record,   Relation,   Remark,   Require,   Reserved,   Resolve,   Rewrite,   Save,   Scope,   Search,   Section,   Show,   Strict,   Structure,   Tactic,   Theorem,   Unset,   Variable,   Variables,   View,   inside,   outside }, morekeywords=[2]{   as,   cofix,   else,   end,   exists,   exists2,   fix,   for,   forall,   fun,   if,   in,   is,   let,   match,   nosimpl,   of,   return,   struct,   then,   vfun,   with }, morekeywords=[3]{Type, Prop, Set, True, False}, morekeywords=[4]{   after,   apply,   assert,   auto,   bool_congr,   case,   change,   clear,   compute,   congr,   cut,   cutrewrite,   destruct,   elim,   field,   fold,   generalize,   have,   heval,    hnf,   induction,   injection,   intro,   intros,   intuition,   inversion,   left,   loss,   move,   nat_congr,   nat_norm,   pattern,   pose,   refine,   rename,   replace,   revert,   rewrite,   right,   ring,   set,   simpl,   split,   subst,   suff,   suffices,   symmetry,   transitivity,   trivial,   unfold,   unlock,   using,   without,   wlog,   autorewrite },         morekeywords=[5]{   assumption,   by,   contradiction,   congruence,   done,   exact,   lia,   gappa,   omega,   reflexivity,   romega,   solve,   tauto,   discriminate,   unsat }, morecomment=[s]{(*}{*)}, morekeywords=[6]{do, first, try, idtac, repeat}, showstringspaces=false, morestring=[b]", tabsize=3,							 extendedchars=true,  		 		 sensitive=true,  breaklines=false, basicstyle=\footnotesize\ttfamily, captionpos=b,							 columns=[l]fullflexible, identifierstyle={\color{black}}, keywordstyle=[1]{\color{violet}}, keywordstyle=[2]{\color{forestgreen}}, keywordstyle=[3]{\color{forestgreen}}, keywordstyle=[4]{\color{blue}}, keywordstyle=[5]{\color{red}}, keywordstyle=[6]{\color{violet}}, stringstyle=, commentstyle=\it\ttfamily\color{brown}, numberstyle=\tiny%
}

\lstdefinestyle{Coq}{language=Coq} 

\newcommand{\RevisionInfo}{\Fix{Time: \today{} at \currenttime{.}}}

\newcommand{\cc}{coding conventions\xspace} \newcommand{\CC}{Coding Conventions\xspace} \newcommand{\Cc}{Coding conventions\xspace}

\newcommand{\LemmaN}{Lemma Names\xspace} \newcommand{\lemman}{lemma names\xspace} \newcommand{\Lemman}{Lemma names\xspace}

\newcommand{\MClibrary}{family of projects\xspace} \newcommand{\Coq}{Coq\xspace} \newcommand{\Tool}{\textsc{Roosterize}\xspace} \newcommand{\tooltype}{tool\-chain\xspace} \newcommand{\Tooltype}{Tool\-chain\xspace} \newcommand{\Title}{Learning to Format Coq Code Using Language Models} \newcommand{\ShortTitle}{\Title}

\newcommand{\xcut}{\trimmed{}\xspace} \newcommand{\bidirectional}{bi-directional\xspace} \newcommand{\dsta}{all tiers\xspace} \newcommand{\Dsta}{All tiers\xspace} \newcommand{\DSta}{All Tiers\xspace} \newcommand{\dsti}{tier 1\xspace} \newcommand{\Dsti}{Tier 1\xspace} \newcommand{\DSti}{Tier 1\xspace} \newcommand{\dstii}{tier 2\xspace} \newcommand{\Dstii}{Tier 2\xspace} \newcommand{\DStii}{Tier 2\xspace} \newcommand{\dstiii}{tier 3\xspace} \newcommand{\Dstiii}{Tier 3\xspace} \newcommand{\DStiii}{Tier 3\xspace} \newcommand{\encdec}{encoder-decoder\xspace} \newcommand{\flatten}{flatten\xspace} \newcommand{\flattened}{flattened\xspace} \newcommand{\flattening}{flattening\xspace} \newcommand{\formatting}{formatting\xspace} \newcommand{\Formatting}{Formatting\xspace} \newcommand{\Ktree}{Kernel tree\xspace} \newcommand{\ktree}{kernel tree\xspace} \newcommand{\ktrees}{kernel trees\xspace} \newcommand{\etree}{elaborated tree\xspace} \newcommand{\etrees}{elaborated trees\xspace} \newcommand{\eterms}{elaborated terms\xspace} \newcommand{\eterm}{elaborated term\xspace} \newcommand{\KTreeAcro}{KnlTree\xspace} \newcommand{\kindcom}{\CodeIn{COM}\xspace} \newcommand{\kindid}{\CodeIn{ID}\xspace} \newcommand{\lemmanaming}{lemma naming\xspace} \newcommand{\Lemmanaming}{Lemma naming\xspace} \newcommand{\LemmaNaming}{Lemma Naming\xspace} \newcommand{\leftoutcorpus}{left-out\xspace} \newcommand{\LeftOutCorpus}{Left-out\xspace} \newcommand{\lemmastmt}{lemma statement\xspace} \newcommand{\lemmastmts}{lemma statements\xspace} \newcommand{\Lemmastmt}{Lemma statement\xspace} \newcommand{\LemmaStmt}{Lemma Statement\xspace} \newcommand{\lms}{language models\xspace} \newcommand{\LMS}{Language Models\xspace} \newcommand{\lstmt}{statement\xspace} \newcommand{\Lstmt}{Statement\xspace} \newcommand{\lstmts}{statements\xspace} \newcommand{\LStmtAcro}{Stmt\xspace} \newcommand{\maincorpus}{main\xspace} \newcommand{\MainCorpus}{Main\xspace} \newcommand{\notation}{notation\xspace} \newcommand{\Notation}{Notation\xspace} \newcommand{\ngram}{n-gram\xspace} \newcommand{\Ngram}{n-gram\xspace} \newcommand{\retrievalbased}{retrieval-based\xspace} \newcommand{\RetrievalBased}{Retrieval-based\xspace} \newcommand{\rnn}{RNN\xspace} \newcommand{\rnns}{RNNs\xspace} \newcommand{\rnnlm}{RNNLM\xspace} \newcommand{\simplification}{chopping\xspace} \newcommand{\Simplification}{\Trimming\xspace} \newcommand{\sexp}{sexp\xspace} \newcommand{\sexps}{sexps\xspace} \newcommand{\stree}{syntax tree\xspace} \newcommand{\strees}{syntax trees\xspace} \newcommand{\STreeAcro}{SynTree\xspace} \newcommand{\sktree}{syntax and kernel tree\xspace} \newcommand{\sktrees}{syntax and kernel trees\xspace} \newcommand{\SKTrees}{Syntax and Kernel Trees\xspace} \newcommand{\subtok}{sub-token\xspace} \newcommand{\SubTok}{Sub-token\xspace} \newcommand{\subtokenize}{sub-tokenize\xspace} \newcommand{\subtokenized}{sub-tokenized\xspace} \newcommand{\subtokenizer}{sub-tokenizer\xspace} \newcommand{\subtokenization}{sub-tok\-eniza\-tion\xspace} \newcommand{\SubTokenization}{Sub-tok\-eniza\-tion\xspace} \newcommand{\taskformatting}{\formatting task\xspace} \newcommand{\Taskformatting}{\Formatting task\xspace} \newcommand{\TaskFormatting}{\Formatting Task\xspace} \newcommand{\tasklemmanaming}{\lemmanaming task\xspace} \newcommand{\Tasklemmanaming}{\Lemmanaming task\xspace} \newcommand{\TaskLemmaNaming}{\LemmaNaming Task\xspace} \newcommand{\test}{testing\xspace} \newcommand{\Test}{Testing\xspace} \newcommand{\train}{training\xspace} \newcommand{\Train}{Training\xspace} \newcommand{\trim}{chop\xspace} \newcommand{\trims}{chops\xspace} \newcommand{\trimmed}{chopped\xspace} %
\newcommand{\TrimmedAcro}{Chop\xspace} \newcommand{\trimmedktree}{\trimmed \ktree} \newcommand{\trimmedktrees}{\trimmed \ktrees} \newcommand{\TrimmedKTreeAcro}{\TrimmedAcro{}\KTreeAcro{}\xspace} \newcommand{\trimmedstree}{\trimmed \stree} \newcommand{\trimmedstrees}{\trimmed \strees} \newcommand{\TrimmedSTreeAcro}{\TrimmedAcro{}\STreeAcro{}\xspace} \newcommand{\trimming}{chopping\xspace} \newcommand{\Trimming}{Chopping\xspace} \newcommand{\val}{validation\xspace} \newcommand{\Val}{Validation\xspace} \newcommand{\vernacular}{vernacular\xspace}

\newcommand{\Gallina}{Gallina\xspace} \newcommand{\Ltac}{L$_{\mathit{tac}}$\xspace} \newcommand{\Vernac}{Vernacular\xspace} \newcommand{\MathematicalComponents}{Mathematical Components\xspace} \newcommand{\MathComp}{MathComp\xspace} \newcommand{\Naturalize}{Naturalize\xspace} \newcommand{\seqtoseq}{\textsc{seq2seq}\xspace} \newcommand{\SerAPI}{SerAPI\xspace} \newcommand{\ssreflect}{SSReflect\xspace} \newcommand{\sertok}{\texttt{sertok}\xspace} \newcommand{\sercomp}{\texttt{sercomp}\xspace} \newcommand{\sername}{\texttt{sername}\xspace} \newcommand{\postproc}{\texttt{postproc}\xspace} \newcommand{\learner}{\texttt{learner}\xspace} \newcommand{\DotV}{\texttt{.v}\xspace} \newcommand{\coqwc}{\texttt{coqwc}\xspace} \newcommand{\NLP}{NLP\xspace} \newcommand{\doc}{file\xspace} %
\newcommand{\docs}{files\xspace} %

\newcommand{\bleu}{BLEU\xspace} \newcommand{\Bleu}{BLEU\xspace} \newcommand{\fragacc}{fragment accuracy\xspace} \newcommand{\Fragacc}{Fragment accuracy\xspace} \newcommand{\topone}{\mbox{top-1}\xspace} \newcommand{\Topone}{\mbox{Top-1}\xspace} \newcommand{\topthree}{\mbox{top-3}\xspace} \newcommand{\Topthree}{\mbox{Top-3}\xspace} \newcommand{\toponeacc}{\topone accuracy\xspace} \newcommand{\Toponeacc}{\Topone accuracy\xspace} \newcommand{\topthreeacc}{\topthree accuracy\xspace} \newcommand{\Topthreeacc}{\Topthree accuracy\xspace}

\newcommand{\NGNLMF}{\Ngram LM\xspace} \newcommand{\NGNLMFfull}{\Ngram Language Model\xspace} \newcommand{\NGLMN}{\Ngram LM\xspace} \newcommand{\NGLMNfull}{\Ngram Language Model\xspace}

\newcommand{\TableCaptionResultsF}{\label{tbl:results-f}     Evaluation of Formatting Suggestions on our MathComp     Corpus.\vspace{-6pt}}

\newcommand{\TableHeadAVG}{\textbf{Avg.}\xspace} \newcommand{\TableHeadSUM}{\textbf{$\Sigma$}\xspace} \newcommand{\TableHeadNA}{N/A\xspace} \newcommand{\TableHeadProject}{\textbf{Project}\xspace} \newcommand{\TableHeadURL}{\textbf{URL} (https://github.com/)\xspace} \newcommand{\TableHeadSHA}{\textbf{SHA}\xspace} \newcommand{\TableHeadFiles}{\textbf{\#Files}\xspace} \newcommand{\TableHeadTextLOC}{LOC\xspace} \newcommand{\TableHeadLOC}{\textbf{\TableHeadTextLOC}\xspace} \newcommand{\TableHeadTextSpecLOC}{Spec.\xspace} \newcommand{\TableHeadSpecLOC}{\textbf{\TableHeadTextSpecLOC}\xspace} \newcommand{\TableHeadTextProofLOC}{Proof\xspace} \newcommand{\TableHeadProofLOC}{\textbf{\TableHeadTextProofLOC}\xspace} \newcommand{\TableHeadTextCommentLOC}{Com.~LOC\xspace} \newcommand{\TableHeadCommentLOC}{\textbf{\TableHeadTextCommentLOC}\xspace} \newcommand{\TableHeadNumSpecSent}{\textbf{\#Spec.~sents.}\xspace} \newcommand{\TableHeadNumProofSent}{\textbf{\#Pr.~sents.}\xspace} \newcommand{\TableHeadTier}{\textbf{Tier}\xspace} \DefMacro{table-head-summary-group-t1}{\makecell[l]{\TableHeadAVG{}\\\TableHeadSUM{}}\xspace} \DefMacro{table-head-summary-group-ta}{\MainCorpus{}\hspace{0.5em}\makecell[l]{\TableHeadAVG{}\\\TableHeadSUM{}}\xspace} \DefMacro{table-head-summary-group-allgroup}{All\hspace{1.5em}\makecell[l]{\TableHeadAVG{}\\\TableHeadSUM{}}\xspace} \newcommand{\TableHeadTextLocPerDoc}{LOC/\doc{}\xspace} \newcommand{\TableHeadLocPerDoc}{\textbf{\TableHeadTextLocPerDoc}\xspace} \newcommand{\TableHeadTextSpecLocPerDoc}{Spec.\xspace} \newcommand{\TableHeadSpecLocPerDoc}{\textbf{\TableHeadTextSpecLocPerDoc}\xspace} \newcommand{\TableHeadTextProofLocPerDoc}{Proof\xspace} \newcommand{\TableHeadProofLocPerDoc}{\textbf{\TableHeadTextProofLocPerDoc}\xspace} \newcommand{\TableHeadSpecSentPerDoc}{\textbf{\#Spec.~sents./doc}\xspace} \newcommand{\TableHeadProofSentPerDoc}{\textbf{\#Pr.~sents./doc}\xspace} \newcommand{\TableHeadTokensPerSpecSent}{\textbf{\#Toks/Spec.~sent}\xspace} \newcommand{\TableHeadTokensPerProofSent}{\textbf{\#Toks/Pr.~sent}\xspace} \newcommand{\TableHeadDSTier}{\xspace} \newcommand{\TableHeadDSGroup}{\xspace} \newcommand{\TableHeadDSSet}{\xspace} \newcommand{\TableHeadTextNumPrj}{\#Prjs\xspace} \newcommand{\TableHeadNumPrj}{\textbf{\TableHeadTextNumPrj}\xspace} \newcommand{\TableHeadTextNumDoc}{\#Files\xspace} \newcommand{\TableHeadNumDoc}{\textbf{\TableHeadTextNumDoc}\xspace} \newcommand{\TableHeadTextNumSent}{\#Sents\xspace} \newcommand{\TableHeadNumSent}{\textbf{\TableHeadTextNumSent}\xspace} \newcommand{\TableHeadTextNumTok}{\#Toks\xspace} \newcommand{\TableHeadNumTok}{\textbf{\TableHeadTextNumTok}\xspace} \newcommand{\TableHeadTextLenOfDoc}{len(Files)\xspace} \newcommand{\TableHeadLenOfDoc}{\textbf{\TableHeadTextLenOfDoc}\xspace} \newcommand{\TableHeadTextLenOfSent}{len(Sents)\xspace} \newcommand{\TableHeadLenOfSent}{\textbf{\TableHeadTextLenOfSent}\xspace} \newcommand{\TableHeadTextGallina}{G\xspace} \newcommand{\TableHeadGallina}{\textbf{\TableHeadTextGallina}\xspace} \newcommand{\TableHeadTextLtac}{L\xspace} \newcommand{\TableHeadLtac}{\textbf{\TableHeadTextLtac}\xspace} \newcommand{\TableHeadTextLtacWGallina}{L+G\xspace} \newcommand{\TableHeadLtacWGallina}{\textbf{\TableHeadTextLtacWGallina}\xspace} \newcommand{\TableHeadTextLtacWOGallina}{L\xspace} \newcommand{\TableHeadLtacWOGallina}{\textbf{\TableHeadTextLtacWOGallina}\xspace} \newcommand{\TableHeadTextVernacWGallina}{V+G\xspace} \newcommand{\TableHeadVernacWGallina}{\textbf{\TableHeadTextVernacWGallina}\xspace} \newcommand{\TableHeadTextVernacWOGallina}{V\xspace} \newcommand{\TableHeadVernacWOGallina}{\textbf{\TableHeadTextVernacWOGallina}\xspace} \newcommand{\TableHeadTextVernac}{V\xspace} \newcommand{\TableHeadVernac}{\textbf{\TableHeadTextVernac}\xspace} \newcommand{\TableHeadLMSet}{\xspace} \newcommand{\TableHeadNumLemma}{\textbf{\#Lemmas}\xspace} \newcommand{\TableHeadNumLemmaFiltered}{\textbf{\#Lemmas}\xspace} \newcommand{\TableHeadLMName}{\textbf{Name}\xspace} \newcommand{\TableHeadLMStmt}{\textbf{\LStmtAcro}\xspace} \newcommand{\TableHeadLMKTree}{\textbf{\KTreeAcro}\xspace} \newcommand{\TableHeadLMKTreeTrimmed}{\textbf{\KTreeAcro{}\TrimmedAcro}\xspace} \newcommand{\TableHeadLMSTree}{\textbf{\STreeAcro}\xspace} \newcommand{\TableHeadLMSTreeTrimmed}{\textbf{\STreeAcro{}\TrimmedAcro}\xspace} \newcommand{\TableHeadNumChar}{\textbf{\#Char}\xspace} \newcommand{\TableHeadNumSubToken}{\textbf{\#SubToks}\xspace} \newcommand{\TableHeadDepth}{\textbf{Depth}\xspace} \newcommand{\TableHeadModel}{\textbf{Model}\xspace} \newcommand{\TableHeadGroup}{\textbf{Group}\xspace} \newcommand{\TablePartMultiAttnCopy}{\makecell{Multi-input\\+attn\\+copy}} \newcommand{\TablePartMonoAttnCopy}{\makecell{Single-input\\+attn\\+copy}} \newcommand{\TablePartMultiAttn}{\makecell{Multi-input\\+attn}} \newcommand{\TablePartMonoAttn}{\makecell{Single-input\\+attn}} \newcommand{\TablePartMulti}{\makecell{Multi-input}} \newcommand{\TablePartMono}{\makecell{Single-input}} \newcommand{\TablePartRB}{\makecell{}} \newcommand{\TableHeadLOTrainSet}{\textbf{\#Lemmas}} \newcommand{\QSStmt}[1]{\textbf{Statement}: & \lstinline[language=Coq,basicstyle=\footnotesize\ttfamily]{#1}} \newcommand{\QSStmtCont}[1]{\multicolumn{2}{l}{\hspace{5.8em}\lstinline[language=Coq,basicstyle=\footnotesize\ttfamily]{#1}}} \newcommand{\QSTruth}[1]{\multicolumn{2}{l}{\textbf{Hand-written}: \lstinline[language=Coq,basicstyle=\footnotesize\ttfamily]{#1}}} \newcommand{\QSPred}[1]{\multicolumn{2}{l}{\textbf{\CoqConvTool}: \lstinline[language=Coq,basicstyle=\footnotesize\ttfamily]{#1}}} \newcommand{\QSTruthPred}[2]{\multicolumn{2}{l}{\textbf{Hand-written}: \lstinline[language=Coq,basicstyle=\footnotesize\ttfamily]{#1}\ \ \textbf{\CoqConvTool}: \lstinline[language=Coq,basicstyle=\footnotesize\ttfamily]{#2}}} \newcommand{\QSComment}[1]{\multicolumn{2}{l}{\makecell[{{p{\columnwidth}}}]{\textbf{Comment}: #1}}} \newcommand{\QSTruthOnly}{\textbf{Hand-written}} \newcommand{\QSPredOnly}{\textbf{\CoqConvTool}} \newcommand{\QSCode}[1]{\lstinline[language=Coq,basicstyle=\footnotesize\ttfamily,keepspaces=true]{#1}}

\DefMacro{vspace-results-f}{-10pt}

\newcommand{\CorpusKLOC}{\UseMacro{corpus-t1-SUM-k-code-loc}k\xspace} \newcommand{\CorpusNumProjects}{\numtoword{\UseMacro{corpus-t1-num-projects}}\xspace}

\newcommand{\CoqVersion}{8.10.2\xspace} \newcommand{\ToolURL}{\url{https://tinyurl.com/spkzytl}} %

\newcommand{\NumExpTrials}{3\xspace} \newcommand{\NumExpTimeout}{12\xspace} \newcommand{\NumSubTokenInspected}{200\xspace} \newcommand{\NumSubTokenAcc}{79.5\%\xspace} \newcommand{\NumQSLemmaNameComments}{150\xspace} \newcommand{\NumQSLemmaNameGood}{17\xspace} \newcommand{\PerQSLemmaNameGood}{11.3\%\xspace} \newcommand{\NumQSLemmaNameNeutral}{13\xspace} \newcommand{\PerQSLemmaNameGoodOrNeutral}{20\%\xspace} \newcommand{\NumQSLemmaNameBad}{120\xspace} \newcommand{\NumQSFormattingomments}{56\xspace}

\newcommand{\NumVocabContent}{10,257\xspace} \newcommand{\NumVocabSpacing}{75\xspace} \newcommand{\NumVocabKind}{10\xspace}

\newcommand{\Pmathcomp}{math-comp\xspace} \newcommand{\Poddorder}{odd-order\xspace} \newcommand{\Pfourcolor}{fourcolor\xspace} \newcommand{\Pfinmap}{finmap\xspace}

\newcommand{\Pbigenough}{bigenough\xspace} \newcommand{\Panalysis}{analysis\xspace} \newcommand{\Prealclosed}{real-closed\xspace} \newcommand{\Probot}{robot\xspace} \newcommand{\Pmultinomials}{multinomials\xspace} \newcommand{\Pellipticcurves}{elliptic-curves\xspace} \newcommand{\Ptwosquare}{two-square\xspace} \newcommand{\Pgrobner}{grobner\xspace} \newcommand{\Pinfotheo}{infotheo\xspace}

\newcommand{\Pfcslpcm}{fcsl-pcm\xspace} \newcommand{\Pdisel}{disel\xspace} \newcommand{\Preglang}{reglang\xspace} \newcommand{\Pcompdecpdl}{comp-dec-pdl\xspace} \newcommand{\Ptoychain}{toychain\xspace} \newcommand{\Pbits}{bits\xspace} \newcommand{\Pmonae}{monae\xspace} \newcommand{\Pgames}{games\xspace}

\newcommand{\PTEXTfcslpcm}{the PCM library\xspace} \newcommand{\PTEXTFcslpcm}{The PCM library\xspace} \newcommand{\PTEXTinfotheo}{Infotheo\xspace}

\newcommand{\mycheckmark}{{\normalsize \checkmark}\xspace} \newcommand{\mycross}{$\mathbin{\tikz [x=1.4ex,y=1.4ex,line width=.2ex] \draw (0,0) -- (1,1) (0,1) -- (1,0);}$\xspace\xspace}

\DefMacro{f-brnn}{Neural\xspace} \DefMacro{f-ngram}{\Ngram{}\xspace} \DefMacro{f-brnn-text}{neural\xspace} \DefMacro{f-ngram-text}{\Ngram{}\xspace} \DefMacro{table-head-results-f-acc-all}{\textbf{\Topone Acc.}\xspace} \DefMacro{table-head-results-f-acc-top-3}{\textbf{\Topthree Acc.}\xspace}

\usepackage{doc}

\title{\Title}

\author{ Pengyu Nie\inst{1} \and Karl Palmskog\inst{2} \and Junyi Jessy Li\inst{1} \and Milos Gligoric\inst{1} }

\institute{   The University of Texas at Austin,   Austin, TX, USA\\ \and    KTH Royal Institute of Technology,    Stockholm, Sweden\\    \email{pynie@utexas.edu, palmskog@acm.org, jessy@austin.utexas.edu, gligoric@utexas.edu} }

\authorrunning{Nie, Palmskog, Li, and Gligoric}

\titlerunning{\ShortTitle}

\begin{document}

\maketitle

\noindent Should the final right bracket in a \CoqIn{Record} declaration be on a separate line? Should arguments to \CoqIn{rewrite} be separated by a single space? Coq code tends to be written in distinct manners by different people and teams. The expressiveness, flexibility, and extensibility of Coq's languages and notations means that Coq projects have a wide variety of recognizable coding styles, sometimes explicitly documented as conventions on naming and formatting. In particular, even inexperienced users can distinguish vernacular using the standard library and plain Ltac from idiomatic vernacular using the Mathematical Components (MathComp) library and SSReflect.

While coding conventions are important for comprehension and maintenance, they are costly to document and enforce. Rule-based formatters, such as Coq's beautifier, have limited flexibility and only capture small fractions of desired conventions in large verification projects. We believe that application of \emph{\lms}---a class of Natural Language Processing (NLP) techniques for capturing regularities in corpora---can provide a solution to this conundrum~\cite{AllamanisETAL18Survey}. More specifically, we believe that an approach based on automatically learning conventions from existing Coq code, and then suggesting idiomatic code to users in the proper context, can be superior to manual approaches and static analysis tools---both in terms of effort and results.

As a first step, we here outline initial models to learn and suggest \emph{space formatting} in Coq files, with a preliminary implementation for Coq 8.10, and evaluated using on a corpus based on MathComp 1.9.0 which comprises 164k lines of Coq code from four core projects~\cite{Nie2020}.

\subsection*{\LMS for Coq Formatting}

Natural language has repeating patterns which can be \emph{predicted} statistically at the level of, say, individual words with high accuracy. Programming languages have similar predictability, usually called \emph{naturalness}, which can be exploited to perform a variety of software engineering tasks~\cite{AllamanisETAL18Survey}. We consider, from this view, the problem of predicting spacing between tokens obtained from Coq's lexer. For example, according to MathComp's contribution guide, there should be no space between the tactic tokens \CoqIn{move} and \CoqIn{=>}, which we can learn by observing the relative locations of the two tokens in a large Coq corpus adhering to the conventions.

\MyPara{\UseMacro{f-ngram} model} We constructed a baseline model based on predicting the next token after observing the $n-1$ previous tokens, as often used in NLP and software engineering. To capture formatting, we inserted special tokens holding spacing information before each token.

\MyPara{\UseMacro{f-brnn} model} We constructed a sophisticated model based on bi-directional recurrent neural networks~\cite{PetersETAL17Semi-supervised}. The model embeds Coq tokens and spacing information into vectors, and predicts token formatting using the embedding vectors of both the left-hand and right-hand context.

\subsection*{Preliminary Implementation and Evaluation}


\DefMacro{results-f-f-brnn-val-AVG-accuracy-G}{98.6\%}
\DefMacro{results-f-f-brnn-val-AVG-accuracy-G-G}{98.7\%}
\DefMacro{results-f-f-brnn-val-AVG-accuracy-G-L}{99.2\%}
\DefMacro{results-f-f-brnn-val-AVG-accuracy-G-V}{99.7\%}
\DefMacro{results-f-f-brnn-val-AVG-accuracy-L}{95.1\%}
\DefMacro{results-f-f-brnn-val-AVG-accuracy-L-G}{99.6\%}
\DefMacro{results-f-f-brnn-val-AVG-accuracy-L-L}{99.2\%}
\DefMacro{results-f-f-brnn-val-AVG-accuracy-L-V}{NaN}
\DefMacro{results-f-f-brnn-val-AVG-accuracy-V}{95.5\%}
\DefMacro{results-f-f-brnn-val-AVG-accuracy-V-G}{96.2\%}
\DefMacro{results-f-f-brnn-val-AVG-accuracy-V-L}{NaN}
\DefMacro{results-f-f-brnn-val-AVG-accuracy-V-V}{99.4\%}
\DefMacro{results-f-f-brnn-val-AVG-accuracy-all}{96.8\%}
\DefMacro{results-f-f-brnn-val-AVG-accuracy-betsent}{73.7\%}
\DefMacro{results-f-f-brnn-val-AVG-accuracy-insent}{99.0\%}
\DefMacro{results-f-f-brnn-val-AVG-accuracy-insent-cs}{98.9\%}
\DefMacro{results-f-f-brnn-val-AVG-accuracy-insent-noncs}{99.0\%}
\DefMacro{results-f-f-brnn-val-AVG-accuracy-top-2}{99.3\%}
\DefMacro{results-f-f-brnn-val-AVG-accuracy-top-3}{99.7\%}
\DefMacro{results-f-f-brnn-val-AVG-accuracy-top-5}{99.9\%}
\DefMacro{results-f-f-brnn-val-AVG-correct-G}{7416266.7\%}
\DefMacro{results-f-f-brnn-val-AVG-correct-G-G}{5585266.7\%}
\DefMacro{results-f-f-brnn-val-AVG-correct-G-L}{1240500.0\%}
\DefMacro{results-f-f-brnn-val-AVG-correct-G-V}{605700.0\%}
\DefMacro{results-f-f-brnn-val-AVG-correct-L}{4924066.7\%}
\DefMacro{results-f-f-brnn-val-AVG-correct-L-G}{1246000.0\%}
\DefMacro{results-f-f-brnn-val-AVG-correct-L-L}{3358733.3\%}
\DefMacro{results-f-f-brnn-val-AVG-correct-L-V}{0.0\%}
\DefMacro{results-f-f-brnn-val-AVG-correct-V}{3663000.0\%}
\DefMacro{results-f-f-brnn-val-AVG-correct-V-G}{585000.0\%}
\DefMacro{results-f-f-brnn-val-AVG-correct-V-L}{0.0\%}
\DefMacro{results-f-f-brnn-val-AVG-correct-V-V}{2338900.0\%}
\DefMacro{results-f-f-brnn-val-AVG-correct-all}{16003333.3\%}
\DefMacro{results-f-f-brnn-val-AVG-correct-betsent}{1043233.3\%}
\DefMacro{results-f-f-brnn-val-AVG-correct-insent}{14960100.0\%}
\DefMacro{results-f-f-brnn-val-AVG-correct-insent-cs}{3677200.0\%}
\DefMacro{results-f-f-brnn-val-AVG-correct-insent-noncs}{11282900.0\%}
\DefMacro{results-f-f-brnn-val-AVG-correct-top-2}{16418300.0\%}
\DefMacro{results-f-f-brnn-val-AVG-correct-top-3}{16488566.7\%}
\DefMacro{results-f-f-brnn-val-AVG-correct-top-5}{16519866.7\%}
\DefMacro{results-f-f-brnn-val-AVG-count-G}{7519200.0\%}
\DefMacro{results-f-f-brnn-val-AVG-count-G-G}{5660400.0\%}
\DefMacro{results-f-f-brnn-val-AVG-count-G-L}{1250500.0\%}
\DefMacro{results-f-f-brnn-val-AVG-count-G-V}{607600.0\%}
\DefMacro{results-f-f-brnn-val-AVG-count-L}{5179400.0\%}
\DefMacro{results-f-f-brnn-val-AVG-count-L-G}{1250600.0\%}
\DefMacro{results-f-f-brnn-val-AVG-count-L-L}{3386800.0\%}
\DefMacro{results-f-f-brnn-val-AVG-count-L-V}{0.0\%}
\DefMacro{results-f-f-brnn-val-AVG-count-V}{3833800.0\%}
\DefMacro{results-f-f-brnn-val-AVG-count-V-G}{608200.0\%}
\DefMacro{results-f-f-brnn-val-AVG-count-V-L}{0.0\%}
\DefMacro{results-f-f-brnn-val-AVG-count-V-V}{2352800.0\%}
\DefMacro{results-f-f-brnn-val-AVG-count-all}{16532400.0\%}
\DefMacro{results-f-f-brnn-val-AVG-count-betsent}{1415500.0\%}
\DefMacro{results-f-f-brnn-val-AVG-count-insent}{15116900.0\%}
\DefMacro{results-f-f-brnn-val-AVG-count-insent-cs}{3716900.0\%}
\DefMacro{results-f-f-brnn-val-AVG-count-insent-noncs}{11400000.0\%}
\DefMacro{results-f-f-brnn-val-AVG-count-top-2}{16532400.0\%}
\DefMacro{results-f-f-brnn-val-AVG-count-top-3}{16532400.0\%}
\DefMacro{results-f-f-brnn-val-AVG-count-top-5}{16532400.0\%}
\DefMacro{results-f-f-brnn-val-MAX-accuracy-G}{98.7\%}
\DefMacro{results-f-f-brnn-val-MAX-accuracy-G-G}{98.8\%}
\DefMacro{results-f-f-brnn-val-MAX-accuracy-G-L}{99.3\%}
\DefMacro{results-f-f-brnn-val-MAX-accuracy-G-V}{99.8\%}
\DefMacro{results-f-f-brnn-val-MAX-accuracy-L}{95.1\%}
\DefMacro{results-f-f-brnn-val-MAX-accuracy-L-G}{99.9\%}
\DefMacro{results-f-f-brnn-val-MAX-accuracy-L-L}{99.2\%}
\DefMacro{results-f-f-brnn-val-MAX-accuracy-L-V}{NaN}
\DefMacro{results-f-f-brnn-val-MAX-accuracy-V}{96.1\%}
\DefMacro{results-f-f-brnn-val-MAX-accuracy-V-G}{96.3\%}
\DefMacro{results-f-f-brnn-val-MAX-accuracy-V-L}{NaN}
\DefMacro{results-f-f-brnn-val-MAX-accuracy-V-V}{99.5\%}
\DefMacro{results-f-f-brnn-val-MAX-accuracy-all}{96.9\%}
\DefMacro{results-f-f-brnn-val-MAX-accuracy-betsent}{75.0\%}
\DefMacro{results-f-f-brnn-val-MAX-accuracy-insent}{99.0\%}
\DefMacro{results-f-f-brnn-val-MAX-accuracy-insent-cs}{99.0\%}
\DefMacro{results-f-f-brnn-val-MAX-accuracy-insent-noncs}{99.0\%}
\DefMacro{results-f-f-brnn-val-MAX-accuracy-top-2}{99.3\%}
\DefMacro{results-f-f-brnn-val-MAX-accuracy-top-3}{99.7\%}
\DefMacro{results-f-f-brnn-val-MAX-accuracy-top-5}{99.9\%}
\DefMacro{results-f-f-brnn-val-MAX-correct-G}{74184}
\DefMacro{results-f-f-brnn-val-MAX-correct-G-G}{55899}
\DefMacro{results-f-f-brnn-val-MAX-correct-G-L}{12414}
\DefMacro{results-f-f-brnn-val-MAX-correct-G-V}{6061}
\DefMacro{results-f-f-brnn-val-MAX-correct-L}{49276}
\DefMacro{results-f-f-brnn-val-MAX-correct-L-G}{12490}
\DefMacro{results-f-f-brnn-val-MAX-correct-L-L}{33603}
\DefMacro{results-f-f-brnn-val-MAX-correct-L-V}{0}
\DefMacro{results-f-f-brnn-val-MAX-correct-V}{36837}
\DefMacro{results-f-f-brnn-val-MAX-correct-V-G}{5856}
\DefMacro{results-f-f-brnn-val-MAX-correct-V-L}{0}
\DefMacro{results-f-f-brnn-val-MAX-correct-V-V}{23399}
\DefMacro{results-f-f-brnn-val-MAX-correct-all}{160249}
\DefMacro{results-f-f-brnn-val-MAX-correct-betsent}{10623}
\DefMacro{results-f-f-brnn-val-MAX-correct-insent}{149632}
\DefMacro{results-f-f-brnn-val-MAX-correct-insent-cs}{36814}
\DefMacro{results-f-f-brnn-val-MAX-correct-insent-noncs}{112880}
\DefMacro{results-f-f-brnn-val-MAX-correct-top-2}{164230}
\DefMacro{results-f-f-brnn-val-MAX-correct-top-3}{164901}
\DefMacro{results-f-f-brnn-val-MAX-correct-top-5}{165206}
\DefMacro{results-f-f-brnn-val-MAX-count-G}{75192}
\DefMacro{results-f-f-brnn-val-MAX-count-G-G}{56604}
\DefMacro{results-f-f-brnn-val-MAX-count-G-L}{12505}
\DefMacro{results-f-f-brnn-val-MAX-count-G-V}{6076}
\DefMacro{results-f-f-brnn-val-MAX-count-L}{51794}
\DefMacro{results-f-f-brnn-val-MAX-count-L-G}{12506}
\DefMacro{results-f-f-brnn-val-MAX-count-L-L}{33868}
\DefMacro{results-f-f-brnn-val-MAX-count-L-V}{0}
\DefMacro{results-f-f-brnn-val-MAX-count-V}{38338}
\DefMacro{results-f-f-brnn-val-MAX-count-V-G}{6082}
\DefMacro{results-f-f-brnn-val-MAX-count-V-L}{0}
\DefMacro{results-f-f-brnn-val-MAX-count-V-V}{23528}
\DefMacro{results-f-f-brnn-val-MAX-count-all}{165324}
\DefMacro{results-f-f-brnn-val-MAX-count-betsent}{14155}
\DefMacro{results-f-f-brnn-val-MAX-count-insent}{151169}
\DefMacro{results-f-f-brnn-val-MAX-count-insent-cs}{37169}
\DefMacro{results-f-f-brnn-val-MAX-count-insent-noncs}{114000}
\DefMacro{results-f-f-brnn-val-MAX-count-top-2}{165324}
\DefMacro{results-f-f-brnn-val-MAX-count-top-3}{165324}
\DefMacro{results-f-f-brnn-val-MAX-count-top-5}{165324}
\DefMacro{results-f-f-brnn-val-MEDIAN-accuracy-G}{98.7\%}
\DefMacro{results-f-f-brnn-val-MEDIAN-accuracy-G-G}{98.6\%}
\DefMacro{results-f-f-brnn-val-MEDIAN-accuracy-G-L}{99.2\%}
\DefMacro{results-f-f-brnn-val-MEDIAN-accuracy-G-V}{99.7\%}
\DefMacro{results-f-f-brnn-val-MEDIAN-accuracy-L}{95.0\%}
\DefMacro{results-f-f-brnn-val-MEDIAN-accuracy-L-G}{99.6\%}
\DefMacro{results-f-f-brnn-val-MEDIAN-accuracy-L-L}{99.2\%}
\DefMacro{results-f-f-brnn-val-MEDIAN-accuracy-L-V}{NaN}
\DefMacro{results-f-f-brnn-val-MEDIAN-accuracy-V}{95.5\%}
\DefMacro{results-f-f-brnn-val-MEDIAN-accuracy-V-G}{96.2\%}
\DefMacro{results-f-f-brnn-val-MEDIAN-accuracy-V-L}{NaN}
\DefMacro{results-f-f-brnn-val-MEDIAN-accuracy-V-V}{99.4\%}
\DefMacro{results-f-f-brnn-val-MEDIAN-accuracy-all}{96.8\%}
\DefMacro{results-f-f-brnn-val-MEDIAN-accuracy-betsent}{73.8\%}
\DefMacro{results-f-f-brnn-val-MEDIAN-accuracy-insent}{99.0\%}
\DefMacro{results-f-f-brnn-val-MEDIAN-accuracy-insent-cs}{98.9\%}
\DefMacro{results-f-f-brnn-val-MEDIAN-accuracy-insent-noncs}{99.0\%}
\DefMacro{results-f-f-brnn-val-MEDIAN-accuracy-top-2}{99.3\%}
\DefMacro{results-f-f-brnn-val-MEDIAN-accuracy-top-3}{99.7\%}
\DefMacro{results-f-f-brnn-val-MEDIAN-accuracy-top-5}{99.9\%}
\DefMacro{results-f-f-brnn-val-MEDIAN-correct-G}{7418100.0\%}
\DefMacro{results-f-f-brnn-val-MEDIAN-correct-G-G}{5583500.0\%}
\DefMacro{results-f-f-brnn-val-MEDIAN-correct-G-L}{1240500.0\%}
\DefMacro{results-f-f-brnn-val-MEDIAN-correct-G-V}{605600.0\%}
\DefMacro{results-f-f-brnn-val-MEDIAN-correct-L}{4922800.0\%}
\DefMacro{results-f-f-brnn-val-MEDIAN-correct-L-G}{1245100.0\%}
\DefMacro{results-f-f-brnn-val-MEDIAN-correct-L-L}{3358200.0\%}
\DefMacro{results-f-f-brnn-val-MEDIAN-correct-L-V}{0.0\%}
\DefMacro{results-f-f-brnn-val-MEDIAN-correct-V}{3662500.0\%}
\DefMacro{results-f-f-brnn-val-MEDIAN-correct-V-G}{584800.0\%}
\DefMacro{results-f-f-brnn-val-MEDIAN-correct-V-L}{0.0\%}
\DefMacro{results-f-f-brnn-val-MEDIAN-correct-V-V}{2338800.0\%}
\DefMacro{results-f-f-brnn-val-MEDIAN-correct-all}{16008200.0\%}
\DefMacro{results-f-f-brnn-val-MEDIAN-correct-betsent}{1045000.0\%}
\DefMacro{results-f-f-brnn-val-MEDIAN-correct-insent}{14962600.0\%}
\DefMacro{results-f-f-brnn-val-MEDIAN-correct-insent-cs}{3675600.0\%}
\DefMacro{results-f-f-brnn-val-MEDIAN-correct-insent-noncs}{11281800.0\%}
\DefMacro{results-f-f-brnn-val-MEDIAN-correct-top-2}{16419900.0\%}
\DefMacro{results-f-f-brnn-val-MEDIAN-correct-top-3}{16489100.0\%}
\DefMacro{results-f-f-brnn-val-MEDIAN-correct-top-5}{16520400.0\%}
\DefMacro{results-f-f-brnn-val-MEDIAN-count-G}{7519200.0\%}
\DefMacro{results-f-f-brnn-val-MEDIAN-count-G-G}{5660400.0\%}
\DefMacro{results-f-f-brnn-val-MEDIAN-count-G-L}{1250500.0\%}
\DefMacro{results-f-f-brnn-val-MEDIAN-count-G-V}{607600.0\%}
\DefMacro{results-f-f-brnn-val-MEDIAN-count-L}{5179400.0\%}
\DefMacro{results-f-f-brnn-val-MEDIAN-count-L-G}{1250600.0\%}
\DefMacro{results-f-f-brnn-val-MEDIAN-count-L-L}{3386800.0\%}
\DefMacro{results-f-f-brnn-val-MEDIAN-count-L-V}{0.0\%}
\DefMacro{results-f-f-brnn-val-MEDIAN-count-V}{3833800.0\%}
\DefMacro{results-f-f-brnn-val-MEDIAN-count-V-G}{608200.0\%}
\DefMacro{results-f-f-brnn-val-MEDIAN-count-V-L}{0.0\%}
\DefMacro{results-f-f-brnn-val-MEDIAN-count-V-V}{2352800.0\%}
\DefMacro{results-f-f-brnn-val-MEDIAN-count-all}{16532400.0\%}
\DefMacro{results-f-f-brnn-val-MEDIAN-count-betsent}{1415500.0\%}
\DefMacro{results-f-f-brnn-val-MEDIAN-count-insent}{15116900.0\%}
\DefMacro{results-f-f-brnn-val-MEDIAN-count-insent-cs}{3716900.0\%}
\DefMacro{results-f-f-brnn-val-MEDIAN-count-insent-noncs}{11400000.0\%}
\DefMacro{results-f-f-brnn-val-MEDIAN-count-top-2}{16532400.0\%}
\DefMacro{results-f-f-brnn-val-MEDIAN-count-top-3}{16532400.0\%}
\DefMacro{results-f-f-brnn-val-MEDIAN-count-top-5}{16532400.0\%}
\DefMacro{results-f-f-brnn-val-MIN-accuracy-G}{98.6\%}
\DefMacro{results-f-f-brnn-val-MIN-accuracy-G-G}{98.6\%}
\DefMacro{results-f-f-brnn-val-MIN-accuracy-G-L}{99.1\%}
\DefMacro{results-f-f-brnn-val-MIN-accuracy-G-V}{99.6\%}
\DefMacro{results-f-f-brnn-val-MIN-accuracy-L}{95.0\%}
\DefMacro{results-f-f-brnn-val-MIN-accuracy-L-G}{99.5\%}
\DefMacro{results-f-f-brnn-val-MIN-accuracy-L-L}{99.1\%}
\DefMacro{results-f-f-brnn-val-MIN-accuracy-L-V}{NaN}
\DefMacro{results-f-f-brnn-val-MIN-accuracy-V}{95.0\%}
\DefMacro{results-f-f-brnn-val-MIN-accuracy-V-G}{96.1\%}
\DefMacro{results-f-f-brnn-val-MIN-accuracy-V-L}{NaN}
\DefMacro{results-f-f-brnn-val-MIN-accuracy-V-V}{99.4\%}
\DefMacro{results-f-f-brnn-val-MIN-accuracy-all}{96.6\%}
\DefMacro{results-f-f-brnn-val-MIN-accuracy-betsent}{72.2\%}
\DefMacro{results-f-f-brnn-val-MIN-accuracy-insent}{98.9\%}
\DefMacro{results-f-f-brnn-val-MIN-accuracy-insent-cs}{98.9\%}
\DefMacro{results-f-f-brnn-val-MIN-accuracy-insent-noncs}{98.9\%}
\DefMacro{results-f-f-brnn-val-MIN-accuracy-top-2}{99.3\%}
\DefMacro{results-f-f-brnn-val-MIN-accuracy-top-3}{99.7\%}
\DefMacro{results-f-f-brnn-val-MIN-accuracy-top-5}{99.9\%}
\DefMacro{results-f-f-brnn-val-MIN-correct-G}{74123}
\DefMacro{results-f-f-brnn-val-MIN-correct-G-G}{55824}
\DefMacro{results-f-f-brnn-val-MIN-correct-G-L}{12396}
\DefMacro{results-f-f-brnn-val-MIN-correct-G-V}{6054}
\DefMacro{results-f-f-brnn-val-MIN-correct-L}{49218}
\DefMacro{results-f-f-brnn-val-MIN-correct-L-G}{12439}
\DefMacro{results-f-f-brnn-val-MIN-correct-L-L}{33577}
\DefMacro{results-f-f-brnn-val-MIN-correct-L-V}{0}
\DefMacro{results-f-f-brnn-val-MIN-correct-V}{36428}
\DefMacro{results-f-f-brnn-val-MIN-correct-V-G}{5846}
\DefMacro{results-f-f-brnn-val-MIN-correct-V-L}{0}
\DefMacro{results-f-f-brnn-val-MIN-correct-V-V}{23380}
\DefMacro{results-f-f-brnn-val-MIN-correct-all}{159769}
\DefMacro{results-f-f-brnn-val-MIN-correct-betsent}{10224}
\DefMacro{results-f-f-brnn-val-MIN-correct-insent}{149545}
\DefMacro{results-f-f-brnn-val-MIN-correct-insent-cs}{36746}
\DefMacro{results-f-f-brnn-val-MIN-correct-insent-noncs}{112789}
\DefMacro{results-f-f-brnn-val-MIN-correct-top-2}{164120}
\DefMacro{results-f-f-brnn-val-MIN-correct-top-3}{164865}
\DefMacro{results-f-f-brnn-val-MIN-correct-top-5}{165186}
\DefMacro{results-f-f-brnn-val-MIN-count-G}{75192}
\DefMacro{results-f-f-brnn-val-MIN-count-G-G}{56604}
\DefMacro{results-f-f-brnn-val-MIN-count-G-L}{12505}
\DefMacro{results-f-f-brnn-val-MIN-count-G-V}{6076}
\DefMacro{results-f-f-brnn-val-MIN-count-L}{51794}
\DefMacro{results-f-f-brnn-val-MIN-count-L-G}{12506}
\DefMacro{results-f-f-brnn-val-MIN-count-L-L}{33868}
\DefMacro{results-f-f-brnn-val-MIN-count-L-V}{0}
\DefMacro{results-f-f-brnn-val-MIN-count-V}{38338}
\DefMacro{results-f-f-brnn-val-MIN-count-V-G}{6082}
\DefMacro{results-f-f-brnn-val-MIN-count-V-L}{0}
\DefMacro{results-f-f-brnn-val-MIN-count-V-V}{23528}
\DefMacro{results-f-f-brnn-val-MIN-count-all}{165324}
\DefMacro{results-f-f-brnn-val-MIN-count-betsent}{14155}
\DefMacro{results-f-f-brnn-val-MIN-count-insent}{151169}
\DefMacro{results-f-f-brnn-val-MIN-count-insent-cs}{37169}
\DefMacro{results-f-f-brnn-val-MIN-count-insent-noncs}{114000}
\DefMacro{results-f-f-brnn-val-MIN-count-top-2}{165324}
\DefMacro{results-f-f-brnn-val-MIN-count-top-3}{165324}
\DefMacro{results-f-f-brnn-val-MIN-count-top-5}{165324}
\DefMacro{results-f-f-brnn-val-STDEV-accuracy-G}{0.0\%}
\DefMacro{results-f-f-brnn-val-STDEV-accuracy-G-G}{0.1\%}
\DefMacro{results-f-f-brnn-val-STDEV-accuracy-G-L}{0.1\%}
\DefMacro{results-f-f-brnn-val-STDEV-accuracy-G-V}{0.0\%}
\DefMacro{results-f-f-brnn-val-STDEV-accuracy-L}{0.0\%}
\DefMacro{results-f-f-brnn-val-STDEV-accuracy-L-G}{0.2\%}
\DefMacro{results-f-f-brnn-val-STDEV-accuracy-L-L}{0.0\%}
\DefMacro{results-f-f-brnn-val-STDEV-accuracy-L-V}{NaN}
\DefMacro{results-f-f-brnn-val-STDEV-accuracy-V}{0.4\%}
\DefMacro{results-f-f-brnn-val-STDEV-accuracy-V-G}{0.1\%}
\DefMacro{results-f-f-brnn-val-STDEV-accuracy-V-L}{NaN}
\DefMacro{results-f-f-brnn-val-STDEV-accuracy-V-V}{0.0\%}
\DefMacro{results-f-f-brnn-val-STDEV-accuracy-all}{0.1\%}
\DefMacro{results-f-f-brnn-val-STDEV-accuracy-betsent}{1.2\%}
\DefMacro{results-f-f-brnn-val-STDEV-accuracy-insent}{0.0\%}
\DefMacro{results-f-f-brnn-val-STDEV-accuracy-insent-cs}{0.1\%}
\DefMacro{results-f-f-brnn-val-STDEV-accuracy-insent-noncs}{0.0\%}
\DefMacro{results-f-f-brnn-val-STDEV-accuracy-top-2}{0.0\%}
\DefMacro{results-f-f-brnn-val-STDEV-accuracy-top-3}{0.0\%}
\DefMacro{results-f-f-brnn-val-STDEV-accuracy-top-5}{0.0\%}
\DefMacro{results-f-f-brnn-val-STDEV-correct-G}{2807.5\%}
\DefMacro{results-f-f-brnn-val-STDEV-correct-G-G}{3306.9\%}
\DefMacro{results-f-f-brnn-val-STDEV-correct-G-L}{734.8\%}
\DefMacro{results-f-f-brnn-val-STDEV-correct-G-V}{294.4\%}
\DefMacro{results-f-f-brnn-val-STDEV-correct-L}{2531.6\%}
\DefMacro{results-f-f-brnn-val-STDEV-correct-L-G}{2177.2\%}
\DefMacro{results-f-f-brnn-val-STDEV-correct-L-L}{1126.4\%}
\DefMacro{results-f-f-brnn-val-STDEV-correct-L-V}{0.0\%}
\DefMacro{results-f-f-brnn-val-STDEV-correct-V}{16701.1\%}
\DefMacro{results-f-f-brnn-val-STDEV-correct-V-G}{432.0\%}
\DefMacro{results-f-f-brnn-val-STDEV-correct-V-L}{0.0\%}
\DefMacro{results-f-f-brnn-val-STDEV-correct-V-V}{778.9\%}
\DefMacro{results-f-f-brnn-val-STDEV-correct-all}{19895.8\%}
\DefMacro{results-f-f-brnn-val-STDEV-correct-betsent}{16336.9\%}
\DefMacro{results-f-f-brnn-val-STDEV-correct-insent}{3967.4\%}
\DefMacro{results-f-f-brnn-val-STDEV-correct-insent-cs}{2997.8\%}
\DefMacro{results-f-f-brnn-val-STDEV-correct-insent-noncs}{3795.6\%}
\DefMacro{results-f-f-brnn-val-STDEV-correct-top-2}{4631.1\%}
\DefMacro{results-f-f-brnn-val-STDEV-correct-top-3}{1517.3\%}
\DefMacro{results-f-f-brnn-val-STDEV-correct-top-5}{899.4\%}
\DefMacro{results-f-f-brnn-val-STDEV-count-G}{0.0\%}
\DefMacro{results-f-f-brnn-val-STDEV-count-G-G}{0.0\%}
\DefMacro{results-f-f-brnn-val-STDEV-count-G-L}{0.0\%}
\DefMacro{results-f-f-brnn-val-STDEV-count-G-V}{0.0\%}
\DefMacro{results-f-f-brnn-val-STDEV-count-L}{0.0\%}
\DefMacro{results-f-f-brnn-val-STDEV-count-L-G}{0.0\%}
\DefMacro{results-f-f-brnn-val-STDEV-count-L-L}{0.0\%}
\DefMacro{results-f-f-brnn-val-STDEV-count-L-V}{0.0\%}
\DefMacro{results-f-f-brnn-val-STDEV-count-V}{0.0\%}
\DefMacro{results-f-f-brnn-val-STDEV-count-V-G}{0.0\%}
\DefMacro{results-f-f-brnn-val-STDEV-count-V-L}{0.0\%}
\DefMacro{results-f-f-brnn-val-STDEV-count-V-V}{0.0\%}
\DefMacro{results-f-f-brnn-val-STDEV-count-all}{0.0\%}
\DefMacro{results-f-f-brnn-val-STDEV-count-betsent}{0.0\%}
\DefMacro{results-f-f-brnn-val-STDEV-count-insent}{0.0\%}
\DefMacro{results-f-f-brnn-val-STDEV-count-insent-cs}{0.0\%}
\DefMacro{results-f-f-brnn-val-STDEV-count-insent-noncs}{0.0\%}
\DefMacro{results-f-f-brnn-val-STDEV-count-top-2}{0.0\%}
\DefMacro{results-f-f-brnn-val-STDEV-count-top-3}{0.0\%}
\DefMacro{results-f-f-brnn-val-STDEV-count-top-5}{0.0\%}
\DefMacro{results-f-f-brnn-val-SUM-accuracy-G}{295.9\%}
\DefMacro{results-f-f-brnn-val-SUM-accuracy-G-G}{296.0\%}
\DefMacro{results-f-f-brnn-val-SUM-accuracy-G-L}{297.6\%}
\DefMacro{results-f-f-brnn-val-SUM-accuracy-G-V}{299.1\%}
\DefMacro{results-f-f-brnn-val-SUM-accuracy-L}{285.2\%}
\DefMacro{results-f-f-brnn-val-SUM-accuracy-L-G}{298.9\%}
\DefMacro{results-f-f-brnn-val-SUM-accuracy-L-L}{297.5\%}
\DefMacro{results-f-f-brnn-val-SUM-accuracy-L-V}{NaN}
\DefMacro{results-f-f-brnn-val-SUM-accuracy-V}{286.6\%}
\DefMacro{results-f-f-brnn-val-SUM-accuracy-V-G}{288.6\%}
\DefMacro{results-f-f-brnn-val-SUM-accuracy-V-L}{NaN}
\DefMacro{results-f-f-brnn-val-SUM-accuracy-V-V}{298.2\%}
\DefMacro{results-f-f-brnn-val-SUM-accuracy-all}{290.4\%}
\DefMacro{results-f-f-brnn-val-SUM-accuracy-betsent}{221.1\%}
\DefMacro{results-f-f-brnn-val-SUM-accuracy-insent}{296.9\%}
\DefMacro{results-f-f-brnn-val-SUM-accuracy-insent-cs}{296.8\%}
\DefMacro{results-f-f-brnn-val-SUM-accuracy-insent-noncs}{296.9\%}
\DefMacro{results-f-f-brnn-val-SUM-accuracy-top-2}{297.9\%}
\DefMacro{results-f-f-brnn-val-SUM-accuracy-top-3}{299.2\%}
\DefMacro{results-f-f-brnn-val-SUM-accuracy-top-5}{299.8\%}
\DefMacro{results-f-f-brnn-val-SUM-correct-G}{222488}
\DefMacro{results-f-f-brnn-val-SUM-correct-G-G}{167558}
\DefMacro{results-f-f-brnn-val-SUM-correct-G-L}{37215}
\DefMacro{results-f-f-brnn-val-SUM-correct-G-V}{18171}
\DefMacro{results-f-f-brnn-val-SUM-correct-L}{147722}
\DefMacro{results-f-f-brnn-val-SUM-correct-L-G}{37380}
\DefMacro{results-f-f-brnn-val-SUM-correct-L-L}{100762}
\DefMacro{results-f-f-brnn-val-SUM-correct-L-V}{0}
\DefMacro{results-f-f-brnn-val-SUM-correct-V}{109890}
\DefMacro{results-f-f-brnn-val-SUM-correct-V-G}{17550}
\DefMacro{results-f-f-brnn-val-SUM-correct-V-L}{0}
\DefMacro{results-f-f-brnn-val-SUM-correct-V-V}{70167}
\DefMacro{results-f-f-brnn-val-SUM-correct-all}{480100}
\DefMacro{results-f-f-brnn-val-SUM-correct-betsent}{31297}
\DefMacro{results-f-f-brnn-val-SUM-correct-insent}{448803}
\DefMacro{results-f-f-brnn-val-SUM-correct-insent-cs}{110316}
\DefMacro{results-f-f-brnn-val-SUM-correct-insent-noncs}{338487}
\DefMacro{results-f-f-brnn-val-SUM-correct-top-2}{492549}
\DefMacro{results-f-f-brnn-val-SUM-correct-top-3}{494657}
\DefMacro{results-f-f-brnn-val-SUM-correct-top-5}{495596}
\DefMacro{results-f-f-brnn-val-SUM-count-G}{225576}
\DefMacro{results-f-f-brnn-val-SUM-count-G-G}{169812}
\DefMacro{results-f-f-brnn-val-SUM-count-G-L}{37515}
\DefMacro{results-f-f-brnn-val-SUM-count-G-V}{18228}
\DefMacro{results-f-f-brnn-val-SUM-count-L}{155382}
\DefMacro{results-f-f-brnn-val-SUM-count-L-G}{37518}
\DefMacro{results-f-f-brnn-val-SUM-count-L-L}{101604}
\DefMacro{results-f-f-brnn-val-SUM-count-L-V}{0}
\DefMacro{results-f-f-brnn-val-SUM-count-V}{115014}
\DefMacro{results-f-f-brnn-val-SUM-count-V-G}{18246}
\DefMacro{results-f-f-brnn-val-SUM-count-V-L}{0}
\DefMacro{results-f-f-brnn-val-SUM-count-V-V}{70584}
\DefMacro{results-f-f-brnn-val-SUM-count-all}{495972}
\DefMacro{results-f-f-brnn-val-SUM-count-betsent}{42465}
\DefMacro{results-f-f-brnn-val-SUM-count-insent}{453507}
\DefMacro{results-f-f-brnn-val-SUM-count-insent-cs}{111507}
\DefMacro{results-f-f-brnn-val-SUM-count-insent-noncs}{342000}
\DefMacro{results-f-f-brnn-val-SUM-count-top-2}{495972}
\DefMacro{results-f-f-brnn-val-SUM-count-top-3}{495972}
\DefMacro{results-f-f-brnn-val-SUM-count-top-5}{495972}
\DefMacro{results-f-f-brnn-test-AVG-accuracy-G}{99.0\%}
\DefMacro{results-f-f-brnn-test-AVG-accuracy-G-G}{99.1\%}
\DefMacro{results-f-f-brnn-test-AVG-accuracy-G-L}{99.1\%}
\DefMacro{results-f-f-brnn-test-AVG-accuracy-G-V}{99.7\%}
\DefMacro{results-f-f-brnn-test-AVG-accuracy-L}{94.5\%}
\DefMacro{results-f-f-brnn-test-AVG-accuracy-L-G}{99.8\%}
\DefMacro{results-f-f-brnn-test-AVG-accuracy-L-L}{99.4\%}
\DefMacro{results-f-f-brnn-test-AVG-accuracy-L-V}{NaN}
\DefMacro{results-f-f-brnn-test-AVG-accuracy-V}{95.0\%}
\DefMacro{results-f-f-brnn-test-AVG-accuracy-V-G}{95.6\%}
\DefMacro{results-f-f-brnn-test-AVG-accuracy-V-L}{NaN}
\DefMacro{results-f-f-brnn-test-AVG-accuracy-V-V}{99.6\%}
\DefMacro{results-f-f-brnn-test-AVG-accuracy-all}{96.8\%}
\DefMacro{results-f-f-brnn-test-AVG-accuracy-betsent}{67.4\%}
\DefMacro{results-f-f-brnn-test-AVG-accuracy-insent}{99.2\%}
\DefMacro{results-f-f-brnn-test-AVG-accuracy-insent-cs}{99.0\%}
\DefMacro{results-f-f-brnn-test-AVG-accuracy-insent-noncs}{99.3\%}
\DefMacro{results-f-f-brnn-test-AVG-accuracy-top-2}{99.2\%}
\DefMacro{results-f-f-brnn-test-AVG-accuracy-top-3}{99.7\%}
\DefMacro{results-f-f-brnn-test-AVG-accuracy-top-5}{99.9\%}
\DefMacro{results-f-f-brnn-test-AVG-correct-G}{9180700.0\%}
\DefMacro{results-f-f-brnn-test-AVG-correct-G-G}{6917300.0\%}
\DefMacro{results-f-f-brnn-test-AVG-correct-G-L}{1748066.7\%}
\DefMacro{results-f-f-brnn-test-AVG-correct-G-V}{524300.0\%}
\DefMacro{results-f-f-brnn-test-AVG-correct-L}{6648800.0\%}
\DefMacro{results-f-f-brnn-test-AVG-correct-L-G}{1760100.0\%}
\DefMacro{results-f-f-brnn-test-AVG-correct-L-L}{4456800.0\%}
\DefMacro{results-f-f-brnn-test-AVG-correct-L-V}{0.0\%}
\DefMacro{results-f-f-brnn-test-AVG-correct-V}{2648466.7\%}
\DefMacro{results-f-f-brnn-test-AVG-correct-V-G}{503300.0\%}
\DefMacro{results-f-f-brnn-test-AVG-correct-V-L}{0.0\%}
\DefMacro{results-f-f-brnn-test-AVG-correct-V-V}{1591366.7\%}
\DefMacro{results-f-f-brnn-test-AVG-correct-all}{18477966.7\%}
\DefMacro{results-f-f-brnn-test-AVG-correct-betsent}{976733.3\%}
\DefMacro{results-f-f-brnn-test-AVG-correct-insent}{17501233.3\%}
\DefMacro{results-f-f-brnn-test-AVG-correct-insent-cs}{4535766.7\%}
\DefMacro{results-f-f-brnn-test-AVG-correct-insent-noncs}{12965466.7\%}
\DefMacro{results-f-f-brnn-test-AVG-correct-top-2}{18937800.0\%}
\DefMacro{results-f-f-brnn-test-AVG-correct-top-3}{19035933.3\%}
\DefMacro{results-f-f-brnn-test-AVG-correct-top-5}{19075166.7\%}
\DefMacro{results-f-f-brnn-test-AVG-count-G}{9271000.0\%}
\DefMacro{results-f-f-brnn-test-AVG-count-G-G}{6980500.0\%}
\DefMacro{results-f-f-brnn-test-AVG-count-G-L}{1763900.0\%}
\DefMacro{results-f-f-brnn-test-AVG-count-G-V}{525900.0\%}
\DefMacro{results-f-f-brnn-test-AVG-count-L}{7033200.0\%}
\DefMacro{results-f-f-brnn-test-AVG-count-L-G}{1764000.0\%}
\DefMacro{results-f-f-brnn-test-AVG-count-L-L}{4483700.0\%}
\DefMacro{results-f-f-brnn-test-AVG-count-L-V}{0.0\%}
\DefMacro{results-f-f-brnn-test-AVG-count-V}{2786900.0\%}
\DefMacro{results-f-f-brnn-test-AVG-count-V-G}{526500.0\%}
\DefMacro{results-f-f-brnn-test-AVG-count-V-L}{0.0\%}
\DefMacro{results-f-f-brnn-test-AVG-count-V-V}{1598500.0\%}
\DefMacro{results-f-f-brnn-test-AVG-count-all}{19091100.0\%}
\DefMacro{results-f-f-brnn-test-AVG-count-betsent}{1448100.0\%}
\DefMacro{results-f-f-brnn-test-AVG-count-insent}{17643000.0\%}
\DefMacro{results-f-f-brnn-test-AVG-count-insent-cs}{4580300.0\%}
\DefMacro{results-f-f-brnn-test-AVG-count-insent-noncs}{13062700.0\%}
\DefMacro{results-f-f-brnn-test-AVG-count-top-2}{19091100.0\%}
\DefMacro{results-f-f-brnn-test-AVG-count-top-3}{19091100.0\%}
\DefMacro{results-f-f-brnn-test-AVG-count-top-5}{19091100.0\%}
\DefMacro{results-f-f-brnn-test-MAX-accuracy-G}{99.1\%}
\DefMacro{results-f-f-brnn-test-MAX-accuracy-G-G}{99.1\%}
\DefMacro{results-f-f-brnn-test-MAX-accuracy-G-L}{99.1\%}
\DefMacro{results-f-f-brnn-test-MAX-accuracy-G-V}{99.8\%}
\DefMacro{results-f-f-brnn-test-MAX-accuracy-L}{94.6\%}
\DefMacro{results-f-f-brnn-test-MAX-accuracy-L-G}{99.8\%}
\DefMacro{results-f-f-brnn-test-MAX-accuracy-L-L}{99.4\%}
\DefMacro{results-f-f-brnn-test-MAX-accuracy-L-V}{NaN}
\DefMacro{results-f-f-brnn-test-MAX-accuracy-V}{95.3\%}
\DefMacro{results-f-f-brnn-test-MAX-accuracy-V-G}{96.0\%}
\DefMacro{results-f-f-brnn-test-MAX-accuracy-V-L}{NaN}
\DefMacro{results-f-f-brnn-test-MAX-accuracy-V-V}{99.6\%}
\DefMacro{results-f-f-brnn-test-MAX-accuracy-all}{96.9\%}
\DefMacro{results-f-f-brnn-test-MAX-accuracy-betsent}{68.3\%}
\DefMacro{results-f-f-brnn-test-MAX-accuracy-insent}{99.2\%}
\DefMacro{results-f-f-brnn-test-MAX-accuracy-insent-cs}{99.1\%}
\DefMacro{results-f-f-brnn-test-MAX-accuracy-insent-noncs}{99.3\%}
\DefMacro{results-f-f-brnn-test-MAX-accuracy-top-2}{99.2\%}
\DefMacro{results-f-f-brnn-test-MAX-accuracy-top-3}{99.7\%}
\DefMacro{results-f-f-brnn-test-MAX-accuracy-top-5}{99.9\%}
\DefMacro{results-f-f-brnn-test-MAX-correct-G}{91854}
\DefMacro{results-f-f-brnn-test-MAX-correct-G-G}{69205}
\DefMacro{results-f-f-brnn-test-MAX-correct-G-L}{17484}
\DefMacro{results-f-f-brnn-test-MAX-correct-G-V}{5246}
\DefMacro{results-f-f-brnn-test-MAX-correct-L}{66530}
\DefMacro{results-f-f-brnn-test-MAX-correct-L-G}{17605}
\DefMacro{results-f-f-brnn-test-MAX-correct-L-L}{44576}
\DefMacro{results-f-f-brnn-test-MAX-correct-L-V}{0}
\DefMacro{results-f-f-brnn-test-MAX-correct-V}{26562}
\DefMacro{results-f-f-brnn-test-MAX-correct-V-G}{5055}
\DefMacro{results-f-f-brnn-test-MAX-correct-V-L}{0}
\DefMacro{results-f-f-brnn-test-MAX-correct-V-V}{15924}
\DefMacro{results-f-f-brnn-test-MAX-correct-all}{184946}
\DefMacro{results-f-f-brnn-test-MAX-correct-betsent}{9884}
\DefMacro{results-f-f-brnn-test-MAX-correct-insent}{175062}
\DefMacro{results-f-f-brnn-test-MAX-correct-insent-cs}{45375}
\DefMacro{results-f-f-brnn-test-MAX-correct-insent-noncs}{129687}
\DefMacro{results-f-f-brnn-test-MAX-correct-top-2}{189406}
\DefMacro{results-f-f-brnn-test-MAX-correct-top-3}{190366}
\DefMacro{results-f-f-brnn-test-MAX-correct-top-5}{190760}
\DefMacro{results-f-f-brnn-test-MAX-count-G}{92710}
\DefMacro{results-f-f-brnn-test-MAX-count-G-G}{69805}
\DefMacro{results-f-f-brnn-test-MAX-count-G-L}{17639}
\DefMacro{results-f-f-brnn-test-MAX-count-G-V}{5259}
\DefMacro{results-f-f-brnn-test-MAX-count-L}{70332}
\DefMacro{results-f-f-brnn-test-MAX-count-L-G}{17640}
\DefMacro{results-f-f-brnn-test-MAX-count-L-L}{44837}
\DefMacro{results-f-f-brnn-test-MAX-count-L-V}{0}
\DefMacro{results-f-f-brnn-test-MAX-count-V}{27869}
\DefMacro{results-f-f-brnn-test-MAX-count-V-G}{5265}
\DefMacro{results-f-f-brnn-test-MAX-count-V-L}{0}
\DefMacro{results-f-f-brnn-test-MAX-count-V-V}{15985}
\DefMacro{results-f-f-brnn-test-MAX-count-all}{190911}
\DefMacro{results-f-f-brnn-test-MAX-count-betsent}{14481}
\DefMacro{results-f-f-brnn-test-MAX-count-insent}{176430}
\DefMacro{results-f-f-brnn-test-MAX-count-insent-cs}{45803}
\DefMacro{results-f-f-brnn-test-MAX-count-insent-noncs}{130627}
\DefMacro{results-f-f-brnn-test-MAX-count-top-2}{190911}
\DefMacro{results-f-f-brnn-test-MAX-count-top-3}{190911}
\DefMacro{results-f-f-brnn-test-MAX-count-top-5}{190911}
\DefMacro{results-f-f-brnn-test-MEDIAN-accuracy-G}{99.0\%}
\DefMacro{results-f-f-brnn-test-MEDIAN-accuracy-G-G}{99.1\%}
\DefMacro{results-f-f-brnn-test-MEDIAN-accuracy-G-L}{99.1\%}
\DefMacro{results-f-f-brnn-test-MEDIAN-accuracy-G-V}{99.7\%}
\DefMacro{results-f-f-brnn-test-MEDIAN-accuracy-L}{94.5\%}
\DefMacro{results-f-f-brnn-test-MEDIAN-accuracy-L-G}{99.8\%}
\DefMacro{results-f-f-brnn-test-MEDIAN-accuracy-L-L}{99.4\%}
\DefMacro{results-f-f-brnn-test-MEDIAN-accuracy-L-V}{NaN}
\DefMacro{results-f-f-brnn-test-MEDIAN-accuracy-V}{95.2\%}
\DefMacro{results-f-f-brnn-test-MEDIAN-accuracy-V-G}{95.6\%}
\DefMacro{results-f-f-brnn-test-MEDIAN-accuracy-V-L}{NaN}
\DefMacro{results-f-f-brnn-test-MEDIAN-accuracy-V-V}{99.5\%}
\DefMacro{results-f-f-brnn-test-MEDIAN-accuracy-all}{96.8\%}
\DefMacro{results-f-f-brnn-test-MEDIAN-accuracy-betsent}{67.4\%}
\DefMacro{results-f-f-brnn-test-MEDIAN-accuracy-insent}{99.2\%}
\DefMacro{results-f-f-brnn-test-MEDIAN-accuracy-insent-cs}{99.0\%}
\DefMacro{results-f-f-brnn-test-MEDIAN-accuracy-insent-noncs}{99.2\%}
\DefMacro{results-f-f-brnn-test-MEDIAN-accuracy-top-2}{99.2\%}
\DefMacro{results-f-f-brnn-test-MEDIAN-accuracy-top-3}{99.7\%}
\DefMacro{results-f-f-brnn-test-MEDIAN-accuracy-top-5}{99.9\%}
\DefMacro{results-f-f-brnn-test-MEDIAN-correct-G}{9179000.0\%}
\DefMacro{results-f-f-brnn-test-MEDIAN-correct-G-G}{6916200.0\%}
\DefMacro{results-f-f-brnn-test-MEDIAN-correct-G-L}{1748000.0\%}
\DefMacro{results-f-f-brnn-test-MEDIAN-correct-G-V}{524200.0\%}
\DefMacro{results-f-f-brnn-test-MEDIAN-correct-L}{6649500.0\%}
\DefMacro{results-f-f-brnn-test-MEDIAN-correct-L-G}{1760400.0\%}
\DefMacro{results-f-f-brnn-test-MEDIAN-correct-L-L}{4457000.0\%}
\DefMacro{results-f-f-brnn-test-MEDIAN-correct-L-V}{0.0\%}
\DefMacro{results-f-f-brnn-test-MEDIAN-correct-V}{2651800.0\%}
\DefMacro{results-f-f-brnn-test-MEDIAN-correct-V-G}{503400.0\%}
\DefMacro{results-f-f-brnn-test-MEDIAN-correct-V-L}{0.0\%}
\DefMacro{results-f-f-brnn-test-MEDIAN-correct-V-V}{1591200.0\%}
\DefMacro{results-f-f-brnn-test-MEDIAN-correct-all}{18473400.0\%}
\DefMacro{results-f-f-brnn-test-MEDIAN-correct-betsent}{975600.0\%}
\DefMacro{results-f-f-brnn-test-MEDIAN-correct-insent}{17499700.0\%}
\DefMacro{results-f-f-brnn-test-MEDIAN-correct-insent-cs}{4535700.0\%}
\DefMacro{results-f-f-brnn-test-MEDIAN-correct-insent-noncs}{12964000.0\%}
\DefMacro{results-f-f-brnn-test-MEDIAN-correct-top-2}{18936500.0\%}
\DefMacro{results-f-f-brnn-test-MEDIAN-correct-top-3}{19036300.0\%}
\DefMacro{results-f-f-brnn-test-MEDIAN-correct-top-5}{19074900.0\%}
\DefMacro{results-f-f-brnn-test-MEDIAN-count-G}{9271000.0\%}
\DefMacro{results-f-f-brnn-test-MEDIAN-count-G-G}{6980500.0\%}
\DefMacro{results-f-f-brnn-test-MEDIAN-count-G-L}{1763900.0\%}
\DefMacro{results-f-f-brnn-test-MEDIAN-count-G-V}{525900.0\%}
\DefMacro{results-f-f-brnn-test-MEDIAN-count-L}{7033200.0\%}
\DefMacro{results-f-f-brnn-test-MEDIAN-count-L-G}{1764000.0\%}
\DefMacro{results-f-f-brnn-test-MEDIAN-count-L-L}{4483700.0\%}
\DefMacro{results-f-f-brnn-test-MEDIAN-count-L-V}{0.0\%}
\DefMacro{results-f-f-brnn-test-MEDIAN-count-V}{2786900.0\%}
\DefMacro{results-f-f-brnn-test-MEDIAN-count-V-G}{526500.0\%}
\DefMacro{results-f-f-brnn-test-MEDIAN-count-V-L}{0.0\%}
\DefMacro{results-f-f-brnn-test-MEDIAN-count-V-V}{1598500.0\%}
\DefMacro{results-f-f-brnn-test-MEDIAN-count-all}{19091100.0\%}
\DefMacro{results-f-f-brnn-test-MEDIAN-count-betsent}{1448100.0\%}
\DefMacro{results-f-f-brnn-test-MEDIAN-count-insent}{17643000.0\%}
\DefMacro{results-f-f-brnn-test-MEDIAN-count-insent-cs}{4580300.0\%}
\DefMacro{results-f-f-brnn-test-MEDIAN-count-insent-noncs}{13062700.0\%}
\DefMacro{results-f-f-brnn-test-MEDIAN-count-top-2}{19091100.0\%}
\DefMacro{results-f-f-brnn-test-MEDIAN-count-top-3}{19091100.0\%}
\DefMacro{results-f-f-brnn-test-MEDIAN-count-top-5}{19091100.0\%}
\DefMacro{results-f-f-brnn-test-MIN-accuracy-G}{99.0\%}
\DefMacro{results-f-f-brnn-test-MIN-accuracy-G-G}{99.1\%}
\DefMacro{results-f-f-brnn-test-MIN-accuracy-G-L}{99.1\%}
\DefMacro{results-f-f-brnn-test-MIN-accuracy-G-V}{99.7\%}
\DefMacro{results-f-f-brnn-test-MIN-accuracy-L}{94.5\%}
\DefMacro{results-f-f-brnn-test-MIN-accuracy-L-G}{99.7\%}
\DefMacro{results-f-f-brnn-test-MIN-accuracy-L-L}{99.4\%}
\DefMacro{results-f-f-brnn-test-MIN-accuracy-L-V}{NaN}
\DefMacro{results-f-f-brnn-test-MIN-accuracy-V}{94.6\%}
\DefMacro{results-f-f-brnn-test-MIN-accuracy-V-G}{95.2\%}
\DefMacro{results-f-f-brnn-test-MIN-accuracy-V-L}{NaN}
\DefMacro{results-f-f-brnn-test-MIN-accuracy-V-V}{99.5\%}
\DefMacro{results-f-f-brnn-test-MIN-accuracy-all}{96.7\%}
\DefMacro{results-f-f-brnn-test-MIN-accuracy-betsent}{66.7\%}
\DefMacro{results-f-f-brnn-test-MIN-accuracy-insent}{99.2\%}
\DefMacro{results-f-f-brnn-test-MIN-accuracy-insent-cs}{99.0\%}
\DefMacro{results-f-f-brnn-test-MIN-accuracy-insent-noncs}{99.2\%}
\DefMacro{results-f-f-brnn-test-MIN-accuracy-top-2}{99.2\%}
\DefMacro{results-f-f-brnn-test-MIN-accuracy-top-3}{99.7\%}
\DefMacro{results-f-f-brnn-test-MIN-accuracy-top-5}{99.9\%}
\DefMacro{results-f-f-brnn-test-MIN-correct-G}{91777}
\DefMacro{results-f-f-brnn-test-MIN-correct-G-G}{69152}
\DefMacro{results-f-f-brnn-test-MIN-correct-G-L}{17478}
\DefMacro{results-f-f-brnn-test-MIN-correct-G-V}{5241}
\DefMacro{results-f-f-brnn-test-MIN-correct-L}{66439}
\DefMacro{results-f-f-brnn-test-MIN-correct-L-G}{17594}
\DefMacro{results-f-f-brnn-test-MIN-correct-L-L}{44558}
\DefMacro{results-f-f-brnn-test-MIN-correct-L-V}{0}
\DefMacro{results-f-f-brnn-test-MIN-correct-V}{26374}
\DefMacro{results-f-f-brnn-test-MIN-correct-V-G}{5010}
\DefMacro{results-f-f-brnn-test-MIN-correct-V-L}{0}
\DefMacro{results-f-f-brnn-test-MIN-correct-V-V}{15905}
\DefMacro{results-f-f-brnn-test-MIN-correct-all}{184659}
\DefMacro{results-f-f-brnn-test-MIN-correct-betsent}{9662}
\DefMacro{results-f-f-brnn-test-MIN-correct-insent}{174978}
\DefMacro{results-f-f-brnn-test-MIN-correct-insent-cs}{45341}
\DefMacro{results-f-f-brnn-test-MIN-correct-insent-noncs}{129637}
\DefMacro{results-f-f-brnn-test-MIN-correct-top-2}{189363}
\DefMacro{results-f-f-brnn-test-MIN-correct-top-3}{190349}
\DefMacro{results-f-f-brnn-test-MIN-correct-top-5}{190746}
\DefMacro{results-f-f-brnn-test-MIN-count-G}{92710}
\DefMacro{results-f-f-brnn-test-MIN-count-G-G}{69805}
\DefMacro{results-f-f-brnn-test-MIN-count-G-L}{17639}
\DefMacro{results-f-f-brnn-test-MIN-count-G-V}{5259}
\DefMacro{results-f-f-brnn-test-MIN-count-L}{70332}
\DefMacro{results-f-f-brnn-test-MIN-count-L-G}{17640}
\DefMacro{results-f-f-brnn-test-MIN-count-L-L}{44837}
\DefMacro{results-f-f-brnn-test-MIN-count-L-V}{0}
\DefMacro{results-f-f-brnn-test-MIN-count-V}{27869}
\DefMacro{results-f-f-brnn-test-MIN-count-V-G}{5265}
\DefMacro{results-f-f-brnn-test-MIN-count-V-L}{0}
\DefMacro{results-f-f-brnn-test-MIN-count-V-V}{15985}
\DefMacro{results-f-f-brnn-test-MIN-count-all}{190911}
\DefMacro{results-f-f-brnn-test-MIN-count-betsent}{14481}
\DefMacro{results-f-f-brnn-test-MIN-count-insent}{176430}
\DefMacro{results-f-f-brnn-test-MIN-count-insent-cs}{45803}
\DefMacro{results-f-f-brnn-test-MIN-count-insent-noncs}{130627}
\DefMacro{results-f-f-brnn-test-MIN-count-top-2}{190911}
\DefMacro{results-f-f-brnn-test-MIN-count-top-3}{190911}
\DefMacro{results-f-f-brnn-test-MIN-count-top-5}{190911}
\DefMacro{results-f-f-brnn-test-STDEV-accuracy-G}{0.0\%}
\DefMacro{results-f-f-brnn-test-STDEV-accuracy-G-G}{0.0\%}
\DefMacro{results-f-f-brnn-test-STDEV-accuracy-G-L}{0.0\%}
\DefMacro{results-f-f-brnn-test-STDEV-accuracy-G-V}{0.0\%}
\DefMacro{results-f-f-brnn-test-STDEV-accuracy-L}{0.1\%}
\DefMacro{results-f-f-brnn-test-STDEV-accuracy-L-G}{0.0\%}
\DefMacro{results-f-f-brnn-test-STDEV-accuracy-L-L}{0.0\%}
\DefMacro{results-f-f-brnn-test-STDEV-accuracy-L-V}{NaN}
\DefMacro{results-f-f-brnn-test-STDEV-accuracy-V}{0.3\%}
\DefMacro{results-f-f-brnn-test-STDEV-accuracy-V-G}{0.3\%}
\DefMacro{results-f-f-brnn-test-STDEV-accuracy-V-L}{NaN}
\DefMacro{results-f-f-brnn-test-STDEV-accuracy-V-V}{0.0\%}
\DefMacro{results-f-f-brnn-test-STDEV-accuracy-all}{0.1\%}
\DefMacro{results-f-f-brnn-test-STDEV-accuracy-betsent}{0.6\%}
\DefMacro{results-f-f-brnn-test-STDEV-accuracy-insent}{0.0\%}
\DefMacro{results-f-f-brnn-test-STDEV-accuracy-insent-cs}{0.0\%}
\DefMacro{results-f-f-brnn-test-STDEV-accuracy-insent-noncs}{0.0\%}
\DefMacro{results-f-f-brnn-test-STDEV-accuracy-top-2}{0.0\%}
\DefMacro{results-f-f-brnn-test-STDEV-accuracy-top-3}{0.0\%}
\DefMacro{results-f-f-brnn-test-STDEV-accuracy-top-5}{0.0\%}
\DefMacro{results-f-f-brnn-test-STDEV-correct-G}{3365.5\%}
\DefMacro{results-f-f-brnn-test-STDEV-correct-G-G}{2299.3\%}
\DefMacro{results-f-f-brnn-test-STDEV-correct-G-L}{249.4\%}
\DefMacro{results-f-f-brnn-test-STDEV-correct-G-V}{216.0\%}
\DefMacro{results-f-f-brnn-test-STDEV-correct-L}{3747.9\%}
\DefMacro{results-f-f-brnn-test-STDEV-correct-L-G}{496.7\%}
\DefMacro{results-f-f-brnn-test-STDEV-correct-L-L}{748.3\%}
\DefMacro{results-f-f-brnn-test-STDEV-correct-L-V}{0.0\%}
\DefMacro{results-f-f-brnn-test-STDEV-correct-V}{8028.8\%}
\DefMacro{results-f-f-brnn-test-STDEV-correct-V-G}{1838.5\%}
\DefMacro{results-f-f-brnn-test-STDEV-correct-V-L}{0.0\%}
\DefMacro{results-f-f-brnn-test-STDEV-correct-V-V}{784.6\%}
\DefMacro{results-f-f-brnn-test-STDEV-correct-all}{12153.6\%}
\DefMacro{results-f-f-brnn-test-STDEV-correct-betsent}{9098.5\%}
\DefMacro{results-f-f-brnn-test-STDEV-correct-insent}{3596.6\%}
\DefMacro{results-f-f-brnn-test-STDEV-correct-insent-cs}{1388.8\%}
\DefMacro{results-f-f-brnn-test-STDEV-correct-insent-noncs}{2289.6\%}
\DefMacro{results-f-f-brnn-test-STDEV-correct-top-2}{1981.6\%}
\DefMacro{results-f-f-brnn-test-STDEV-correct-top-3}{740.9\%}
\DefMacro{results-f-f-brnn-test-STDEV-correct-top-5}{601.8\%}
\DefMacro{results-f-f-brnn-test-STDEV-count-G}{0.0\%}
\DefMacro{results-f-f-brnn-test-STDEV-count-G-G}{0.0\%}
\DefMacro{results-f-f-brnn-test-STDEV-count-G-L}{0.0\%}
\DefMacro{results-f-f-brnn-test-STDEV-count-G-V}{0.0\%}
\DefMacro{results-f-f-brnn-test-STDEV-count-L}{0.0\%}
\DefMacro{results-f-f-brnn-test-STDEV-count-L-G}{0.0\%}
\DefMacro{results-f-f-brnn-test-STDEV-count-L-L}{0.0\%}
\DefMacro{results-f-f-brnn-test-STDEV-count-L-V}{0.0\%}
\DefMacro{results-f-f-brnn-test-STDEV-count-V}{0.0\%}
\DefMacro{results-f-f-brnn-test-STDEV-count-V-G}{0.0\%}
\DefMacro{results-f-f-brnn-test-STDEV-count-V-L}{0.0\%}
\DefMacro{results-f-f-brnn-test-STDEV-count-V-V}{0.0\%}
\DefMacro{results-f-f-brnn-test-STDEV-count-all}{0.0\%}
\DefMacro{results-f-f-brnn-test-STDEV-count-betsent}{0.0\%}
\DefMacro{results-f-f-brnn-test-STDEV-count-insent}{0.0\%}
\DefMacro{results-f-f-brnn-test-STDEV-count-insent-cs}{0.0\%}
\DefMacro{results-f-f-brnn-test-STDEV-count-insent-noncs}{0.0\%}
\DefMacro{results-f-f-brnn-test-STDEV-count-top-2}{0.0\%}
\DefMacro{results-f-f-brnn-test-STDEV-count-top-3}{0.0\%}
\DefMacro{results-f-f-brnn-test-STDEV-count-top-5}{0.0\%}
\DefMacro{results-f-f-brnn-test-SUM-accuracy-G}{297.1\%}
\DefMacro{results-f-f-brnn-test-SUM-accuracy-G-G}{297.3\%}
\DefMacro{results-f-f-brnn-test-SUM-accuracy-G-L}{297.3\%}
\DefMacro{results-f-f-brnn-test-SUM-accuracy-G-V}{299.1\%}
\DefMacro{results-f-f-brnn-test-SUM-accuracy-L}{283.6\%}
\DefMacro{results-f-f-brnn-test-SUM-accuracy-L-G}{299.3\%}
\DefMacro{results-f-f-brnn-test-SUM-accuracy-L-L}{298.2\%}
\DefMacro{results-f-f-brnn-test-SUM-accuracy-L-V}{NaN}
\DefMacro{results-f-f-brnn-test-SUM-accuracy-V}{285.1\%}
\DefMacro{results-f-f-brnn-test-SUM-accuracy-V-G}{286.8\%}
\DefMacro{results-f-f-brnn-test-SUM-accuracy-V-L}{NaN}
\DefMacro{results-f-f-brnn-test-SUM-accuracy-V-V}{298.7\%}
\DefMacro{results-f-f-brnn-test-SUM-accuracy-all}{290.4\%}
\DefMacro{results-f-f-brnn-test-SUM-accuracy-betsent}{202.3\%}
\DefMacro{results-f-f-brnn-test-SUM-accuracy-insent}{297.6\%}
\DefMacro{results-f-f-brnn-test-SUM-accuracy-insent-cs}{297.1\%}
\DefMacro{results-f-f-brnn-test-SUM-accuracy-insent-noncs}{297.8\%}
\DefMacro{results-f-f-brnn-test-SUM-accuracy-top-2}{297.6\%}
\DefMacro{results-f-f-brnn-test-SUM-accuracy-top-3}{299.1\%}
\DefMacro{results-f-f-brnn-test-SUM-accuracy-top-5}{299.7\%}
\DefMacro{results-f-f-brnn-test-SUM-correct-G}{275421}
\DefMacro{results-f-f-brnn-test-SUM-correct-G-G}{207519}
\DefMacro{results-f-f-brnn-test-SUM-correct-G-L}{52442}
\DefMacro{results-f-f-brnn-test-SUM-correct-G-V}{15729}
\DefMacro{results-f-f-brnn-test-SUM-correct-L}{199464}
\DefMacro{results-f-f-brnn-test-SUM-correct-L-G}{52803}
\DefMacro{results-f-f-brnn-test-SUM-correct-L-L}{133704}
\DefMacro{results-f-f-brnn-test-SUM-correct-L-V}{0}
\DefMacro{results-f-f-brnn-test-SUM-correct-V}{79454}
\DefMacro{results-f-f-brnn-test-SUM-correct-V-G}{15099}
\DefMacro{results-f-f-brnn-test-SUM-correct-V-L}{0}
\DefMacro{results-f-f-brnn-test-SUM-correct-V-V}{47741}
\DefMacro{results-f-f-brnn-test-SUM-correct-all}{554339}
\DefMacro{results-f-f-brnn-test-SUM-correct-betsent}{29302}
\DefMacro{results-f-f-brnn-test-SUM-correct-insent}{525037}
\DefMacro{results-f-f-brnn-test-SUM-correct-insent-cs}{136073}
\DefMacro{results-f-f-brnn-test-SUM-correct-insent-noncs}{388964}
\DefMacro{results-f-f-brnn-test-SUM-correct-top-2}{568134}
\DefMacro{results-f-f-brnn-test-SUM-correct-top-3}{571078}
\DefMacro{results-f-f-brnn-test-SUM-correct-top-5}{572255}
\DefMacro{results-f-f-brnn-test-SUM-count-G}{278130}
\DefMacro{results-f-f-brnn-test-SUM-count-G-G}{209415}
\DefMacro{results-f-f-brnn-test-SUM-count-G-L}{52917}
\DefMacro{results-f-f-brnn-test-SUM-count-G-V}{15777}
\DefMacro{results-f-f-brnn-test-SUM-count-L}{210996}
\DefMacro{results-f-f-brnn-test-SUM-count-L-G}{52920}
\DefMacro{results-f-f-brnn-test-SUM-count-L-L}{134511}
\DefMacro{results-f-f-brnn-test-SUM-count-L-V}{0}
\DefMacro{results-f-f-brnn-test-SUM-count-V}{83607}
\DefMacro{results-f-f-brnn-test-SUM-count-V-G}{15795}
\DefMacro{results-f-f-brnn-test-SUM-count-V-L}{0}
\DefMacro{results-f-f-brnn-test-SUM-count-V-V}{47955}
\DefMacro{results-f-f-brnn-test-SUM-count-all}{572733}
\DefMacro{results-f-f-brnn-test-SUM-count-betsent}{43443}
\DefMacro{results-f-f-brnn-test-SUM-count-insent}{529290}
\DefMacro{results-f-f-brnn-test-SUM-count-insent-cs}{137409}
\DefMacro{results-f-f-brnn-test-SUM-count-insent-noncs}{391881}
\DefMacro{results-f-f-brnn-test-SUM-count-top-2}{572733}
\DefMacro{results-f-f-brnn-test-SUM-count-top-3}{572733}
\DefMacro{results-f-f-brnn-test-SUM-count-top-5}{572733}
\DefMacro{results-f-f-rnn-val-AVG-accuracy-G}{98.4\%}
\DefMacro{results-f-f-rnn-val-AVG-accuracy-G-G}{98.5\%}
\DefMacro{results-f-f-rnn-val-AVG-accuracy-G-L}{99.1\%}
\DefMacro{results-f-f-rnn-val-AVG-accuracy-G-V}{99.6\%}
\DefMacro{results-f-f-rnn-val-AVG-accuracy-L}{93.7\%}
\DefMacro{results-f-f-rnn-val-AVG-accuracy-L-G}{99.9\%}
\DefMacro{results-f-f-rnn-val-AVG-accuracy-L-L}{98.9\%}
\DefMacro{results-f-f-rnn-val-AVG-accuracy-L-V}{NaN}
\DefMacro{results-f-f-rnn-val-AVG-accuracy-V}{94.0\%}
\DefMacro{results-f-f-rnn-val-AVG-accuracy-V-G}{94.1\%}
\DefMacro{results-f-f-rnn-val-AVG-accuracy-V-L}{NaN}
\DefMacro{results-f-f-rnn-val-AVG-accuracy-V-V}{99.2\%}
\DefMacro{results-f-f-rnn-val-AVG-accuracy-all}{95.9\%}
\DefMacro{results-f-f-rnn-val-AVG-accuracy-betsent}{65.7\%}
\DefMacro{results-f-f-rnn-val-AVG-accuracy-insent}{98.7\%}
\DefMacro{results-f-f-rnn-val-AVG-accuracy-insent-cs}{98.6\%}
\DefMacro{results-f-f-rnn-val-AVG-accuracy-insent-noncs}{98.8\%}
\DefMacro{results-f-f-rnn-val-AVG-accuracy-top-2}{99.0\%}
\DefMacro{results-f-f-rnn-val-AVG-accuracy-top-3}{99.7\%}
\DefMacro{results-f-f-rnn-val-AVG-accuracy-top-5}{99.9\%}
\DefMacro{results-f-f-rnn-val-AVG-correct-G}{7398633.3\%}
\DefMacro{results-f-f-rnn-val-AVG-correct-G-G}{5577433.3\%}
\DefMacro{results-f-f-rnn-val-AVG-correct-G-L}{1239533.3\%}
\DefMacro{results-f-f-rnn-val-AVG-correct-G-V}{605033.3\%}
\DefMacro{results-f-f-rnn-val-AVG-correct-L}{4853200.0\%}
\DefMacro{results-f-f-rnn-val-AVG-correct-L-G}{1248800.0\%}
\DefMacro{results-f-f-rnn-val-AVG-correct-L-L}{3348966.7\%}
\DefMacro{results-f-f-rnn-val-AVG-correct-L-V}{0.0\%}
\DefMacro{results-f-f-rnn-val-AVG-correct-V}{3603800.0\%}
\DefMacro{results-f-f-rnn-val-AVG-correct-V-G}{572400.0\%}
\DefMacro{results-f-f-rnn-val-AVG-correct-V-L}{0.0\%}
\DefMacro{results-f-f-rnn-val-AVG-correct-V-V}{2333133.3\%}
\DefMacro{results-f-f-rnn-val-AVG-correct-all}{15855633.3\%}
\DefMacro{results-f-f-rnn-val-AVG-correct-betsent}{930333.3\%}
\DefMacro{results-f-f-rnn-val-AVG-correct-insent}{14925300.0\%}
\DefMacro{results-f-f-rnn-val-AVG-correct-insent-cs}{3665766.7\%}
\DefMacro{results-f-f-rnn-val-AVG-correct-insent-noncs}{11259533.3\%}
\DefMacro{results-f-f-rnn-val-AVG-correct-top-2}{16360466.7\%}
\DefMacro{results-f-f-rnn-val-AVG-correct-top-3}{16478466.7\%}
\DefMacro{results-f-f-rnn-val-AVG-correct-top-5}{16518866.7\%}
\DefMacro{results-f-f-rnn-val-AVG-count-G}{7519200.0\%}
\DefMacro{results-f-f-rnn-val-AVG-count-G-G}{5660400.0\%}
\DefMacro{results-f-f-rnn-val-AVG-count-G-L}{1250500.0\%}
\DefMacro{results-f-f-rnn-val-AVG-count-G-V}{607600.0\%}
\DefMacro{results-f-f-rnn-val-AVG-count-L}{5179400.0\%}
\DefMacro{results-f-f-rnn-val-AVG-count-L-G}{1250600.0\%}
\DefMacro{results-f-f-rnn-val-AVG-count-L-L}{3386800.0\%}
\DefMacro{results-f-f-rnn-val-AVG-count-L-V}{0.0\%}
\DefMacro{results-f-f-rnn-val-AVG-count-V}{3833800.0\%}
\DefMacro{results-f-f-rnn-val-AVG-count-V-G}{608200.0\%}
\DefMacro{results-f-f-rnn-val-AVG-count-V-L}{0.0\%}
\DefMacro{results-f-f-rnn-val-AVG-count-V-V}{2352800.0\%}
\DefMacro{results-f-f-rnn-val-AVG-count-all}{16532400.0\%}
\DefMacro{results-f-f-rnn-val-AVG-count-betsent}{1415500.0\%}
\DefMacro{results-f-f-rnn-val-AVG-count-insent}{15116900.0\%}
\DefMacro{results-f-f-rnn-val-AVG-count-insent-cs}{3716900.0\%}
\DefMacro{results-f-f-rnn-val-AVG-count-insent-noncs}{11400000.0\%}
\DefMacro{results-f-f-rnn-val-AVG-count-top-2}{16532400.0\%}
\DefMacro{results-f-f-rnn-val-AVG-count-top-3}{16532400.0\%}
\DefMacro{results-f-f-rnn-val-AVG-count-top-5}{16532400.0\%}
\DefMacro{results-f-f-rnn-val-MAX-accuracy-G}{98.4\%}
\DefMacro{results-f-f-rnn-val-MAX-accuracy-G-G}{98.6\%}
\DefMacro{results-f-f-rnn-val-MAX-accuracy-G-L}{99.2\%}
\DefMacro{results-f-f-rnn-val-MAX-accuracy-G-V}{99.6\%}
\DefMacro{results-f-f-rnn-val-MAX-accuracy-L}{93.8\%}
\DefMacro{results-f-f-rnn-val-MAX-accuracy-L-G}{99.9\%}
\DefMacro{results-f-f-rnn-val-MAX-accuracy-L-L}{99.0\%}
\DefMacro{results-f-f-rnn-val-MAX-accuracy-L-V}{NaN}
\DefMacro{results-f-f-rnn-val-MAX-accuracy-V}{95.6\%}
\DefMacro{results-f-f-rnn-val-MAX-accuracy-V-G}{94.2\%}
\DefMacro{results-f-f-rnn-val-MAX-accuracy-V-L}{NaN}
\DefMacro{results-f-f-rnn-val-MAX-accuracy-V-V}{99.2\%}
\DefMacro{results-f-f-rnn-val-MAX-accuracy-all}{96.3\%}
\DefMacro{results-f-f-rnn-val-MAX-accuracy-betsent}{70.0\%}
\DefMacro{results-f-f-rnn-val-MAX-accuracy-insent}{98.8\%}
\DefMacro{results-f-f-rnn-val-MAX-accuracy-insent-cs}{98.7\%}
\DefMacro{results-f-f-rnn-val-MAX-accuracy-insent-noncs}{98.8\%}
\DefMacro{results-f-f-rnn-val-MAX-accuracy-top-2}{99.1\%}
\DefMacro{results-f-f-rnn-val-MAX-accuracy-top-3}{99.7\%}
\DefMacro{results-f-f-rnn-val-MAX-accuracy-top-5}{99.9\%}
\DefMacro{results-f-f-rnn-val-MAX-correct-G}{74014}
\DefMacro{results-f-f-rnn-val-MAX-correct-G-G}{55800}
\DefMacro{results-f-f-rnn-val-MAX-correct-G-L}{12410}
\DefMacro{results-f-f-rnn-val-MAX-correct-G-V}{6053}
\DefMacro{results-f-f-rnn-val-MAX-correct-L}{48557}
\DefMacro{results-f-f-rnn-val-MAX-correct-L-G}{12492}
\DefMacro{results-f-f-rnn-val-MAX-correct-L-L}{33513}
\DefMacro{results-f-f-rnn-val-MAX-correct-L-V}{0}
\DefMacro{results-f-f-rnn-val-MAX-correct-V}{36643}
\DefMacro{results-f-f-rnn-val-MAX-correct-V-G}{5732}
\DefMacro{results-f-f-rnn-val-MAX-correct-V-L}{0}
\DefMacro{results-f-f-rnn-val-MAX-correct-V-V}{23342}
\DefMacro{results-f-f-rnn-val-MAX-correct-all}{159133}
\DefMacro{results-f-f-rnn-val-MAX-correct-betsent}{9911}
\DefMacro{results-f-f-rnn-val-MAX-correct-insent}{149297}
\DefMacro{results-f-f-rnn-val-MAX-correct-insent-cs}{36681}
\DefMacro{results-f-f-rnn-val-MAX-correct-insent-noncs}{112629}
\DefMacro{results-f-f-rnn-val-MAX-correct-top-2}{163822}
\DefMacro{results-f-f-rnn-val-MAX-correct-top-3}{164808}
\DefMacro{results-f-f-rnn-val-MAX-correct-top-5}{165197}
\DefMacro{results-f-f-rnn-val-MAX-count-G}{75192}
\DefMacro{results-f-f-rnn-val-MAX-count-G-G}{56604}
\DefMacro{results-f-f-rnn-val-MAX-count-G-L}{12505}
\DefMacro{results-f-f-rnn-val-MAX-count-G-V}{6076}
\DefMacro{results-f-f-rnn-val-MAX-count-L}{51794}
\DefMacro{results-f-f-rnn-val-MAX-count-L-G}{12506}
\DefMacro{results-f-f-rnn-val-MAX-count-L-L}{33868}
\DefMacro{results-f-f-rnn-val-MAX-count-L-V}{0}
\DefMacro{results-f-f-rnn-val-MAX-count-V}{38338}
\DefMacro{results-f-f-rnn-val-MAX-count-V-G}{6082}
\DefMacro{results-f-f-rnn-val-MAX-count-V-L}{0}
\DefMacro{results-f-f-rnn-val-MAX-count-V-V}{23528}
\DefMacro{results-f-f-rnn-val-MAX-count-all}{165324}
\DefMacro{results-f-f-rnn-val-MAX-count-betsent}{14155}
\DefMacro{results-f-f-rnn-val-MAX-count-insent}{151169}
\DefMacro{results-f-f-rnn-val-MAX-count-insent-cs}{37169}
\DefMacro{results-f-f-rnn-val-MAX-count-insent-noncs}{114000}
\DefMacro{results-f-f-rnn-val-MAX-count-top-2}{165324}
\DefMacro{results-f-f-rnn-val-MAX-count-top-3}{165324}
\DefMacro{results-f-f-rnn-val-MAX-count-top-5}{165324}
\DefMacro{results-f-f-rnn-val-MEDIAN-accuracy-G}{98.4\%}
\DefMacro{results-f-f-rnn-val-MEDIAN-accuracy-G-G}{98.5\%}
\DefMacro{results-f-f-rnn-val-MEDIAN-accuracy-G-L}{99.2\%}
\DefMacro{results-f-f-rnn-val-MEDIAN-accuracy-G-V}{99.6\%}
\DefMacro{results-f-f-rnn-val-MEDIAN-accuracy-L}{93.7\%}
\DefMacro{results-f-f-rnn-val-MEDIAN-accuracy-L-G}{99.9\%}
\DefMacro{results-f-f-rnn-val-MEDIAN-accuracy-L-L}{98.9\%}
\DefMacro{results-f-f-rnn-val-MEDIAN-accuracy-L-V}{NaN}
\DefMacro{results-f-f-rnn-val-MEDIAN-accuracy-V}{93.7\%}
\DefMacro{results-f-f-rnn-val-MEDIAN-accuracy-V-G}{94.1\%}
\DefMacro{results-f-f-rnn-val-MEDIAN-accuracy-V-L}{NaN}
\DefMacro{results-f-f-rnn-val-MEDIAN-accuracy-V-V}{99.2\%}
\DefMacro{results-f-f-rnn-val-MEDIAN-accuracy-all}{95.9\%}
\DefMacro{results-f-f-rnn-val-MEDIAN-accuracy-betsent}{64.8\%}
\DefMacro{results-f-f-rnn-val-MEDIAN-accuracy-insent}{98.7\%}
\DefMacro{results-f-f-rnn-val-MEDIAN-accuracy-insent-cs}{98.7\%}
\DefMacro{results-f-f-rnn-val-MEDIAN-accuracy-insent-noncs}{98.8\%}
\DefMacro{results-f-f-rnn-val-MEDIAN-accuracy-top-2}{98.9\%}
\DefMacro{results-f-f-rnn-val-MEDIAN-accuracy-top-3}{99.7\%}
\DefMacro{results-f-f-rnn-val-MEDIAN-accuracy-top-5}{99.9\%}
\DefMacro{results-f-f-rnn-val-MEDIAN-correct-G}{7397600.0\%}
\DefMacro{results-f-f-rnn-val-MEDIAN-correct-G-G}{5577100.0\%}
\DefMacro{results-f-f-rnn-val-MEDIAN-correct-G-L}{1240300.0\%}
\DefMacro{results-f-f-rnn-val-MEDIAN-correct-G-V}{605100.0\%}
\DefMacro{results-f-f-rnn-val-MEDIAN-correct-L}{4852500.0\%}
\DefMacro{results-f-f-rnn-val-MEDIAN-correct-L-G}{1248800.0\%}
\DefMacro{results-f-f-rnn-val-MEDIAN-correct-L-L}{3348700.0\%}
\DefMacro{results-f-f-rnn-val-MEDIAN-correct-L-V}{0.0\%}
\DefMacro{results-f-f-rnn-val-MEDIAN-correct-V}{3592600.0\%}
\DefMacro{results-f-f-rnn-val-MEDIAN-correct-V-G}{572600.0\%}
\DefMacro{results-f-f-rnn-val-MEDIAN-correct-V-L}{0.0\%}
\DefMacro{results-f-f-rnn-val-MEDIAN-correct-V-V}{2333200.0\%}
\DefMacro{results-f-f-rnn-val-MEDIAN-correct-all}{15846500.0\%}
\DefMacro{results-f-f-rnn-val-MEDIAN-correct-betsent}{916800.0\%}
\DefMacro{results-f-f-rnn-val-MEDIAN-correct-insent}{14924000.0\%}
\DefMacro{results-f-f-rnn-val-MEDIAN-correct-insent-cs}{3666800.0\%}
\DefMacro{results-f-f-rnn-val-MEDIAN-correct-insent-noncs}{11261600.0\%}
\DefMacro{results-f-f-rnn-val-MEDIAN-correct-top-2}{16351500.0\%}
\DefMacro{results-f-f-rnn-val-MEDIAN-correct-top-3}{16478700.0\%}
\DefMacro{results-f-f-rnn-val-MEDIAN-correct-top-5}{16519000.0\%}
\DefMacro{results-f-f-rnn-val-MEDIAN-count-G}{7519200.0\%}
\DefMacro{results-f-f-rnn-val-MEDIAN-count-G-G}{5660400.0\%}
\DefMacro{results-f-f-rnn-val-MEDIAN-count-G-L}{1250500.0\%}
\DefMacro{results-f-f-rnn-val-MEDIAN-count-G-V}{607600.0\%}
\DefMacro{results-f-f-rnn-val-MEDIAN-count-L}{5179400.0\%}
\DefMacro{results-f-f-rnn-val-MEDIAN-count-L-G}{1250600.0\%}
\DefMacro{results-f-f-rnn-val-MEDIAN-count-L-L}{3386800.0\%}
\DefMacro{results-f-f-rnn-val-MEDIAN-count-L-V}{0.0\%}
\DefMacro{results-f-f-rnn-val-MEDIAN-count-V}{3833800.0\%}
\DefMacro{results-f-f-rnn-val-MEDIAN-count-V-G}{608200.0\%}
\DefMacro{results-f-f-rnn-val-MEDIAN-count-V-L}{0.0\%}
\DefMacro{results-f-f-rnn-val-MEDIAN-count-V-V}{2352800.0\%}
\DefMacro{results-f-f-rnn-val-MEDIAN-count-all}{16532400.0\%}
\DefMacro{results-f-f-rnn-val-MEDIAN-count-betsent}{1415500.0\%}
\DefMacro{results-f-f-rnn-val-MEDIAN-count-insent}{15116900.0\%}
\DefMacro{results-f-f-rnn-val-MEDIAN-count-insent-cs}{3716900.0\%}
\DefMacro{results-f-f-rnn-val-MEDIAN-count-insent-noncs}{11400000.0\%}
\DefMacro{results-f-f-rnn-val-MEDIAN-count-top-2}{16532400.0\%}
\DefMacro{results-f-f-rnn-val-MEDIAN-count-top-3}{16532400.0\%}
\DefMacro{results-f-f-rnn-val-MEDIAN-count-top-5}{16532400.0\%}
\DefMacro{results-f-f-rnn-val-MIN-accuracy-G}{98.4\%}
\DefMacro{results-f-f-rnn-val-MIN-accuracy-G-G}{98.5\%}
\DefMacro{results-f-f-rnn-val-MIN-accuracy-G-L}{98.9\%}
\DefMacro{results-f-f-rnn-val-MIN-accuracy-G-V}{99.5\%}
\DefMacro{results-f-f-rnn-val-MIN-accuracy-L}{93.7\%}
\DefMacro{results-f-f-rnn-val-MIN-accuracy-L-G}{99.8\%}
\DefMacro{results-f-f-rnn-val-MIN-accuracy-L-L}{98.8\%}
\DefMacro{results-f-f-rnn-val-MIN-accuracy-L-V}{NaN}
\DefMacro{results-f-f-rnn-val-MIN-accuracy-V}{92.7\%}
\DefMacro{results-f-f-rnn-val-MIN-accuracy-V-G}{93.9\%}
\DefMacro{results-f-f-rnn-val-MIN-accuracy-V-L}{NaN}
\DefMacro{results-f-f-rnn-val-MIN-accuracy-V-V}{99.1\%}
\DefMacro{results-f-f-rnn-val-MIN-accuracy-all}{95.6\%}
\DefMacro{results-f-f-rnn-val-MIN-accuracy-betsent}{62.4\%}
\DefMacro{results-f-f-rnn-val-MIN-accuracy-insent}{98.7\%}
\DefMacro{results-f-f-rnn-val-MIN-accuracy-insent-cs}{98.5\%}
\DefMacro{results-f-f-rnn-val-MIN-accuracy-insent-noncs}{98.7\%}
\DefMacro{results-f-f-rnn-val-MIN-accuracy-top-2}{98.9\%}
\DefMacro{results-f-f-rnn-val-MIN-accuracy-top-3}{99.7\%}
\DefMacro{results-f-f-rnn-val-MIN-accuracy-top-5}{99.9\%}
\DefMacro{results-f-f-rnn-val-MIN-correct-G}{73969}
\DefMacro{results-f-f-rnn-val-MIN-correct-G-G}{55752}
\DefMacro{results-f-f-rnn-val-MIN-correct-G-L}{12373}
\DefMacro{results-f-f-rnn-val-MIN-correct-G-V}{6047}
\DefMacro{results-f-f-rnn-val-MIN-correct-L}{48514}
\DefMacro{results-f-f-rnn-val-MIN-correct-L-G}{12484}
\DefMacro{results-f-f-rnn-val-MIN-correct-L-L}{33469}
\DefMacro{results-f-f-rnn-val-MIN-correct-L-V}{0}
\DefMacro{results-f-f-rnn-val-MIN-correct-V}{35545}
\DefMacro{results-f-f-rnn-val-MIN-correct-V-G}{5714}
\DefMacro{results-f-f-rnn-val-MIN-correct-V-L}{0}
\DefMacro{results-f-f-rnn-val-MIN-correct-V-V}{23320}
\DefMacro{results-f-f-rnn-val-MIN-correct-all}{158071}
\DefMacro{results-f-f-rnn-val-MIN-correct-betsent}{8831}
\DefMacro{results-f-f-rnn-val-MIN-correct-insent}{149222}
\DefMacro{results-f-f-rnn-val-MIN-correct-insent-cs}{36624}
\DefMacro{results-f-f-rnn-val-MIN-correct-insent-noncs}{112541}
\DefMacro{results-f-f-rnn-val-MIN-correct-top-2}{163477}
\DefMacro{results-f-f-rnn-val-MIN-correct-top-3}{164759}
\DefMacro{results-f-f-rnn-val-MIN-correct-top-5}{165179}
\DefMacro{results-f-f-rnn-val-MIN-count-G}{75192}
\DefMacro{results-f-f-rnn-val-MIN-count-G-G}{56604}
\DefMacro{results-f-f-rnn-val-MIN-count-G-L}{12505}
\DefMacro{results-f-f-rnn-val-MIN-count-G-V}{6076}
\DefMacro{results-f-f-rnn-val-MIN-count-L}{51794}
\DefMacro{results-f-f-rnn-val-MIN-count-L-G}{12506}
\DefMacro{results-f-f-rnn-val-MIN-count-L-L}{33868}
\DefMacro{results-f-f-rnn-val-MIN-count-L-V}{0}
\DefMacro{results-f-f-rnn-val-MIN-count-V}{38338}
\DefMacro{results-f-f-rnn-val-MIN-count-V-G}{6082}
\DefMacro{results-f-f-rnn-val-MIN-count-V-L}{0}
\DefMacro{results-f-f-rnn-val-MIN-count-V-V}{23528}
\DefMacro{results-f-f-rnn-val-MIN-count-all}{165324}
\DefMacro{results-f-f-rnn-val-MIN-count-betsent}{14155}
\DefMacro{results-f-f-rnn-val-MIN-count-insent}{151169}
\DefMacro{results-f-f-rnn-val-MIN-count-insent-cs}{37169}
\DefMacro{results-f-f-rnn-val-MIN-count-insent-noncs}{114000}
\DefMacro{results-f-f-rnn-val-MIN-count-top-2}{165324}
\DefMacro{results-f-f-rnn-val-MIN-count-top-3}{165324}
\DefMacro{results-f-f-rnn-val-MIN-count-top-5}{165324}
\DefMacro{results-f-f-rnn-val-STDEV-accuracy-G}{0.0\%}
\DefMacro{results-f-f-rnn-val-STDEV-accuracy-G-G}{0.0\%}
\DefMacro{results-f-f-rnn-val-STDEV-accuracy-G-L}{0.1\%}
\DefMacro{results-f-f-rnn-val-STDEV-accuracy-G-V}{0.0\%}
\DefMacro{results-f-f-rnn-val-STDEV-accuracy-L}{0.0\%}
\DefMacro{results-f-f-rnn-val-STDEV-accuracy-L-G}{0.0\%}
\DefMacro{results-f-f-rnn-val-STDEV-accuracy-L-L}{0.1\%}
\DefMacro{results-f-f-rnn-val-STDEV-accuracy-L-V}{NaN}
\DefMacro{results-f-f-rnn-val-STDEV-accuracy-V}{1.2\%}
\DefMacro{results-f-f-rnn-val-STDEV-accuracy-V-G}{0.1\%}
\DefMacro{results-f-f-rnn-val-STDEV-accuracy-V-L}{NaN}
\DefMacro{results-f-f-rnn-val-STDEV-accuracy-V-V}{0.0\%}
\DefMacro{results-f-f-rnn-val-STDEV-accuracy-all}{0.3\%}
\DefMacro{results-f-f-rnn-val-STDEV-accuracy-betsent}{3.2\%}
\DefMacro{results-f-f-rnn-val-STDEV-accuracy-insent}{0.0\%}
\DefMacro{results-f-f-rnn-val-STDEV-accuracy-insent-cs}{0.1\%}
\DefMacro{results-f-f-rnn-val-STDEV-accuracy-insent-noncs}{0.0\%}
\DefMacro{results-f-f-rnn-val-STDEV-accuracy-top-2}{0.1\%}
\DefMacro{results-f-f-rnn-val-STDEV-accuracy-top-3}{0.0\%}
\DefMacro{results-f-f-rnn-val-STDEV-accuracy-top-5}{0.0\%}
\DefMacro{results-f-f-rnn-val-STDEV-correct-G}{1977.1\%}
\DefMacro{results-f-f-rnn-val-STDEV-correct-G-G}{1973.7\%}
\DefMacro{results-f-f-rnn-val-STDEV-correct-G-L}{1604.9\%}
\DefMacro{results-f-f-rnn-val-STDEV-correct-G-V}{249.4\%}
\DefMacro{results-f-f-rnn-val-STDEV-correct-L}{1823.9\%}
\DefMacro{results-f-f-rnn-val-STDEV-correct-L-G}{326.6\%}
\DefMacro{results-f-f-rnn-val-STDEV-correct-L-L}{1806.2\%}
\DefMacro{results-f-f-rnn-val-STDEV-correct-L-V}{0.0\%}
\DefMacro{results-f-f-rnn-val-STDEV-correct-V}{45519.9\%}
\DefMacro{results-f-f-rnn-val-STDEV-correct-V-G}{748.3\%}
\DefMacro{results-f-f-rnn-val-STDEV-correct-V-L}{0.0\%}
\DefMacro{results-f-f-rnn-val-STDEV-correct-V-V}{899.4\%}
\DefMacro{results-f-f-rnn-val-STDEV-correct-all}{43834.3\%}
\DefMacro{results-f-f-rnn-val-STDEV-correct-betsent}{45117.4\%}
\DefMacro{results-f-f-rnn-val-STDEV-correct-insent}{3196.9\%}
\DefMacro{results-f-f-rnn-val-STDEV-correct-insent-cs}{2439.0\%}
\DefMacro{results-f-f-rnn-val-STDEV-correct-insent-noncs}{3878.4\%}
\DefMacro{results-f-f-rnn-val-STDEV-correct-top-2}{15445.9\%}
\DefMacro{results-f-f-rnn-val-STDEV-correct-top-3}{2007.2\%}
\DefMacro{results-f-f-rnn-val-STDEV-correct-top-5}{740.9\%}
\DefMacro{results-f-f-rnn-val-STDEV-count-G}{0.0\%}
\DefMacro{results-f-f-rnn-val-STDEV-count-G-G}{0.0\%}
\DefMacro{results-f-f-rnn-val-STDEV-count-G-L}{0.0\%}
\DefMacro{results-f-f-rnn-val-STDEV-count-G-V}{0.0\%}
\DefMacro{results-f-f-rnn-val-STDEV-count-L}{0.0\%}
\DefMacro{results-f-f-rnn-val-STDEV-count-L-G}{0.0\%}
\DefMacro{results-f-f-rnn-val-STDEV-count-L-L}{0.0\%}
\DefMacro{results-f-f-rnn-val-STDEV-count-L-V}{0.0\%}
\DefMacro{results-f-f-rnn-val-STDEV-count-V}{0.0\%}
\DefMacro{results-f-f-rnn-val-STDEV-count-V-G}{0.0\%}
\DefMacro{results-f-f-rnn-val-STDEV-count-V-L}{0.0\%}
\DefMacro{results-f-f-rnn-val-STDEV-count-V-V}{0.0\%}
\DefMacro{results-f-f-rnn-val-STDEV-count-all}{0.0\%}
\DefMacro{results-f-f-rnn-val-STDEV-count-betsent}{0.0\%}
\DefMacro{results-f-f-rnn-val-STDEV-count-insent}{0.0\%}
\DefMacro{results-f-f-rnn-val-STDEV-count-insent-cs}{0.0\%}
\DefMacro{results-f-f-rnn-val-STDEV-count-insent-noncs}{0.0\%}
\DefMacro{results-f-f-rnn-val-STDEV-count-top-2}{0.0\%}
\DefMacro{results-f-f-rnn-val-STDEV-count-top-3}{0.0\%}
\DefMacro{results-f-f-rnn-val-STDEV-count-top-5}{0.0\%}
\DefMacro{results-f-f-rnn-val-SUM-accuracy-G}{295.2\%}
\DefMacro{results-f-f-rnn-val-SUM-accuracy-G-G}{295.6\%}
\DefMacro{results-f-f-rnn-val-SUM-accuracy-G-L}{297.4\%}
\DefMacro{results-f-f-rnn-val-SUM-accuracy-G-V}{298.7\%}
\DefMacro{results-f-f-rnn-val-SUM-accuracy-L}{281.1\%}
\DefMacro{results-f-f-rnn-val-SUM-accuracy-L-G}{299.6\%}
\DefMacro{results-f-f-rnn-val-SUM-accuracy-L-L}{296.6\%}
\DefMacro{results-f-f-rnn-val-SUM-accuracy-L-V}{NaN}
\DefMacro{results-f-f-rnn-val-SUM-accuracy-V}{282.0\%}
\DefMacro{results-f-f-rnn-val-SUM-accuracy-V-G}{282.3\%}
\DefMacro{results-f-f-rnn-val-SUM-accuracy-V-L}{NaN}
\DefMacro{results-f-f-rnn-val-SUM-accuracy-V-V}{297.5\%}
\DefMacro{results-f-f-rnn-val-SUM-accuracy-all}{287.7\%}
\DefMacro{results-f-f-rnn-val-SUM-accuracy-betsent}{197.2\%}
\DefMacro{results-f-f-rnn-val-SUM-accuracy-insent}{296.2\%}
\DefMacro{results-f-f-rnn-val-SUM-accuracy-insent-cs}{295.9\%}
\DefMacro{results-f-f-rnn-val-SUM-accuracy-insent-noncs}{296.3\%}
\DefMacro{results-f-f-rnn-val-SUM-accuracy-top-2}{296.9\%}
\DefMacro{results-f-f-rnn-val-SUM-accuracy-top-3}{299.0\%}
\DefMacro{results-f-f-rnn-val-SUM-accuracy-top-5}{299.8\%}
\DefMacro{results-f-f-rnn-val-SUM-correct-G}{221959}
\DefMacro{results-f-f-rnn-val-SUM-correct-G-G}{167323}
\DefMacro{results-f-f-rnn-val-SUM-correct-G-L}{37186}
\DefMacro{results-f-f-rnn-val-SUM-correct-G-V}{18151}
\DefMacro{results-f-f-rnn-val-SUM-correct-L}{145596}
\DefMacro{results-f-f-rnn-val-SUM-correct-L-G}{37464}
\DefMacro{results-f-f-rnn-val-SUM-correct-L-L}{100469}
\DefMacro{results-f-f-rnn-val-SUM-correct-L-V}{0}
\DefMacro{results-f-f-rnn-val-SUM-correct-V}{108114}
\DefMacro{results-f-f-rnn-val-SUM-correct-V-G}{17172}
\DefMacro{results-f-f-rnn-val-SUM-correct-V-L}{0}
\DefMacro{results-f-f-rnn-val-SUM-correct-V-V}{69994}
\DefMacro{results-f-f-rnn-val-SUM-correct-all}{475669}
\DefMacro{results-f-f-rnn-val-SUM-correct-betsent}{27910}
\DefMacro{results-f-f-rnn-val-SUM-correct-insent}{447759}
\DefMacro{results-f-f-rnn-val-SUM-correct-insent-cs}{109973}
\DefMacro{results-f-f-rnn-val-SUM-correct-insent-noncs}{337786}
\DefMacro{results-f-f-rnn-val-SUM-correct-top-2}{490814}
\DefMacro{results-f-f-rnn-val-SUM-correct-top-3}{494354}
\DefMacro{results-f-f-rnn-val-SUM-correct-top-5}{495566}
\DefMacro{results-f-f-rnn-val-SUM-count-G}{225576}
\DefMacro{results-f-f-rnn-val-SUM-count-G-G}{169812}
\DefMacro{results-f-f-rnn-val-SUM-count-G-L}{37515}
\DefMacro{results-f-f-rnn-val-SUM-count-G-V}{18228}
\DefMacro{results-f-f-rnn-val-SUM-count-L}{155382}
\DefMacro{results-f-f-rnn-val-SUM-count-L-G}{37518}
\DefMacro{results-f-f-rnn-val-SUM-count-L-L}{101604}
\DefMacro{results-f-f-rnn-val-SUM-count-L-V}{0}
\DefMacro{results-f-f-rnn-val-SUM-count-V}{115014}
\DefMacro{results-f-f-rnn-val-SUM-count-V-G}{18246}
\DefMacro{results-f-f-rnn-val-SUM-count-V-L}{0}
\DefMacro{results-f-f-rnn-val-SUM-count-V-V}{70584}
\DefMacro{results-f-f-rnn-val-SUM-count-all}{495972}
\DefMacro{results-f-f-rnn-val-SUM-count-betsent}{42465}
\DefMacro{results-f-f-rnn-val-SUM-count-insent}{453507}
\DefMacro{results-f-f-rnn-val-SUM-count-insent-cs}{111507}
\DefMacro{results-f-f-rnn-val-SUM-count-insent-noncs}{342000}
\DefMacro{results-f-f-rnn-val-SUM-count-top-2}{495972}
\DefMacro{results-f-f-rnn-val-SUM-count-top-3}{495972}
\DefMacro{results-f-f-rnn-val-SUM-count-top-5}{495972}
\DefMacro{results-f-f-rnn-test-AVG-accuracy-G}{98.8\%}
\DefMacro{results-f-f-rnn-test-AVG-accuracy-G-G}{99.0\%}
\DefMacro{results-f-f-rnn-test-AVG-accuracy-G-L}{99.0\%}
\DefMacro{results-f-f-rnn-test-AVG-accuracy-G-V}{99.5\%}
\DefMacro{results-f-f-rnn-test-AVG-accuracy-L}{93.3\%}
\DefMacro{results-f-f-rnn-test-AVG-accuracy-L-G}{99.7\%}
\DefMacro{results-f-f-rnn-test-AVG-accuracy-L-L}{99.1\%}
\DefMacro{results-f-f-rnn-test-AVG-accuracy-L-V}{NaN}
\DefMacro{results-f-f-rnn-test-AVG-accuracy-V}{94.6\%}
\DefMacro{results-f-f-rnn-test-AVG-accuracy-V-G}{92.8\%}
\DefMacro{results-f-f-rnn-test-AVG-accuracy-V-L}{NaN}
\DefMacro{results-f-f-rnn-test-AVG-accuracy-V-V}{99.3\%}
\DefMacro{results-f-f-rnn-test-AVG-accuracy-all}{96.2\%}
\DefMacro{results-f-f-rnn-test-AVG-accuracy-betsent}{62.2\%}
\DefMacro{results-f-f-rnn-test-AVG-accuracy-insent}{99.0\%}
\DefMacro{results-f-f-rnn-test-AVG-accuracy-insent-cs}{98.6\%}
\DefMacro{results-f-f-rnn-test-AVG-accuracy-insent-noncs}{99.1\%}
\DefMacro{results-f-f-rnn-test-AVG-accuracy-top-2}{98.9\%}
\DefMacro{results-f-f-rnn-test-AVG-accuracy-top-3}{99.6\%}
\DefMacro{results-f-f-rnn-test-AVG-accuracy-top-5}{99.9\%}
\DefMacro{results-f-f-rnn-test-AVG-correct-G}{9157933.3\%}
\DefMacro{results-f-f-rnn-test-AVG-correct-G-G}{6910266.7\%}
\DefMacro{results-f-f-rnn-test-AVG-correct-G-L}{1746666.7\%}
\DefMacro{results-f-f-rnn-test-AVG-correct-G-V}{523033.3\%}
\DefMacro{results-f-f-rnn-test-AVG-correct-L}{6565466.7\%}
\DefMacro{results-f-f-rnn-test-AVG-correct-L-G}{1759200.0\%}
\DefMacro{results-f-f-rnn-test-AVG-correct-L-L}{4443766.7\%}
\DefMacro{results-f-f-rnn-test-AVG-correct-L-V}{0.0\%}
\DefMacro{results-f-f-rnn-test-AVG-correct-V}{2635966.7\%}
\DefMacro{results-f-f-rnn-test-AVG-correct-V-G}{488466.7\%}
\DefMacro{results-f-f-rnn-test-AVG-correct-V-L}{0.0\%}
\DefMacro{results-f-f-rnn-test-AVG-correct-V-V}{1586733.3\%}
\DefMacro{results-f-f-rnn-test-AVG-correct-all}{18359366.7\%}
\DefMacro{results-f-f-rnn-test-AVG-correct-betsent}{901233.3\%}
\DefMacro{results-f-f-rnn-test-AVG-correct-insent}{17458133.3\%}
\DefMacro{results-f-f-rnn-test-AVG-correct-insent-cs}{4517366.7\%}
\DefMacro{results-f-f-rnn-test-AVG-correct-insent-noncs}{12940766.7\%}
\DefMacro{results-f-f-rnn-test-AVG-correct-top-2}{18872866.7\%}
\DefMacro{results-f-f-rnn-test-AVG-correct-top-3}{19018300.0\%}
\DefMacro{results-f-f-rnn-test-AVG-correct-top-5}{19073500.0\%}
\DefMacro{results-f-f-rnn-test-AVG-count-G}{9271000.0\%}
\DefMacro{results-f-f-rnn-test-AVG-count-G-G}{6980500.0\%}
\DefMacro{results-f-f-rnn-test-AVG-count-G-L}{1763900.0\%}
\DefMacro{results-f-f-rnn-test-AVG-count-G-V}{525900.0\%}
\DefMacro{results-f-f-rnn-test-AVG-count-L}{7033200.0\%}
\DefMacro{results-f-f-rnn-test-AVG-count-L-G}{1764000.0\%}
\DefMacro{results-f-f-rnn-test-AVG-count-L-L}{4483700.0\%}
\DefMacro{results-f-f-rnn-test-AVG-count-L-V}{0.0\%}
\DefMacro{results-f-f-rnn-test-AVG-count-V}{2786900.0\%}
\DefMacro{results-f-f-rnn-test-AVG-count-V-G}{526500.0\%}
\DefMacro{results-f-f-rnn-test-AVG-count-V-L}{0.0\%}
\DefMacro{results-f-f-rnn-test-AVG-count-V-V}{1598500.0\%}
\DefMacro{results-f-f-rnn-test-AVG-count-all}{19091100.0\%}
\DefMacro{results-f-f-rnn-test-AVG-count-betsent}{1448100.0\%}
\DefMacro{results-f-f-rnn-test-AVG-count-insent}{17643000.0\%}
\DefMacro{results-f-f-rnn-test-AVG-count-insent-cs}{4580300.0\%}
\DefMacro{results-f-f-rnn-test-AVG-count-insent-noncs}{13062700.0\%}
\DefMacro{results-f-f-rnn-test-AVG-count-top-2}{19091100.0\%}
\DefMacro{results-f-f-rnn-test-AVG-count-top-3}{19091100.0\%}
\DefMacro{results-f-f-rnn-test-AVG-count-top-5}{19091100.0\%}
\DefMacro{results-f-f-rnn-test-MAX-accuracy-G}{98.8\%}
\DefMacro{results-f-f-rnn-test-MAX-accuracy-G-G}{99.0\%}
\DefMacro{results-f-f-rnn-test-MAX-accuracy-G-L}{99.1\%}
\DefMacro{results-f-f-rnn-test-MAX-accuracy-G-V}{99.5\%}
\DefMacro{results-f-f-rnn-test-MAX-accuracy-L}{93.4\%}
\DefMacro{results-f-f-rnn-test-MAX-accuracy-L-G}{99.8\%}
\DefMacro{results-f-f-rnn-test-MAX-accuracy-L-L}{99.1\%}
\DefMacro{results-f-f-rnn-test-MAX-accuracy-L-V}{NaN}
\DefMacro{results-f-f-rnn-test-MAX-accuracy-V}{94.9\%}
\DefMacro{results-f-f-rnn-test-MAX-accuracy-V-G}{93.1\%}
\DefMacro{results-f-f-rnn-test-MAX-accuracy-V-L}{NaN}
\DefMacro{results-f-f-rnn-test-MAX-accuracy-V-V}{99.3\%}
\DefMacro{results-f-f-rnn-test-MAX-accuracy-all}{96.2\%}
\DefMacro{results-f-f-rnn-test-MAX-accuracy-betsent}{63.0\%}
\DefMacro{results-f-f-rnn-test-MAX-accuracy-insent}{99.0\%}
\DefMacro{results-f-f-rnn-test-MAX-accuracy-insent-cs}{98.7\%}
\DefMacro{results-f-f-rnn-test-MAX-accuracy-insent-noncs}{99.1\%}
\DefMacro{results-f-f-rnn-test-MAX-accuracy-top-2}{98.9\%}
\DefMacro{results-f-f-rnn-test-MAX-accuracy-top-3}{99.6\%}
\DefMacro{results-f-f-rnn-test-MAX-accuracy-top-5}{99.9\%}
\DefMacro{results-f-f-rnn-test-MAX-correct-G}{91625}
\DefMacro{results-f-f-rnn-test-MAX-correct-G-G}{69141}
\DefMacro{results-f-f-rnn-test-MAX-correct-G-L}{17472}
\DefMacro{results-f-f-rnn-test-MAX-correct-G-V}{5231}
\DefMacro{results-f-f-rnn-test-MAX-correct-L}{65671}
\DefMacro{results-f-f-rnn-test-MAX-correct-L-G}{17599}
\DefMacro{results-f-f-rnn-test-MAX-correct-L-L}{44452}
\DefMacro{results-f-f-rnn-test-MAX-correct-L-V}{0}
\DefMacro{results-f-f-rnn-test-MAX-correct-V}{26442}
\DefMacro{results-f-f-rnn-test-MAX-correct-V-G}{4902}
\DefMacro{results-f-f-rnn-test-MAX-correct-V-L}{0}
\DefMacro{results-f-f-rnn-test-MAX-correct-V-V}{15877}
\DefMacro{results-f-f-rnn-test-MAX-correct-all}{183714}
\DefMacro{results-f-f-rnn-test-MAX-correct-betsent}{9117}
\DefMacro{results-f-f-rnn-test-MAX-correct-insent}{174597}
\DefMacro{results-f-f-rnn-test-MAX-correct-insent-cs}{45204}
\DefMacro{results-f-f-rnn-test-MAX-correct-insent-noncs}{129456}
\DefMacro{results-f-f-rnn-test-MAX-correct-top-2}{188815}
\DefMacro{results-f-f-rnn-test-MAX-correct-top-3}{190201}
\DefMacro{results-f-f-rnn-test-MAX-correct-top-5}{190751}
\DefMacro{results-f-f-rnn-test-MAX-count-G}{92710}
\DefMacro{results-f-f-rnn-test-MAX-count-G-G}{69805}
\DefMacro{results-f-f-rnn-test-MAX-count-G-L}{17639}
\DefMacro{results-f-f-rnn-test-MAX-count-G-V}{5259}
\DefMacro{results-f-f-rnn-test-MAX-count-L}{70332}
\DefMacro{results-f-f-rnn-test-MAX-count-L-G}{17640}
\DefMacro{results-f-f-rnn-test-MAX-count-L-L}{44837}
\DefMacro{results-f-f-rnn-test-MAX-count-L-V}{0}
\DefMacro{results-f-f-rnn-test-MAX-count-V}{27869}
\DefMacro{results-f-f-rnn-test-MAX-count-V-G}{5265}
\DefMacro{results-f-f-rnn-test-MAX-count-V-L}{0}
\DefMacro{results-f-f-rnn-test-MAX-count-V-V}{15985}
\DefMacro{results-f-f-rnn-test-MAX-count-all}{190911}
\DefMacro{results-f-f-rnn-test-MAX-count-betsent}{14481}
\DefMacro{results-f-f-rnn-test-MAX-count-insent}{176430}
\DefMacro{results-f-f-rnn-test-MAX-count-insent-cs}{45803}
\DefMacro{results-f-f-rnn-test-MAX-count-insent-noncs}{130627}
\DefMacro{results-f-f-rnn-test-MAX-count-top-2}{190911}
\DefMacro{results-f-f-rnn-test-MAX-count-top-3}{190911}
\DefMacro{results-f-f-rnn-test-MAX-count-top-5}{190911}
\DefMacro{results-f-f-rnn-test-MEDIAN-accuracy-G}{98.8\%}
\DefMacro{results-f-f-rnn-test-MEDIAN-accuracy-G-G}{99.0\%}
\DefMacro{results-f-f-rnn-test-MEDIAN-accuracy-G-L}{99.0\%}
\DefMacro{results-f-f-rnn-test-MEDIAN-accuracy-G-V}{99.5\%}
\DefMacro{results-f-f-rnn-test-MEDIAN-accuracy-L}{93.3\%}
\DefMacro{results-f-f-rnn-test-MEDIAN-accuracy-L-G}{99.8\%}
\DefMacro{results-f-f-rnn-test-MEDIAN-accuracy-L-L}{99.1\%}
\DefMacro{results-f-f-rnn-test-MEDIAN-accuracy-L-V}{NaN}
\DefMacro{results-f-f-rnn-test-MEDIAN-accuracy-V}{94.6\%}
\DefMacro{results-f-f-rnn-test-MEDIAN-accuracy-V-G}{92.8\%}
\DefMacro{results-f-f-rnn-test-MEDIAN-accuracy-V-L}{NaN}
\DefMacro{results-f-f-rnn-test-MEDIAN-accuracy-V-V}{99.3\%}
\DefMacro{results-f-f-rnn-test-MEDIAN-accuracy-all}{96.2\%}
\DefMacro{results-f-f-rnn-test-MEDIAN-accuracy-betsent}{62.2\%}
\DefMacro{results-f-f-rnn-test-MEDIAN-accuracy-insent}{99.0\%}
\DefMacro{results-f-f-rnn-test-MEDIAN-accuracy-insent-cs}{98.6\%}
\DefMacro{results-f-f-rnn-test-MEDIAN-accuracy-insent-noncs}{99.1\%}
\DefMacro{results-f-f-rnn-test-MEDIAN-accuracy-top-2}{98.8\%}
\DefMacro{results-f-f-rnn-test-MEDIAN-accuracy-top-3}{99.6\%}
\DefMacro{results-f-f-rnn-test-MEDIAN-accuracy-top-5}{99.9\%}
\DefMacro{results-f-f-rnn-test-MEDIAN-correct-G}{9157200.0\%}
\DefMacro{results-f-f-rnn-test-MEDIAN-correct-G-G}{6912700.0\%}
\DefMacro{results-f-f-rnn-test-MEDIAN-correct-G-L}{1746400.0\%}
\DefMacro{results-f-f-rnn-test-MEDIAN-correct-G-V}{523100.0\%}
\DefMacro{results-f-f-rnn-test-MEDIAN-correct-L}{6564700.0\%}
\DefMacro{results-f-f-rnn-test-MEDIAN-correct-L-G}{1759800.0\%}
\DefMacro{results-f-f-rnn-test-MEDIAN-correct-L-L}{4444000.0\%}
\DefMacro{results-f-f-rnn-test-MEDIAN-correct-L-V}{0.0\%}
\DefMacro{results-f-f-rnn-test-MEDIAN-correct-V}{2635100.0\%}
\DefMacro{results-f-f-rnn-test-MEDIAN-correct-V-G}{488600.0\%}
\DefMacro{results-f-f-rnn-test-MEDIAN-correct-V-L}{0.0\%}
\DefMacro{results-f-f-rnn-test-MEDIAN-correct-V-V}{1586700.0\%}
\DefMacro{results-f-f-rnn-test-MEDIAN-correct-all}{18356300.0\%}
\DefMacro{results-f-f-rnn-test-MEDIAN-correct-betsent}{901200.0\%}
\DefMacro{results-f-f-rnn-test-MEDIAN-correct-insent}{17459600.0\%}
\DefMacro{results-f-f-rnn-test-MEDIAN-correct-insent-cs}{4517700.0\%}
\DefMacro{results-f-f-rnn-test-MEDIAN-correct-insent-noncs}{12942000.0\%}
\DefMacro{results-f-f-rnn-test-MEDIAN-correct-top-2}{18870300.0\%}
\DefMacro{results-f-f-rnn-test-MEDIAN-correct-top-3}{19018800.0\%}
\DefMacro{results-f-f-rnn-test-MEDIAN-correct-top-5}{19074800.0\%}
\DefMacro{results-f-f-rnn-test-MEDIAN-count-G}{9271000.0\%}
\DefMacro{results-f-f-rnn-test-MEDIAN-count-G-G}{6980500.0\%}
\DefMacro{results-f-f-rnn-test-MEDIAN-count-G-L}{1763900.0\%}
\DefMacro{results-f-f-rnn-test-MEDIAN-count-G-V}{525900.0\%}
\DefMacro{results-f-f-rnn-test-MEDIAN-count-L}{7033200.0\%}
\DefMacro{results-f-f-rnn-test-MEDIAN-count-L-G}{1764000.0\%}
\DefMacro{results-f-f-rnn-test-MEDIAN-count-L-L}{4483700.0\%}
\DefMacro{results-f-f-rnn-test-MEDIAN-count-L-V}{0.0\%}
\DefMacro{results-f-f-rnn-test-MEDIAN-count-V}{2786900.0\%}
\DefMacro{results-f-f-rnn-test-MEDIAN-count-V-G}{526500.0\%}
\DefMacro{results-f-f-rnn-test-MEDIAN-count-V-L}{0.0\%}
\DefMacro{results-f-f-rnn-test-MEDIAN-count-V-V}{1598500.0\%}
\DefMacro{results-f-f-rnn-test-MEDIAN-count-all}{19091100.0\%}
\DefMacro{results-f-f-rnn-test-MEDIAN-count-betsent}{1448100.0\%}
\DefMacro{results-f-f-rnn-test-MEDIAN-count-insent}{17643000.0\%}
\DefMacro{results-f-f-rnn-test-MEDIAN-count-insent-cs}{4580300.0\%}
\DefMacro{results-f-f-rnn-test-MEDIAN-count-insent-noncs}{13062700.0\%}
\DefMacro{results-f-f-rnn-test-MEDIAN-count-top-2}{19091100.0\%}
\DefMacro{results-f-f-rnn-test-MEDIAN-count-top-3}{19091100.0\%}
\DefMacro{results-f-f-rnn-test-MEDIAN-count-top-5}{19091100.0\%}
\DefMacro{results-f-f-rnn-test-MIN-accuracy-G}{98.7\%}
\DefMacro{results-f-f-rnn-test-MIN-accuracy-G-G}{98.9\%}
\DefMacro{results-f-f-rnn-test-MIN-accuracy-G-L}{99.0\%}
\DefMacro{results-f-f-rnn-test-MIN-accuracy-G-V}{99.4\%}
\DefMacro{results-f-f-rnn-test-MIN-accuracy-L}{93.3\%}
\DefMacro{results-f-f-rnn-test-MIN-accuracy-L-G}{99.7\%}
\DefMacro{results-f-f-rnn-test-MIN-accuracy-L-L}{99.1\%}
\DefMacro{results-f-f-rnn-test-MIN-accuracy-L-V}{NaN}
\DefMacro{results-f-f-rnn-test-MIN-accuracy-V}{94.3\%}
\DefMacro{results-f-f-rnn-test-MIN-accuracy-V-G}{92.4\%}
\DefMacro{results-f-f-rnn-test-MIN-accuracy-V-L}{NaN}
\DefMacro{results-f-f-rnn-test-MIN-accuracy-V-V}{99.2\%}
\DefMacro{results-f-f-rnn-test-MIN-accuracy-all}{96.1\%}
\DefMacro{results-f-f-rnn-test-MIN-accuracy-betsent}{61.5\%}
\DefMacro{results-f-f-rnn-test-MIN-accuracy-insent}{98.9\%}
\DefMacro{results-f-f-rnn-test-MIN-accuracy-insent-cs}{98.6\%}
\DefMacro{results-f-f-rnn-test-MIN-accuracy-insent-noncs}{99.0\%}
\DefMacro{results-f-f-rnn-test-MIN-accuracy-top-2}{98.8\%}
\DefMacro{results-f-f-rnn-test-MIN-accuracy-top-3}{99.6\%}
\DefMacro{results-f-f-rnn-test-MIN-accuracy-top-5}{99.9\%}
\DefMacro{results-f-f-rnn-test-MIN-correct-G}{91541}
\DefMacro{results-f-f-rnn-test-MIN-correct-G-G}{69040}
\DefMacro{results-f-f-rnn-test-MIN-correct-G-L}{17464}
\DefMacro{results-f-f-rnn-test-MIN-correct-G-V}{5229}
\DefMacro{results-f-f-rnn-test-MIN-correct-L}{65646}
\DefMacro{results-f-f-rnn-test-MIN-correct-L-G}{17579}
\DefMacro{results-f-f-rnn-test-MIN-correct-L-L}{44421}
\DefMacro{results-f-f-rnn-test-MIN-correct-L-V}{0}
\DefMacro{results-f-f-rnn-test-MIN-correct-V}{26286}
\DefMacro{results-f-f-rnn-test-MIN-correct-V-G}{4866}
\DefMacro{results-f-f-rnn-test-MIN-correct-V-L}{0}
\DefMacro{results-f-f-rnn-test-MIN-correct-V-V}{15858}
\DefMacro{results-f-f-rnn-test-MIN-correct-all}{183504}
\DefMacro{results-f-f-rnn-test-MIN-correct-betsent}{8908}
\DefMacro{results-f-f-rnn-test-MIN-correct-insent}{174551}
\DefMacro{results-f-f-rnn-test-MIN-correct-insent-cs}{45140}
\DefMacro{results-f-f-rnn-test-MIN-correct-insent-noncs}{129347}
\DefMacro{results-f-f-rnn-test-MIN-correct-top-2}{188668}
\DefMacro{results-f-f-rnn-test-MIN-correct-top-3}{190160}
\DefMacro{results-f-f-rnn-test-MIN-correct-top-5}{190706}
\DefMacro{results-f-f-rnn-test-MIN-count-G}{92710}
\DefMacro{results-f-f-rnn-test-MIN-count-G-G}{69805}
\DefMacro{results-f-f-rnn-test-MIN-count-G-L}{17639}
\DefMacro{results-f-f-rnn-test-MIN-count-G-V}{5259}
\DefMacro{results-f-f-rnn-test-MIN-count-L}{70332}
\DefMacro{results-f-f-rnn-test-MIN-count-L-G}{17640}
\DefMacro{results-f-f-rnn-test-MIN-count-L-L}{44837}
\DefMacro{results-f-f-rnn-test-MIN-count-L-V}{0}
\DefMacro{results-f-f-rnn-test-MIN-count-V}{27869}
\DefMacro{results-f-f-rnn-test-MIN-count-V-G}{5265}
\DefMacro{results-f-f-rnn-test-MIN-count-V-L}{0}
\DefMacro{results-f-f-rnn-test-MIN-count-V-V}{15985}
\DefMacro{results-f-f-rnn-test-MIN-count-all}{190911}
\DefMacro{results-f-f-rnn-test-MIN-count-betsent}{14481}
\DefMacro{results-f-f-rnn-test-MIN-count-insent}{176430}
\DefMacro{results-f-f-rnn-test-MIN-count-insent-cs}{45803}
\DefMacro{results-f-f-rnn-test-MIN-count-insent-noncs}{130627}
\DefMacro{results-f-f-rnn-test-MIN-count-top-2}{190911}
\DefMacro{results-f-f-rnn-test-MIN-count-top-3}{190911}
\DefMacro{results-f-f-rnn-test-MIN-count-top-5}{190911}
\DefMacro{results-f-f-rnn-test-STDEV-accuracy-G}{0.0\%}
\DefMacro{results-f-f-rnn-test-STDEV-accuracy-G-G}{0.1\%}
\DefMacro{results-f-f-rnn-test-STDEV-accuracy-G-L}{0.0\%}
\DefMacro{results-f-f-rnn-test-STDEV-accuracy-G-V}{0.0\%}
\DefMacro{results-f-f-rnn-test-STDEV-accuracy-L}{0.0\%}
\DefMacro{results-f-f-rnn-test-STDEV-accuracy-L-G}{0.1\%}
\DefMacro{results-f-f-rnn-test-STDEV-accuracy-L-L}{0.0\%}
\DefMacro{results-f-f-rnn-test-STDEV-accuracy-L-V}{NaN}
\DefMacro{results-f-f-rnn-test-STDEV-accuracy-V}{0.2\%}
\DefMacro{results-f-f-rnn-test-STDEV-accuracy-V-G}{0.3\%}
\DefMacro{results-f-f-rnn-test-STDEV-accuracy-V-L}{NaN}
\DefMacro{results-f-f-rnn-test-STDEV-accuracy-V-V}{0.0\%}
\DefMacro{results-f-f-rnn-test-STDEV-accuracy-all}{0.0\%}
\DefMacro{results-f-f-rnn-test-STDEV-accuracy-betsent}{0.6\%}
\DefMacro{results-f-f-rnn-test-STDEV-accuracy-insent}{0.0\%}
\DefMacro{results-f-f-rnn-test-STDEV-accuracy-insent-cs}{0.1\%}
\DefMacro{results-f-f-rnn-test-STDEV-accuracy-insent-noncs}{0.0\%}
\DefMacro{results-f-f-rnn-test-STDEV-accuracy-top-2}{0.0\%}
\DefMacro{results-f-f-rnn-test-STDEV-accuracy-top-3}{0.0\%}
\DefMacro{results-f-f-rnn-test-STDEV-accuracy-top-5}{0.0\%}
\DefMacro{results-f-f-rnn-test-STDEV-correct-G}{3468.3\%}
\DefMacro{results-f-f-rnn-test-STDEV-correct-G-G}{4467.9\%}
\DefMacro{results-f-f-rnn-test-STDEV-correct-G-L}{377.1\%}
\DefMacro{results-f-f-rnn-test-STDEV-correct-G-V}{94.3\%}
\DefMacro{results-f-f-rnn-test-STDEV-correct-L}{1155.7\%}
\DefMacro{results-f-f-rnn-test-STDEV-correct-L-G}{920.1\%}
\DefMacro{results-f-f-rnn-test-STDEV-correct-L-L}{1276.3\%}
\DefMacro{results-f-f-rnn-test-STDEV-correct-L-V}{0.0\%}
\DefMacro{results-f-f-rnn-test-STDEV-correct-V}{6398.1\%}
\DefMacro{results-f-f-rnn-test-STDEV-correct-V-G}{1472.7\%}
\DefMacro{results-f-f-rnn-test-STDEV-correct-V-L}{0.0\%}
\DefMacro{results-f-f-rnn-test-STDEV-correct-V-V}{776.0\%}
\DefMacro{results-f-f-rnn-test-STDEV-correct-all}{8843.2\%}
\DefMacro{results-f-f-rnn-test-STDEV-correct-betsent}{8532.4\%}
\DefMacro{results-f-f-rnn-test-STDEV-correct-insent}{2145.3\%}
\DefMacro{results-f-f-rnn-test-STDEV-correct-insent-cs}{2623.4\%}
\DefMacro{results-f-f-rnn-test-STDEV-correct-insent-noncs}{4534.6\%}
\DefMacro{results-f-f-rnn-test-STDEV-correct-top-2}{6269.7\%}
\DefMacro{results-f-f-rnn-test-STDEV-correct-top-3}{1710.8\%}
\DefMacro{results-f-f-rnn-test-STDEV-correct-top-5}{2054.3\%}
\DefMacro{results-f-f-rnn-test-STDEV-count-G}{0.0\%}
\DefMacro{results-f-f-rnn-test-STDEV-count-G-G}{0.0\%}
\DefMacro{results-f-f-rnn-test-STDEV-count-G-L}{0.0\%}
\DefMacro{results-f-f-rnn-test-STDEV-count-G-V}{0.0\%}
\DefMacro{results-f-f-rnn-test-STDEV-count-L}{0.0\%}
\DefMacro{results-f-f-rnn-test-STDEV-count-L-G}{0.0\%}
\DefMacro{results-f-f-rnn-test-STDEV-count-L-L}{0.0\%}
\DefMacro{results-f-f-rnn-test-STDEV-count-L-V}{0.0\%}
\DefMacro{results-f-f-rnn-test-STDEV-count-V}{0.0\%}
\DefMacro{results-f-f-rnn-test-STDEV-count-V-G}{0.0\%}
\DefMacro{results-f-f-rnn-test-STDEV-count-V-L}{0.0\%}
\DefMacro{results-f-f-rnn-test-STDEV-count-V-V}{0.0\%}
\DefMacro{results-f-f-rnn-test-STDEV-count-all}{0.0\%}
\DefMacro{results-f-f-rnn-test-STDEV-count-betsent}{0.0\%}
\DefMacro{results-f-f-rnn-test-STDEV-count-insent}{0.0\%}
\DefMacro{results-f-f-rnn-test-STDEV-count-insent-cs}{0.0\%}
\DefMacro{results-f-f-rnn-test-STDEV-count-insent-noncs}{0.0\%}
\DefMacro{results-f-f-rnn-test-STDEV-count-top-2}{0.0\%}
\DefMacro{results-f-f-rnn-test-STDEV-count-top-3}{0.0\%}
\DefMacro{results-f-f-rnn-test-STDEV-count-top-5}{0.0\%}
\DefMacro{results-f-f-rnn-test-SUM-accuracy-G}{296.3\%}
\DefMacro{results-f-f-rnn-test-SUM-accuracy-G-G}{297.0\%}
\DefMacro{results-f-f-rnn-test-SUM-accuracy-G-L}{297.1\%}
\DefMacro{results-f-f-rnn-test-SUM-accuracy-G-V}{298.4\%}
\DefMacro{results-f-f-rnn-test-SUM-accuracy-L}{280.0\%}
\DefMacro{results-f-f-rnn-test-SUM-accuracy-L-G}{299.2\%}
\DefMacro{results-f-f-rnn-test-SUM-accuracy-L-L}{297.3\%}
\DefMacro{results-f-f-rnn-test-SUM-accuracy-L-V}{NaN}
\DefMacro{results-f-f-rnn-test-SUM-accuracy-V}{283.8\%}
\DefMacro{results-f-f-rnn-test-SUM-accuracy-V-G}{278.3\%}
\DefMacro{results-f-f-rnn-test-SUM-accuracy-V-L}{NaN}
\DefMacro{results-f-f-rnn-test-SUM-accuracy-V-V}{297.8\%}
\DefMacro{results-f-f-rnn-test-SUM-accuracy-all}{288.5\%}
\DefMacro{results-f-f-rnn-test-SUM-accuracy-betsent}{186.7\%}
\DefMacro{results-f-f-rnn-test-SUM-accuracy-insent}{296.9\%}
\DefMacro{results-f-f-rnn-test-SUM-accuracy-insent-cs}{295.9\%}
\DefMacro{results-f-f-rnn-test-SUM-accuracy-insent-noncs}{297.2\%}
\DefMacro{results-f-f-rnn-test-SUM-accuracy-top-2}{296.6\%}
\DefMacro{results-f-f-rnn-test-SUM-accuracy-top-3}{298.9\%}
\DefMacro{results-f-f-rnn-test-SUM-accuracy-top-5}{299.7\%}
\DefMacro{results-f-f-rnn-test-SUM-correct-G}{274738}
\DefMacro{results-f-f-rnn-test-SUM-correct-G-G}{207308}
\DefMacro{results-f-f-rnn-test-SUM-correct-G-L}{52400}
\DefMacro{results-f-f-rnn-test-SUM-correct-G-V}{15691}
\DefMacro{results-f-f-rnn-test-SUM-correct-L}{196964}
\DefMacro{results-f-f-rnn-test-SUM-correct-L-G}{52776}
\DefMacro{results-f-f-rnn-test-SUM-correct-L-L}{133313}
\DefMacro{results-f-f-rnn-test-SUM-correct-L-V}{0}
\DefMacro{results-f-f-rnn-test-SUM-correct-V}{79079}
\DefMacro{results-f-f-rnn-test-SUM-correct-V-G}{14654}
\DefMacro{results-f-f-rnn-test-SUM-correct-V-L}{0}
\DefMacro{results-f-f-rnn-test-SUM-correct-V-V}{47602}
\DefMacro{results-f-f-rnn-test-SUM-correct-all}{550781}
\DefMacro{results-f-f-rnn-test-SUM-correct-betsent}{27037}
\DefMacro{results-f-f-rnn-test-SUM-correct-insent}{523744}
\DefMacro{results-f-f-rnn-test-SUM-correct-insent-cs}{135521}
\DefMacro{results-f-f-rnn-test-SUM-correct-insent-noncs}{388223}
\DefMacro{results-f-f-rnn-test-SUM-correct-top-2}{566186}
\DefMacro{results-f-f-rnn-test-SUM-correct-top-3}{570549}
\DefMacro{results-f-f-rnn-test-SUM-correct-top-5}{572205}
\DefMacro{results-f-f-rnn-test-SUM-count-G}{278130}
\DefMacro{results-f-f-rnn-test-SUM-count-G-G}{209415}
\DefMacro{results-f-f-rnn-test-SUM-count-G-L}{52917}
\DefMacro{results-f-f-rnn-test-SUM-count-G-V}{15777}
\DefMacro{results-f-f-rnn-test-SUM-count-L}{210996}
\DefMacro{results-f-f-rnn-test-SUM-count-L-G}{52920}
\DefMacro{results-f-f-rnn-test-SUM-count-L-L}{134511}
\DefMacro{results-f-f-rnn-test-SUM-count-L-V}{0}
\DefMacro{results-f-f-rnn-test-SUM-count-V}{83607}
\DefMacro{results-f-f-rnn-test-SUM-count-V-G}{15795}
\DefMacro{results-f-f-rnn-test-SUM-count-V-L}{0}
\DefMacro{results-f-f-rnn-test-SUM-count-V-V}{47955}
\DefMacro{results-f-f-rnn-test-SUM-count-all}{572733}
\DefMacro{results-f-f-rnn-test-SUM-count-betsent}{43443}
\DefMacro{results-f-f-rnn-test-SUM-count-insent}{529290}
\DefMacro{results-f-f-rnn-test-SUM-count-insent-cs}{137409}
\DefMacro{results-f-f-rnn-test-SUM-count-insent-noncs}{391881}
\DefMacro{results-f-f-rnn-test-SUM-count-top-2}{572733}
\DefMacro{results-f-f-rnn-test-SUM-count-top-3}{572733}
\DefMacro{results-f-f-rnn-test-SUM-count-top-5}{572733}
\DefMacro{results-f-f-rrnn-val-AVG-accuracy-G}{89.8\%}
\DefMacro{results-f-f-rrnn-val-AVG-accuracy-G-G}{92.5\%}
\DefMacro{results-f-f-rrnn-val-AVG-accuracy-G-L}{93.4\%}
\DefMacro{results-f-f-rrnn-val-AVG-accuracy-G-V}{97.7\%}
\DefMacro{results-f-f-rrnn-val-AVG-accuracy-L}{86.7\%}
\DefMacro{results-f-f-rrnn-val-AVG-accuracy-L-G}{77.4\%}
\DefMacro{results-f-f-rrnn-val-AVG-accuracy-L-L}{89.6\%}
\DefMacro{results-f-f-rrnn-val-AVG-accuracy-L-V}{NaN}
\DefMacro{results-f-f-rrnn-val-AVG-accuracy-V}{93.2\%}
\DefMacro{results-f-f-rrnn-val-AVG-accuracy-V-G}{90.5\%}
\DefMacro{results-f-f-rrnn-val-AVG-accuracy-V-L}{NaN}
\DefMacro{results-f-f-rrnn-val-AVG-accuracy-V-V}{96.5\%}
\DefMacro{results-f-f-rrnn-val-AVG-accuracy-all}{89.7\%}
\DefMacro{results-f-f-rrnn-val-AVG-accuracy-betsent}{70.6\%}
\DefMacro{results-f-f-rrnn-val-AVG-accuracy-insent}{91.4\%}
\DefMacro{results-f-f-rrnn-val-AVG-accuracy-insent-cs}{88.3\%}
\DefMacro{results-f-f-rrnn-val-AVG-accuracy-insent-noncs}{92.5\%}
\DefMacro{results-f-f-rrnn-val-AVG-accuracy-top-2}{98.7\%}
\DefMacro{results-f-f-rrnn-val-AVG-accuracy-top-3}{99.5\%}
\DefMacro{results-f-f-rrnn-val-AVG-accuracy-top-5}{99.9\%}
\DefMacro{results-f-f-rrnn-val-AVG-correct-G}{6755433.3\%}
\DefMacro{results-f-f-rrnn-val-AVG-correct-G-G}{5236533.3\%}
\DefMacro{results-f-f-rrnn-val-AVG-correct-G-L}{1168500.0\%}
\DefMacro{results-f-f-rrnn-val-AVG-correct-G-V}{593366.7\%}
\DefMacro{results-f-f-rrnn-val-AVG-correct-L}{4492633.3\%}
\DefMacro{results-f-f-rrnn-val-AVG-correct-L-G}{968466.7\%}
\DefMacro{results-f-f-rrnn-val-AVG-correct-L-L}{3034333.3\%}
\DefMacro{results-f-f-rrnn-val-AVG-correct-L-V}{0.0\%}
\DefMacro{results-f-f-rrnn-val-AVG-correct-V}{3573366.7\%}
\DefMacro{results-f-f-rrnn-val-AVG-correct-V-G}{550433.3\%}
\DefMacro{results-f-f-rrnn-val-AVG-correct-V-L}{0.0\%}
\DefMacro{results-f-f-rrnn-val-AVG-correct-V-V}{2270500.0\%}
\DefMacro{results-f-f-rrnn-val-AVG-correct-all}{14821433.3\%}
\DefMacro{results-f-f-rrnn-val-AVG-correct-betsent}{999300.0\%}
\DefMacro{results-f-f-rrnn-val-AVG-correct-insent}{13822133.3\%}
\DefMacro{results-f-f-rrnn-val-AVG-correct-insent-cs}{3280766.7\%}
\DefMacro{results-f-f-rrnn-val-AVG-correct-insent-noncs}{10541366.7\%}
\DefMacro{results-f-f-rrnn-val-AVG-correct-top-2}{16320533.3\%}
\DefMacro{results-f-f-rrnn-val-AVG-correct-top-3}{16451066.7\%}
\DefMacro{results-f-f-rrnn-val-AVG-correct-top-5}{16513066.7\%}
\DefMacro{results-f-f-rrnn-val-AVG-count-G}{7519200.0\%}
\DefMacro{results-f-f-rrnn-val-AVG-count-G-G}{5660400.0\%}
\DefMacro{results-f-f-rrnn-val-AVG-count-G-L}{1250500.0\%}
\DefMacro{results-f-f-rrnn-val-AVG-count-G-V}{607600.0\%}
\DefMacro{results-f-f-rrnn-val-AVG-count-L}{5179400.0\%}
\DefMacro{results-f-f-rrnn-val-AVG-count-L-G}{1250600.0\%}
\DefMacro{results-f-f-rrnn-val-AVG-count-L-L}{3386800.0\%}
\DefMacro{results-f-f-rrnn-val-AVG-count-L-V}{0.0\%}
\DefMacro{results-f-f-rrnn-val-AVG-count-V}{3833800.0\%}
\DefMacro{results-f-f-rrnn-val-AVG-count-V-G}{608200.0\%}
\DefMacro{results-f-f-rrnn-val-AVG-count-V-L}{0.0\%}
\DefMacro{results-f-f-rrnn-val-AVG-count-V-V}{2352800.0\%}
\DefMacro{results-f-f-rrnn-val-AVG-count-all}{16532400.0\%}
\DefMacro{results-f-f-rrnn-val-AVG-count-betsent}{1415500.0\%}
\DefMacro{results-f-f-rrnn-val-AVG-count-insent}{15116900.0\%}
\DefMacro{results-f-f-rrnn-val-AVG-count-insent-cs}{3716900.0\%}
\DefMacro{results-f-f-rrnn-val-AVG-count-insent-noncs}{11400000.0\%}
\DefMacro{results-f-f-rrnn-val-AVG-count-top-2}{16532400.0\%}
\DefMacro{results-f-f-rrnn-val-AVG-count-top-3}{16532400.0\%}
\DefMacro{results-f-f-rrnn-val-AVG-count-top-5}{16532400.0\%}
\DefMacro{results-f-f-rrnn-val-MAX-accuracy-G}{90.0\%}
\DefMacro{results-f-f-rrnn-val-MAX-accuracy-G-G}{92.6\%}
\DefMacro{results-f-f-rrnn-val-MAX-accuracy-G-L}{93.6\%}
\DefMacro{results-f-f-rrnn-val-MAX-accuracy-G-V}{98.1\%}
\DefMacro{results-f-f-rrnn-val-MAX-accuracy-L}{86.9\%}
\DefMacro{results-f-f-rrnn-val-MAX-accuracy-L-G}{77.9\%}
\DefMacro{results-f-f-rrnn-val-MAX-accuracy-L-L}{89.9\%}
\DefMacro{results-f-f-rrnn-val-MAX-accuracy-L-V}{NaN}
\DefMacro{results-f-f-rrnn-val-MAX-accuracy-V}{93.7\%}
\DefMacro{results-f-f-rrnn-val-MAX-accuracy-V-G}{90.7\%}
\DefMacro{results-f-f-rrnn-val-MAX-accuracy-V-L}{NaN}
\DefMacro{results-f-f-rrnn-val-MAX-accuracy-V-V}{96.6\%}
\DefMacro{results-f-f-rrnn-val-MAX-accuracy-all}{89.8\%}
\DefMacro{results-f-f-rrnn-val-MAX-accuracy-betsent}{72.7\%}
\DefMacro{results-f-f-rrnn-val-MAX-accuracy-insent}{91.5\%}
\DefMacro{results-f-f-rrnn-val-MAX-accuracy-insent-cs}{88.3\%}
\DefMacro{results-f-f-rrnn-val-MAX-accuracy-insent-noncs}{92.5\%}
\DefMacro{results-f-f-rrnn-val-MAX-accuracy-top-2}{98.8\%}
\DefMacro{results-f-f-rrnn-val-MAX-accuracy-top-3}{99.5\%}
\DefMacro{results-f-f-rrnn-val-MAX-accuracy-top-5}{99.9\%}
\DefMacro{results-f-f-rrnn-val-MAX-correct-G}{67637}
\DefMacro{results-f-f-rrnn-val-MAX-correct-G-G}{52396}
\DefMacro{results-f-f-rrnn-val-MAX-correct-G-L}{11710}
\DefMacro{results-f-f-rrnn-val-MAX-correct-G-V}{5960}
\DefMacro{results-f-f-rrnn-val-MAX-correct-L}{44996}
\DefMacro{results-f-f-rrnn-val-MAX-correct-L-G}{9739}
\DefMacro{results-f-f-rrnn-val-MAX-correct-L-L}{30433}
\DefMacro{results-f-f-rrnn-val-MAX-correct-L-V}{0}
\DefMacro{results-f-f-rrnn-val-MAX-correct-V}{35904}
\DefMacro{results-f-f-rrnn-val-MAX-correct-V-G}{5517}
\DefMacro{results-f-f-rrnn-val-MAX-correct-V-L}{0}
\DefMacro{results-f-f-rrnn-val-MAX-correct-V-V}{22720}
\DefMacro{results-f-f-rrnn-val-MAX-correct-all}{148466}
\DefMacro{results-f-f-rrnn-val-MAX-correct-betsent}{10297}
\DefMacro{results-f-f-rrnn-val-MAX-correct-insent}{138252}
\DefMacro{results-f-f-rrnn-val-MAX-correct-insent-cs}{32826}
\DefMacro{results-f-f-rrnn-val-MAX-correct-insent-noncs}{105460}
\DefMacro{results-f-f-rrnn-val-MAX-correct-top-2}{163299}
\DefMacro{results-f-f-rrnn-val-MAX-correct-top-3}{164578}
\DefMacro{results-f-f-rrnn-val-MAX-correct-top-5}{165136}
\DefMacro{results-f-f-rrnn-val-MAX-count-G}{75192}
\DefMacro{results-f-f-rrnn-val-MAX-count-G-G}{56604}
\DefMacro{results-f-f-rrnn-val-MAX-count-G-L}{12505}
\DefMacro{results-f-f-rrnn-val-MAX-count-G-V}{6076}
\DefMacro{results-f-f-rrnn-val-MAX-count-L}{51794}
\DefMacro{results-f-f-rrnn-val-MAX-count-L-G}{12506}
\DefMacro{results-f-f-rrnn-val-MAX-count-L-L}{33868}
\DefMacro{results-f-f-rrnn-val-MAX-count-L-V}{0}
\DefMacro{results-f-f-rrnn-val-MAX-count-V}{38338}
\DefMacro{results-f-f-rrnn-val-MAX-count-V-G}{6082}
\DefMacro{results-f-f-rrnn-val-MAX-count-V-L}{0}
\DefMacro{results-f-f-rrnn-val-MAX-count-V-V}{23528}
\DefMacro{results-f-f-rrnn-val-MAX-count-all}{165324}
\DefMacro{results-f-f-rrnn-val-MAX-count-betsent}{14155}
\DefMacro{results-f-f-rrnn-val-MAX-count-insent}{151169}
\DefMacro{results-f-f-rrnn-val-MAX-count-insent-cs}{37169}
\DefMacro{results-f-f-rrnn-val-MAX-count-insent-noncs}{114000}
\DefMacro{results-f-f-rrnn-val-MAX-count-top-2}{165324}
\DefMacro{results-f-f-rrnn-val-MAX-count-top-3}{165324}
\DefMacro{results-f-f-rrnn-val-MAX-count-top-5}{165324}
\DefMacro{results-f-f-rrnn-val-MEDIAN-accuracy-G}{89.9\%}
\DefMacro{results-f-f-rrnn-val-MEDIAN-accuracy-G-G}{92.6\%}
\DefMacro{results-f-f-rrnn-val-MEDIAN-accuracy-G-L}{93.4\%}
\DefMacro{results-f-f-rrnn-val-MEDIAN-accuracy-G-V}{97.7\%}
\DefMacro{results-f-f-rrnn-val-MEDIAN-accuracy-L}{86.7\%}
\DefMacro{results-f-f-rrnn-val-MEDIAN-accuracy-L-G}{77.3\%}
\DefMacro{results-f-f-rrnn-val-MEDIAN-accuracy-L-L}{89.5\%}
\DefMacro{results-f-f-rrnn-val-MEDIAN-accuracy-L-V}{NaN}
\DefMacro{results-f-f-rrnn-val-MEDIAN-accuracy-V}{93.5\%}
\DefMacro{results-f-f-rrnn-val-MEDIAN-accuracy-V-G}{90.6\%}
\DefMacro{results-f-f-rrnn-val-MEDIAN-accuracy-V-L}{NaN}
\DefMacro{results-f-f-rrnn-val-MEDIAN-accuracy-V-V}{96.5\%}
\DefMacro{results-f-f-rrnn-val-MEDIAN-accuracy-all}{89.7\%}
\DefMacro{results-f-f-rrnn-val-MEDIAN-accuracy-betsent}{71.0\%}
\DefMacro{results-f-f-rrnn-val-MEDIAN-accuracy-insent}{91.4\%}
\DefMacro{results-f-f-rrnn-val-MEDIAN-accuracy-insent-cs}{88.3\%}
\DefMacro{results-f-f-rrnn-val-MEDIAN-accuracy-insent-noncs}{92.5\%}
\DefMacro{results-f-f-rrnn-val-MEDIAN-accuracy-top-2}{98.7\%}
\DefMacro{results-f-f-rrnn-val-MEDIAN-accuracy-top-3}{99.5\%}
\DefMacro{results-f-f-rrnn-val-MEDIAN-accuracy-top-5}{99.9\%}
\DefMacro{results-f-f-rrnn-val-MEDIAN-correct-G}{6757900.0\%}
\DefMacro{results-f-f-rrnn-val-MEDIAN-correct-G-G}{5238800.0\%}
\DefMacro{results-f-f-rrnn-val-MEDIAN-correct-G-L}{1167400.0\%}
\DefMacro{results-f-f-rrnn-val-MEDIAN-correct-G-V}{593800.0\%}
\DefMacro{results-f-f-rrnn-val-MEDIAN-correct-L}{4492500.0\%}
\DefMacro{results-f-f-rrnn-val-MEDIAN-correct-L-G}{966600.0\%}
\DefMacro{results-f-f-rrnn-val-MEDIAN-correct-L-L}{3032200.0\%}
\DefMacro{results-f-f-rrnn-val-MEDIAN-correct-L-V}{0.0\%}
\DefMacro{results-f-f-rrnn-val-MEDIAN-correct-V}{3584800.0\%}
\DefMacro{results-f-f-rrnn-val-MEDIAN-correct-V-G}{551000.0\%}
\DefMacro{results-f-f-rrnn-val-MEDIAN-correct-V-L}{0.0\%}
\DefMacro{results-f-f-rrnn-val-MEDIAN-correct-V-V}{2271500.0\%}
\DefMacro{results-f-f-rrnn-val-MEDIAN-correct-all}{14829100.0\%}
\DefMacro{results-f-f-rrnn-val-MEDIAN-correct-betsent}{1004800.0\%}
\DefMacro{results-f-f-rrnn-val-MEDIAN-correct-insent}{13824300.0\%}
\DefMacro{results-f-f-rrnn-val-MEDIAN-correct-insent-cs}{3281400.0\%}
\DefMacro{results-f-f-rrnn-val-MEDIAN-correct-insent-noncs}{10543800.0\%}
\DefMacro{results-f-f-rrnn-val-MEDIAN-correct-top-2}{16317400.0\%}
\DefMacro{results-f-f-rrnn-val-MEDIAN-correct-top-3}{16448500.0\%}
\DefMacro{results-f-f-rrnn-val-MEDIAN-correct-top-5}{16512800.0\%}
\DefMacro{results-f-f-rrnn-val-MEDIAN-count-G}{7519200.0\%}
\DefMacro{results-f-f-rrnn-val-MEDIAN-count-G-G}{5660400.0\%}
\DefMacro{results-f-f-rrnn-val-MEDIAN-count-G-L}{1250500.0\%}
\DefMacro{results-f-f-rrnn-val-MEDIAN-count-G-V}{607600.0\%}
\DefMacro{results-f-f-rrnn-val-MEDIAN-count-L}{5179400.0\%}
\DefMacro{results-f-f-rrnn-val-MEDIAN-count-L-G}{1250600.0\%}
\DefMacro{results-f-f-rrnn-val-MEDIAN-count-L-L}{3386800.0\%}
\DefMacro{results-f-f-rrnn-val-MEDIAN-count-L-V}{0.0\%}
\DefMacro{results-f-f-rrnn-val-MEDIAN-count-V}{3833800.0\%}
\DefMacro{results-f-f-rrnn-val-MEDIAN-count-V-G}{608200.0\%}
\DefMacro{results-f-f-rrnn-val-MEDIAN-count-V-L}{0.0\%}
\DefMacro{results-f-f-rrnn-val-MEDIAN-count-V-V}{2352800.0\%}
\DefMacro{results-f-f-rrnn-val-MEDIAN-count-all}{16532400.0\%}
\DefMacro{results-f-f-rrnn-val-MEDIAN-count-betsent}{1415500.0\%}
\DefMacro{results-f-f-rrnn-val-MEDIAN-count-insent}{15116900.0\%}
\DefMacro{results-f-f-rrnn-val-MEDIAN-count-insent-cs}{3716900.0\%}
\DefMacro{results-f-f-rrnn-val-MEDIAN-count-insent-noncs}{11400000.0\%}
\DefMacro{results-f-f-rrnn-val-MEDIAN-count-top-2}{16532400.0\%}
\DefMacro{results-f-f-rrnn-val-MEDIAN-count-top-3}{16532400.0\%}
\DefMacro{results-f-f-rrnn-val-MEDIAN-count-top-5}{16532400.0\%}
\DefMacro{results-f-f-rrnn-val-MIN-accuracy-G}{89.7\%}
\DefMacro{results-f-f-rrnn-val-MIN-accuracy-G-G}{92.4\%}
\DefMacro{results-f-f-rrnn-val-MIN-accuracy-G-L}{93.3\%}
\DefMacro{results-f-f-rrnn-val-MIN-accuracy-G-V}{97.2\%}
\DefMacro{results-f-f-rrnn-val-MIN-accuracy-L}{86.6\%}
\DefMacro{results-f-f-rrnn-val-MIN-accuracy-L-G}{77.2\%}
\DefMacro{results-f-f-rrnn-val-MIN-accuracy-L-L}{89.4\%}
\DefMacro{results-f-f-rrnn-val-MIN-accuracy-L-V}{NaN}
\DefMacro{results-f-f-rrnn-val-MIN-accuracy-V}{92.5\%}
\DefMacro{results-f-f-rrnn-val-MIN-accuracy-V-G}{90.2\%}
\DefMacro{results-f-f-rrnn-val-MIN-accuracy-V-L}{NaN}
\DefMacro{results-f-f-rrnn-val-MIN-accuracy-V-V}{96.4\%}
\DefMacro{results-f-f-rrnn-val-MIN-accuracy-all}{89.5\%}
\DefMacro{results-f-f-rrnn-val-MIN-accuracy-betsent}{68.1\%}
\DefMacro{results-f-f-rrnn-val-MIN-accuracy-insent}{91.4\%}
\DefMacro{results-f-f-rrnn-val-MIN-accuracy-insent-cs}{88.2\%}
\DefMacro{results-f-f-rrnn-val-MIN-accuracy-insent-noncs}{92.4\%}
\DefMacro{results-f-f-rrnn-val-MIN-accuracy-top-2}{98.7\%}
\DefMacro{results-f-f-rrnn-val-MIN-accuracy-top-3}{99.5\%}
\DefMacro{results-f-f-rrnn-val-MIN-accuracy-top-5}{99.9\%}
\DefMacro{results-f-f-rrnn-val-MIN-correct-G}{67447}
\DefMacro{results-f-f-rrnn-val-MIN-correct-G-G}{52312}
\DefMacro{results-f-f-rrnn-val-MIN-correct-G-L}{11671}
\DefMacro{results-f-f-rrnn-val-MIN-correct-G-V}{5903}
\DefMacro{results-f-f-rrnn-val-MIN-correct-L}{44858}
\DefMacro{results-f-f-rrnn-val-MIN-correct-L-G}{9649}
\DefMacro{results-f-f-rrnn-val-MIN-correct-L-L}{30275}
\DefMacro{results-f-f-rrnn-val-MIN-correct-L-V}{0}
\DefMacro{results-f-f-rrnn-val-MIN-correct-V}{35449}
\DefMacro{results-f-f-rrnn-val-MIN-correct-V-G}{5486}
\DefMacro{results-f-f-rrnn-val-MIN-correct-V-L}{0}
\DefMacro{results-f-f-rrnn-val-MIN-correct-V-V}{22680}
\DefMacro{results-f-f-rrnn-val-MIN-correct-all}{147886}
\DefMacro{results-f-f-rrnn-val-MIN-correct-betsent}{9634}
\DefMacro{results-f-f-rrnn-val-MIN-correct-insent}{138169}
\DefMacro{results-f-f-rrnn-val-MIN-correct-insent-cs}{32783}
\DefMacro{results-f-f-rrnn-val-MIN-correct-insent-noncs}{105343}
\DefMacro{results-f-f-rrnn-val-MIN-correct-top-2}{163143}
\DefMacro{results-f-f-rrnn-val-MIN-correct-top-3}{164469}
\DefMacro{results-f-f-rrnn-val-MIN-correct-top-5}{165128}
\DefMacro{results-f-f-rrnn-val-MIN-count-G}{75192}
\DefMacro{results-f-f-rrnn-val-MIN-count-G-G}{56604}
\DefMacro{results-f-f-rrnn-val-MIN-count-G-L}{12505}
\DefMacro{results-f-f-rrnn-val-MIN-count-G-V}{6076}
\DefMacro{results-f-f-rrnn-val-MIN-count-L}{51794}
\DefMacro{results-f-f-rrnn-val-MIN-count-L-G}{12506}
\DefMacro{results-f-f-rrnn-val-MIN-count-L-L}{33868}
\DefMacro{results-f-f-rrnn-val-MIN-count-L-V}{0}
\DefMacro{results-f-f-rrnn-val-MIN-count-V}{38338}
\DefMacro{results-f-f-rrnn-val-MIN-count-V-G}{6082}
\DefMacro{results-f-f-rrnn-val-MIN-count-V-L}{0}
\DefMacro{results-f-f-rrnn-val-MIN-count-V-V}{23528}
\DefMacro{results-f-f-rrnn-val-MIN-count-all}{165324}
\DefMacro{results-f-f-rrnn-val-MIN-count-betsent}{14155}
\DefMacro{results-f-f-rrnn-val-MIN-count-insent}{151169}
\DefMacro{results-f-f-rrnn-val-MIN-count-insent-cs}{37169}
\DefMacro{results-f-f-rrnn-val-MIN-count-insent-noncs}{114000}
\DefMacro{results-f-f-rrnn-val-MIN-count-top-2}{165324}
\DefMacro{results-f-f-rrnn-val-MIN-count-top-3}{165324}
\DefMacro{results-f-f-rrnn-val-MIN-count-top-5}{165324}
\DefMacro{results-f-f-rrnn-val-STDEV-accuracy-G}{0.1\%}
\DefMacro{results-f-f-rrnn-val-STDEV-accuracy-G-G}{0.1\%}
\DefMacro{results-f-f-rrnn-val-STDEV-accuracy-G-L}{0.1\%}
\DefMacro{results-f-f-rrnn-val-STDEV-accuracy-G-V}{0.4\%}
\DefMacro{results-f-f-rrnn-val-STDEV-accuracy-L}{0.1\%}
\DefMacro{results-f-f-rrnn-val-STDEV-accuracy-L-G}{0.3\%}
\DefMacro{results-f-f-rrnn-val-STDEV-accuracy-L-L}{0.2\%}
\DefMacro{results-f-f-rrnn-val-STDEV-accuracy-L-V}{NaN}
\DefMacro{results-f-f-rrnn-val-STDEV-accuracy-V}{0.5\%}
\DefMacro{results-f-f-rrnn-val-STDEV-accuracy-V-G}{0.2\%}
\DefMacro{results-f-f-rrnn-val-STDEV-accuracy-V-L}{NaN}
\DefMacro{results-f-f-rrnn-val-STDEV-accuracy-V-V}{0.1\%}
\DefMacro{results-f-f-rrnn-val-STDEV-accuracy-all}{0.1\%}
\DefMacro{results-f-f-rrnn-val-STDEV-accuracy-betsent}{1.9\%}
\DefMacro{results-f-f-rrnn-val-STDEV-accuracy-insent}{0.0\%}
\DefMacro{results-f-f-rrnn-val-STDEV-accuracy-insent-cs}{0.0\%}
\DefMacro{results-f-f-rrnn-val-STDEV-accuracy-insent-noncs}{0.0\%}
\DefMacro{results-f-f-rrnn-val-STDEV-accuracy-top-2}{0.0\%}
\DefMacro{results-f-f-rrnn-val-STDEV-accuracy-top-3}{0.0\%}
\DefMacro{results-f-f-rrnn-val-STDEV-accuracy-top-5}{0.0\%}
\DefMacro{results-f-f-rrnn-val-STDEV-correct-G}{7950.4\%}
\DefMacro{results-f-f-rrnn-val-STDEV-correct-G-G}{3785.4\%}
\DefMacro{results-f-f-rrnn-val-STDEV-correct-G-L}{1772.0\%}
\DefMacro{results-f-f-rrnn-val-STDEV-correct-G-V}{2347.1\%}
\DefMacro{results-f-f-rrnn-val-STDEV-correct-L}{5634.6\%}
\DefMacro{results-f-f-rrnn-val-STDEV-correct-L-G}{3904.1\%}
\DefMacro{results-f-f-rrnn-val-STDEV-correct-L-L}{6624.4\%}
\DefMacro{results-f-f-rrnn-val-STDEV-correct-L-V}{0.0\%}
\DefMacro{results-f-f-rrnn-val-STDEV-correct-V}{20258.4\%}
\DefMacro{results-f-f-rrnn-val-STDEV-correct-V-G}{1327.5\%}
\DefMacro{results-f-f-rrnn-val-STDEV-correct-V-L}{0.0\%}
\DefMacro{results-f-f-rrnn-val-STDEV-correct-V-V}{1779.5\%}
\DefMacro{results-f-f-rrnn-val-STDEV-correct-all}{24291.1\%}
\DefMacro{results-f-f-rrnn-val-STDEV-correct-betsent}{27344.8\%}
\DefMacro{results-f-f-rrnn-val-STDEV-correct-insent}{3718.7\%}
\DefMacro{results-f-f-rrnn-val-STDEV-correct-insent-cs}{1811.7\%}
\DefMacro{results-f-f-rrnn-val-STDEV-correct-insent-noncs}{5077.0\%}
\DefMacro{results-f-f-rrnn-val-STDEV-correct-top-2}{6743.1\%}
\DefMacro{results-f-f-rrnn-val-STDEV-correct-top-3}{4805.8\%}
\DefMacro{results-f-f-rrnn-val-STDEV-correct-top-5}{377.1\%}
\DefMacro{results-f-f-rrnn-val-STDEV-count-G}{0.0\%}
\DefMacro{results-f-f-rrnn-val-STDEV-count-G-G}{0.0\%}
\DefMacro{results-f-f-rrnn-val-STDEV-count-G-L}{0.0\%}
\DefMacro{results-f-f-rrnn-val-STDEV-count-G-V}{0.0\%}
\DefMacro{results-f-f-rrnn-val-STDEV-count-L}{0.0\%}
\DefMacro{results-f-f-rrnn-val-STDEV-count-L-G}{0.0\%}
\DefMacro{results-f-f-rrnn-val-STDEV-count-L-L}{0.0\%}
\DefMacro{results-f-f-rrnn-val-STDEV-count-L-V}{0.0\%}
\DefMacro{results-f-f-rrnn-val-STDEV-count-V}{0.0\%}
\DefMacro{results-f-f-rrnn-val-STDEV-count-V-G}{0.0\%}
\DefMacro{results-f-f-rrnn-val-STDEV-count-V-L}{0.0\%}
\DefMacro{results-f-f-rrnn-val-STDEV-count-V-V}{0.0\%}
\DefMacro{results-f-f-rrnn-val-STDEV-count-all}{0.0\%}
\DefMacro{results-f-f-rrnn-val-STDEV-count-betsent}{0.0\%}
\DefMacro{results-f-f-rrnn-val-STDEV-count-insent}{0.0\%}
\DefMacro{results-f-f-rrnn-val-STDEV-count-insent-cs}{0.0\%}
\DefMacro{results-f-f-rrnn-val-STDEV-count-insent-noncs}{0.0\%}
\DefMacro{results-f-f-rrnn-val-STDEV-count-top-2}{0.0\%}
\DefMacro{results-f-f-rrnn-val-STDEV-count-top-3}{0.0\%}
\DefMacro{results-f-f-rrnn-val-STDEV-count-top-5}{0.0\%}
\DefMacro{results-f-f-rrnn-val-SUM-accuracy-G}{269.5\%}
\DefMacro{results-f-f-rrnn-val-SUM-accuracy-G-G}{277.5\%}
\DefMacro{results-f-f-rrnn-val-SUM-accuracy-G-L}{280.3\%}
\DefMacro{results-f-f-rrnn-val-SUM-accuracy-G-V}{293.0\%}
\DefMacro{results-f-f-rrnn-val-SUM-accuracy-L}{260.2\%}
\DefMacro{results-f-f-rrnn-val-SUM-accuracy-L-G}{232.3\%}
\DefMacro{results-f-f-rrnn-val-SUM-accuracy-L-L}{268.8\%}
\DefMacro{results-f-f-rrnn-val-SUM-accuracy-L-V}{NaN}
\DefMacro{results-f-f-rrnn-val-SUM-accuracy-V}{279.6\%}
\DefMacro{results-f-f-rrnn-val-SUM-accuracy-V-G}{271.5\%}
\DefMacro{results-f-f-rrnn-val-SUM-accuracy-V-L}{NaN}
\DefMacro{results-f-f-rrnn-val-SUM-accuracy-V-V}{289.5\%}
\DefMacro{results-f-f-rrnn-val-SUM-accuracy-all}{269.0\%}
\DefMacro{results-f-f-rrnn-val-SUM-accuracy-betsent}{211.8\%}
\DefMacro{results-f-f-rrnn-val-SUM-accuracy-insent}{274.3\%}
\DefMacro{results-f-f-rrnn-val-SUM-accuracy-insent-cs}{264.8\%}
\DefMacro{results-f-f-rrnn-val-SUM-accuracy-insent-noncs}{277.4\%}
\DefMacro{results-f-f-rrnn-val-SUM-accuracy-top-2}{296.2\%}
\DefMacro{results-f-f-rrnn-val-SUM-accuracy-top-3}{298.5\%}
\DefMacro{results-f-f-rrnn-val-SUM-accuracy-top-5}{299.6\%}
\DefMacro{results-f-f-rrnn-val-SUM-correct-G}{202663}
\DefMacro{results-f-f-rrnn-val-SUM-correct-G-G}{157096}
\DefMacro{results-f-f-rrnn-val-SUM-correct-G-L}{35055}
\DefMacro{results-f-f-rrnn-val-SUM-correct-G-V}{17801}
\DefMacro{results-f-f-rrnn-val-SUM-correct-L}{134779}
\DefMacro{results-f-f-rrnn-val-SUM-correct-L-G}{29054}
\DefMacro{results-f-f-rrnn-val-SUM-correct-L-L}{91030}
\DefMacro{results-f-f-rrnn-val-SUM-correct-L-V}{0}
\DefMacro{results-f-f-rrnn-val-SUM-correct-V}{107201}
\DefMacro{results-f-f-rrnn-val-SUM-correct-V-G}{16513}
\DefMacro{results-f-f-rrnn-val-SUM-correct-V-L}{0}
\DefMacro{results-f-f-rrnn-val-SUM-correct-V-V}{68115}
\DefMacro{results-f-f-rrnn-val-SUM-correct-all}{444643}
\DefMacro{results-f-f-rrnn-val-SUM-correct-betsent}{29979}
\DefMacro{results-f-f-rrnn-val-SUM-correct-insent}{414664}
\DefMacro{results-f-f-rrnn-val-SUM-correct-insent-cs}{98423}
\DefMacro{results-f-f-rrnn-val-SUM-correct-insent-noncs}{316241}
\DefMacro{results-f-f-rrnn-val-SUM-correct-top-2}{489616}
\DefMacro{results-f-f-rrnn-val-SUM-correct-top-3}{493532}
\DefMacro{results-f-f-rrnn-val-SUM-correct-top-5}{495392}
\DefMacro{results-f-f-rrnn-val-SUM-count-G}{225576}
\DefMacro{results-f-f-rrnn-val-SUM-count-G-G}{169812}
\DefMacro{results-f-f-rrnn-val-SUM-count-G-L}{37515}
\DefMacro{results-f-f-rrnn-val-SUM-count-G-V}{18228}
\DefMacro{results-f-f-rrnn-val-SUM-count-L}{155382}
\DefMacro{results-f-f-rrnn-val-SUM-count-L-G}{37518}
\DefMacro{results-f-f-rrnn-val-SUM-count-L-L}{101604}
\DefMacro{results-f-f-rrnn-val-SUM-count-L-V}{0}
\DefMacro{results-f-f-rrnn-val-SUM-count-V}{115014}
\DefMacro{results-f-f-rrnn-val-SUM-count-V-G}{18246}
\DefMacro{results-f-f-rrnn-val-SUM-count-V-L}{0}
\DefMacro{results-f-f-rrnn-val-SUM-count-V-V}{70584}
\DefMacro{results-f-f-rrnn-val-SUM-count-all}{495972}
\DefMacro{results-f-f-rrnn-val-SUM-count-betsent}{42465}
\DefMacro{results-f-f-rrnn-val-SUM-count-insent}{453507}
\DefMacro{results-f-f-rrnn-val-SUM-count-insent-cs}{111507}
\DefMacro{results-f-f-rrnn-val-SUM-count-insent-noncs}{342000}
\DefMacro{results-f-f-rrnn-val-SUM-count-top-2}{495972}
\DefMacro{results-f-f-rrnn-val-SUM-count-top-3}{495972}
\DefMacro{results-f-f-rrnn-val-SUM-count-top-5}{495972}
\DefMacro{results-f-f-rrnn-test-AVG-accuracy-G}{90.1\%}
\DefMacro{results-f-f-rrnn-test-AVG-accuracy-G-G}{92.6\%}
\DefMacro{results-f-f-rrnn-test-AVG-accuracy-G-L}{93.9\%}
\DefMacro{results-f-f-rrnn-test-AVG-accuracy-G-V}{96.7\%}
\DefMacro{results-f-f-rrnn-test-AVG-accuracy-L}{86.5\%}
\DefMacro{results-f-f-rrnn-test-AVG-accuracy-L-G}{80.1\%}
\DefMacro{results-f-f-rrnn-test-AVG-accuracy-L-L}{90.2\%}
\DefMacro{results-f-f-rrnn-test-AVG-accuracy-L-V}{NaN}
\DefMacro{results-f-f-rrnn-test-AVG-accuracy-V}{92.6\%}
\DefMacro{results-f-f-rrnn-test-AVG-accuracy-V-G}{90.8\%}
\DefMacro{results-f-f-rrnn-test-AVG-accuracy-V-L}{NaN}
\DefMacro{results-f-f-rrnn-test-AVG-accuracy-V-V}{96.1\%}
\DefMacro{results-f-f-rrnn-test-AVG-accuracy-all}{89.1\%}
\DefMacro{results-f-f-rrnn-test-AVG-accuracy-betsent}{63.4\%}
\DefMacro{results-f-f-rrnn-test-AVG-accuracy-insent}{91.2\%}
\DefMacro{results-f-f-rrnn-test-AVG-accuracy-insent-cs}{88.5\%}
\DefMacro{results-f-f-rrnn-test-AVG-accuracy-insent-noncs}{92.2\%}
\DefMacro{results-f-f-rrnn-test-AVG-accuracy-top-2}{98.5\%}
\DefMacro{results-f-f-rrnn-test-AVG-accuracy-top-3}{99.4\%}
\DefMacro{results-f-f-rrnn-test-AVG-accuracy-top-5}{99.9\%}
\DefMacro{results-f-f-rrnn-test-AVG-correct-G}{8352233.3\%}
\DefMacro{results-f-f-rrnn-test-AVG-correct-G-G}{6461200.0\%}
\DefMacro{results-f-f-rrnn-test-AVG-correct-G-L}{1655800.0\%}
\DefMacro{results-f-f-rrnn-test-AVG-correct-G-V}{508366.7\%}
\DefMacro{results-f-f-rrnn-test-AVG-correct-L}{6080566.7\%}
\DefMacro{results-f-f-rrnn-test-AVG-correct-L-G}{1412900.0\%}
\DefMacro{results-f-f-rrnn-test-AVG-correct-L-L}{4042266.7\%}
\DefMacro{results-f-f-rrnn-test-AVG-correct-L-V}{0.0\%}
\DefMacro{results-f-f-rrnn-test-AVG-correct-V}{2579300.0\%}
\DefMacro{results-f-f-rrnn-test-AVG-correct-V-G}{478133.3\%}
\DefMacro{results-f-f-rrnn-test-AVG-correct-V-L}{0.0\%}
\DefMacro{results-f-f-rrnn-test-AVG-correct-V-V}{1535566.7\%}
\DefMacro{results-f-f-rrnn-test-AVG-correct-all}{17012100.0\%}
\DefMacro{results-f-f-rrnn-test-AVG-correct-betsent}{917866.7\%}
\DefMacro{results-f-f-rrnn-test-AVG-correct-insent}{16094233.3\%}
\DefMacro{results-f-f-rrnn-test-AVG-correct-insent-cs}{4055200.0\%}
\DefMacro{results-f-f-rrnn-test-AVG-correct-insent-noncs}{12039033.3\%}
\DefMacro{results-f-f-rrnn-test-AVG-correct-top-2}{18813566.7\%}
\DefMacro{results-f-f-rrnn-test-AVG-correct-top-3}{18980533.3\%}
\DefMacro{results-f-f-rrnn-test-AVG-correct-top-5}{19068900.0\%}
\DefMacro{results-f-f-rrnn-test-AVG-count-G}{9271000.0\%}
\DefMacro{results-f-f-rrnn-test-AVG-count-G-G}{6980500.0\%}
\DefMacro{results-f-f-rrnn-test-AVG-count-G-L}{1763900.0\%}
\DefMacro{results-f-f-rrnn-test-AVG-count-G-V}{525900.0\%}
\DefMacro{results-f-f-rrnn-test-AVG-count-L}{7033200.0\%}
\DefMacro{results-f-f-rrnn-test-AVG-count-L-G}{1764000.0\%}
\DefMacro{results-f-f-rrnn-test-AVG-count-L-L}{4483700.0\%}
\DefMacro{results-f-f-rrnn-test-AVG-count-L-V}{0.0\%}
\DefMacro{results-f-f-rrnn-test-AVG-count-V}{2786900.0\%}
\DefMacro{results-f-f-rrnn-test-AVG-count-V-G}{526500.0\%}
\DefMacro{results-f-f-rrnn-test-AVG-count-V-L}{0.0\%}
\DefMacro{results-f-f-rrnn-test-AVG-count-V-V}{1598500.0\%}
\DefMacro{results-f-f-rrnn-test-AVG-count-all}{19091100.0\%}
\DefMacro{results-f-f-rrnn-test-AVG-count-betsent}{1448100.0\%}
\DefMacro{results-f-f-rrnn-test-AVG-count-insent}{17643000.0\%}
\DefMacro{results-f-f-rrnn-test-AVG-count-insent-cs}{4580300.0\%}
\DefMacro{results-f-f-rrnn-test-AVG-count-insent-noncs}{13062700.0\%}
\DefMacro{results-f-f-rrnn-test-AVG-count-top-2}{19091100.0\%}
\DefMacro{results-f-f-rrnn-test-AVG-count-top-3}{19091100.0\%}
\DefMacro{results-f-f-rrnn-test-AVG-count-top-5}{19091100.0\%}
\DefMacro{results-f-f-rrnn-test-MAX-accuracy-G}{90.2\%}
\DefMacro{results-f-f-rrnn-test-MAX-accuracy-G-G}{92.7\%}
\DefMacro{results-f-f-rrnn-test-MAX-accuracy-G-L}{94.1\%}
\DefMacro{results-f-f-rrnn-test-MAX-accuracy-G-V}{97.4\%}
\DefMacro{results-f-f-rrnn-test-MAX-accuracy-L}{86.6\%}
\DefMacro{results-f-f-rrnn-test-MAX-accuracy-L-G}{80.5\%}
\DefMacro{results-f-f-rrnn-test-MAX-accuracy-L-L}{90.2\%}
\DefMacro{results-f-f-rrnn-test-MAX-accuracy-L-V}{NaN}
\DefMacro{results-f-f-rrnn-test-MAX-accuracy-V}{92.7\%}
\DefMacro{results-f-f-rrnn-test-MAX-accuracy-V-G}{91.5\%}
\DefMacro{results-f-f-rrnn-test-MAX-accuracy-V-L}{NaN}
\DefMacro{results-f-f-rrnn-test-MAX-accuracy-V-V}{96.1\%}
\DefMacro{results-f-f-rrnn-test-MAX-accuracy-all}{89.2\%}
\DefMacro{results-f-f-rrnn-test-MAX-accuracy-betsent}{64.2\%}
\DefMacro{results-f-f-rrnn-test-MAX-accuracy-insent}{91.2\%}
\DefMacro{results-f-f-rrnn-test-MAX-accuracy-insent-cs}{88.8\%}
\DefMacro{results-f-f-rrnn-test-MAX-accuracy-insent-noncs}{92.2\%}
\DefMacro{results-f-f-rrnn-test-MAX-accuracy-top-2}{98.6\%}
\DefMacro{results-f-f-rrnn-test-MAX-accuracy-top-3}{99.5\%}
\DefMacro{results-f-f-rrnn-test-MAX-accuracy-top-5}{99.9\%}
\DefMacro{results-f-f-rrnn-test-MAX-correct-G}{83605}
\DefMacro{results-f-f-rrnn-test-MAX-correct-G-G}{64702}
\DefMacro{results-f-f-rrnn-test-MAX-correct-G-L}{16607}
\DefMacro{results-f-f-rrnn-test-MAX-correct-G-V}{5123}
\DefMacro{results-f-f-rrnn-test-MAX-correct-L}{60907}
\DefMacro{results-f-f-rrnn-test-MAX-correct-L-G}{14206}
\DefMacro{results-f-f-rrnn-test-MAX-correct-L-L}{40448}
\DefMacro{results-f-f-rrnn-test-MAX-correct-L-V}{0}
\DefMacro{results-f-f-rrnn-test-MAX-correct-V}{25824}
\DefMacro{results-f-f-rrnn-test-MAX-correct-V-G}{4817}
\DefMacro{results-f-f-rrnn-test-MAX-correct-V-L}{0}
\DefMacro{results-f-f-rrnn-test-MAX-correct-V-V}{15365}
\DefMacro{results-f-f-rrnn-test-MAX-correct-all}{170279}
\DefMacro{results-f-f-rrnn-test-MAX-correct-betsent}{9292}
\DefMacro{results-f-f-rrnn-test-MAX-correct-insent}{160987}
\DefMacro{results-f-f-rrnn-test-MAX-correct-insent-cs}{40689}
\DefMacro{results-f-f-rrnn-test-MAX-correct-insent-noncs}{120500}
\DefMacro{results-f-f-rrnn-test-MAX-correct-top-2}{188211}
\DefMacro{results-f-f-rrnn-test-MAX-correct-top-3}{189863}
\DefMacro{results-f-f-rrnn-test-MAX-correct-top-5}{190693}
\DefMacro{results-f-f-rrnn-test-MAX-count-G}{92710}
\DefMacro{results-f-f-rrnn-test-MAX-count-G-G}{69805}
\DefMacro{results-f-f-rrnn-test-MAX-count-G-L}{17639}
\DefMacro{results-f-f-rrnn-test-MAX-count-G-V}{5259}
\DefMacro{results-f-f-rrnn-test-MAX-count-L}{70332}
\DefMacro{results-f-f-rrnn-test-MAX-count-L-G}{17640}
\DefMacro{results-f-f-rrnn-test-MAX-count-L-L}{44837}
\DefMacro{results-f-f-rrnn-test-MAX-count-L-V}{0}
\DefMacro{results-f-f-rrnn-test-MAX-count-V}{27869}
\DefMacro{results-f-f-rrnn-test-MAX-count-V-G}{5265}
\DefMacro{results-f-f-rrnn-test-MAX-count-V-L}{0}
\DefMacro{results-f-f-rrnn-test-MAX-count-V-V}{15985}
\DefMacro{results-f-f-rrnn-test-MAX-count-all}{190911}
\DefMacro{results-f-f-rrnn-test-MAX-count-betsent}{14481}
\DefMacro{results-f-f-rrnn-test-MAX-count-insent}{176430}
\DefMacro{results-f-f-rrnn-test-MAX-count-insent-cs}{45803}
\DefMacro{results-f-f-rrnn-test-MAX-count-insent-noncs}{130627}
\DefMacro{results-f-f-rrnn-test-MAX-count-top-2}{190911}
\DefMacro{results-f-f-rrnn-test-MAX-count-top-3}{190911}
\DefMacro{results-f-f-rrnn-test-MAX-count-top-5}{190911}
\DefMacro{results-f-f-rrnn-test-MEDIAN-accuracy-G}{90.1\%}
\DefMacro{results-f-f-rrnn-test-MEDIAN-accuracy-G-G}{92.7\%}
\DefMacro{results-f-f-rrnn-test-MEDIAN-accuracy-G-L}{94.1\%}
\DefMacro{results-f-f-rrnn-test-MEDIAN-accuracy-G-V}{96.5\%}
\DefMacro{results-f-f-rrnn-test-MEDIAN-accuracy-L}{86.4\%}
\DefMacro{results-f-f-rrnn-test-MEDIAN-accuracy-L-G}{79.9\%}
\DefMacro{results-f-f-rrnn-test-MEDIAN-accuracy-L-L}{90.2\%}
\DefMacro{results-f-f-rrnn-test-MEDIAN-accuracy-L-V}{NaN}
\DefMacro{results-f-f-rrnn-test-MEDIAN-accuracy-V}{92.5\%}
\DefMacro{results-f-f-rrnn-test-MEDIAN-accuracy-V-G}{91.2\%}
\DefMacro{results-f-f-rrnn-test-MEDIAN-accuracy-V-L}{NaN}
\DefMacro{results-f-f-rrnn-test-MEDIAN-accuracy-V-V}{96.0\%}
\DefMacro{results-f-f-rrnn-test-MEDIAN-accuracy-all}{89.1\%}
\DefMacro{results-f-f-rrnn-test-MEDIAN-accuracy-betsent}{63.3\%}
\DefMacro{results-f-f-rrnn-test-MEDIAN-accuracy-insent}{91.2\%}
\DefMacro{results-f-f-rrnn-test-MEDIAN-accuracy-insent-cs}{88.5\%}
\DefMacro{results-f-f-rrnn-test-MEDIAN-accuracy-insent-noncs}{92.2\%}
\DefMacro{results-f-f-rrnn-test-MEDIAN-accuracy-top-2}{98.5\%}
\DefMacro{results-f-f-rrnn-test-MEDIAN-accuracy-top-3}{99.4\%}
\DefMacro{results-f-f-rrnn-test-MEDIAN-accuracy-top-5}{99.9\%}
\DefMacro{results-f-f-rrnn-test-MEDIAN-correct-G}{8351900.0\%}
\DefMacro{results-f-f-rrnn-test-MEDIAN-correct-G-G}{6469700.0\%}
\DefMacro{results-f-f-rrnn-test-MEDIAN-correct-G-L}{1659300.0\%}
\DefMacro{results-f-f-rrnn-test-MEDIAN-correct-G-V}{507600.0\%}
\DefMacro{results-f-f-rrnn-test-MEDIAN-correct-L}{6077700.0\%}
\DefMacro{results-f-f-rrnn-test-MEDIAN-correct-L-G}{1409100.0\%}
\DefMacro{results-f-f-rrnn-test-MEDIAN-correct-L-L}{4043300.0\%}
\DefMacro{results-f-f-rrnn-test-MEDIAN-correct-L-V}{0.0\%}
\DefMacro{results-f-f-rrnn-test-MEDIAN-correct-V}{2578800.0\%}
\DefMacro{results-f-f-rrnn-test-MEDIAN-correct-V-G}{480000.0\%}
\DefMacro{results-f-f-rrnn-test-MEDIAN-correct-V-L}{0.0\%}
\DefMacro{results-f-f-rrnn-test-MEDIAN-correct-V-V}{1535200.0\%}
\DefMacro{results-f-f-rrnn-test-MEDIAN-correct-all}{17007600.0\%}
\DefMacro{results-f-f-rrnn-test-MEDIAN-correct-betsent}{916200.0\%}
\DefMacro{results-f-f-rrnn-test-MEDIAN-correct-insent}{16092600.0\%}
\DefMacro{results-f-f-rrnn-test-MEDIAN-correct-insent-cs}{4055300.0\%}
\DefMacro{results-f-f-rrnn-test-MEDIAN-correct-insent-noncs}{12043400.0\%}
\DefMacro{results-f-f-rrnn-test-MEDIAN-correct-top-2}{18810600.0\%}
\DefMacro{results-f-f-rrnn-test-MEDIAN-correct-top-3}{18978100.0\%}
\DefMacro{results-f-f-rrnn-test-MEDIAN-correct-top-5}{19069100.0\%}
\DefMacro{results-f-f-rrnn-test-MEDIAN-count-G}{9271000.0\%}
\DefMacro{results-f-f-rrnn-test-MEDIAN-count-G-G}{6980500.0\%}
\DefMacro{results-f-f-rrnn-test-MEDIAN-count-G-L}{1763900.0\%}
\DefMacro{results-f-f-rrnn-test-MEDIAN-count-G-V}{525900.0\%}
\DefMacro{results-f-f-rrnn-test-MEDIAN-count-L}{7033200.0\%}
\DefMacro{results-f-f-rrnn-test-MEDIAN-count-L-G}{1764000.0\%}
\DefMacro{results-f-f-rrnn-test-MEDIAN-count-L-L}{4483700.0\%}
\DefMacro{results-f-f-rrnn-test-MEDIAN-count-L-V}{0.0\%}
\DefMacro{results-f-f-rrnn-test-MEDIAN-count-V}{2786900.0\%}
\DefMacro{results-f-f-rrnn-test-MEDIAN-count-V-G}{526500.0\%}
\DefMacro{results-f-f-rrnn-test-MEDIAN-count-V-L}{0.0\%}
\DefMacro{results-f-f-rrnn-test-MEDIAN-count-V-V}{1598500.0\%}
\DefMacro{results-f-f-rrnn-test-MEDIAN-count-all}{19091100.0\%}
\DefMacro{results-f-f-rrnn-test-MEDIAN-count-betsent}{1448100.0\%}
\DefMacro{results-f-f-rrnn-test-MEDIAN-count-insent}{17643000.0\%}
\DefMacro{results-f-f-rrnn-test-MEDIAN-count-insent-cs}{4580300.0\%}
\DefMacro{results-f-f-rrnn-test-MEDIAN-count-insent-noncs}{13062700.0\%}
\DefMacro{results-f-f-rrnn-test-MEDIAN-count-top-2}{19091100.0\%}
\DefMacro{results-f-f-rrnn-test-MEDIAN-count-top-3}{19091100.0\%}
\DefMacro{results-f-f-rrnn-test-MEDIAN-count-top-5}{19091100.0\%}
\DefMacro{results-f-f-rrnn-test-MIN-accuracy-G}{90.0\%}
\DefMacro{results-f-f-rrnn-test-MIN-accuracy-G-G}{92.3\%}
\DefMacro{results-f-f-rrnn-test-MIN-accuracy-G-L}{93.4\%}
\DefMacro{results-f-f-rrnn-test-MIN-accuracy-G-V}{96.1\%}
\DefMacro{results-f-f-rrnn-test-MIN-accuracy-L}{86.4\%}
\DefMacro{results-f-f-rrnn-test-MIN-accuracy-L-G}{79.9\%}
\DefMacro{results-f-f-rrnn-test-MIN-accuracy-L-L}{90.1\%}
\DefMacro{results-f-f-rrnn-test-MIN-accuracy-L-V}{NaN}
\DefMacro{results-f-f-rrnn-test-MIN-accuracy-V}{92.5\%}
\DefMacro{results-f-f-rrnn-test-MIN-accuracy-V-G}{89.8\%}
\DefMacro{results-f-f-rrnn-test-MIN-accuracy-V-L}{NaN}
\DefMacro{results-f-f-rrnn-test-MIN-accuracy-V-V}{96.0\%}
\DefMacro{results-f-f-rrnn-test-MIN-accuracy-all}{89.1\%}
\DefMacro{results-f-f-rrnn-test-MIN-accuracy-betsent}{62.7\%}
\DefMacro{results-f-f-rrnn-test-MIN-accuracy-insent}{91.2\%}
\DefMacro{results-f-f-rrnn-test-MIN-accuracy-insent-cs}{88.2\%}
\DefMacro{results-f-f-rrnn-test-MIN-accuracy-insent-noncs}{92.0\%}
\DefMacro{results-f-f-rrnn-test-MIN-accuracy-top-2}{98.5\%}
\DefMacro{results-f-f-rrnn-test-MIN-accuracy-top-3}{99.4\%}
\DefMacro{results-f-f-rrnn-test-MIN-accuracy-top-5}{99.9\%}
\DefMacro{results-f-f-rrnn-test-MIN-correct-G}{83443}
\DefMacro{results-f-f-rrnn-test-MIN-correct-G-G}{64437}
\DefMacro{results-f-f-rrnn-test-MIN-correct-G-L}{16474}
\DefMacro{results-f-f-rrnn-test-MIN-correct-G-V}{5052}
\DefMacro{results-f-f-rrnn-test-MIN-correct-L}{60733}
\DefMacro{results-f-f-rrnn-test-MIN-correct-L-G}{14090}
\DefMacro{results-f-f-rrnn-test-MIN-correct-L-L}{40387}
\DefMacro{results-f-f-rrnn-test-MIN-correct-L-V}{0}
\DefMacro{results-f-f-rrnn-test-MIN-correct-V}{25767}
\DefMacro{results-f-f-rrnn-test-MIN-correct-V-G}{4727}
\DefMacro{results-f-f-rrnn-test-MIN-correct-V-L}{0}
\DefMacro{results-f-f-rrnn-test-MIN-correct-V-V}{15350}
\DefMacro{results-f-f-rrnn-test-MIN-correct-all}{170008}
\DefMacro{results-f-f-rrnn-test-MIN-correct-betsent}{9082}
\DefMacro{results-f-f-rrnn-test-MIN-correct-insent}{160914}
\DefMacro{results-f-f-rrnn-test-MIN-correct-insent-cs}{40414}
\DefMacro{results-f-f-rrnn-test-MIN-correct-insent-noncs}{120237}
\DefMacro{results-f-f-rrnn-test-MIN-correct-top-2}{188090}
\DefMacro{results-f-f-rrnn-test-MIN-correct-top-3}{189772}
\DefMacro{results-f-f-rrnn-test-MIN-correct-top-5}{190683}
\DefMacro{results-f-f-rrnn-test-MIN-count-G}{92710}
\DefMacro{results-f-f-rrnn-test-MIN-count-G-G}{69805}
\DefMacro{results-f-f-rrnn-test-MIN-count-G-L}{17639}
\DefMacro{results-f-f-rrnn-test-MIN-count-G-V}{5259}
\DefMacro{results-f-f-rrnn-test-MIN-count-L}{70332}
\DefMacro{results-f-f-rrnn-test-MIN-count-L-G}{17640}
\DefMacro{results-f-f-rrnn-test-MIN-count-L-L}{44837}
\DefMacro{results-f-f-rrnn-test-MIN-count-L-V}{0}
\DefMacro{results-f-f-rrnn-test-MIN-count-V}{27869}
\DefMacro{results-f-f-rrnn-test-MIN-count-V-G}{5265}
\DefMacro{results-f-f-rrnn-test-MIN-count-V-L}{0}
\DefMacro{results-f-f-rrnn-test-MIN-count-V-V}{15985}
\DefMacro{results-f-f-rrnn-test-MIN-count-all}{190911}
\DefMacro{results-f-f-rrnn-test-MIN-count-betsent}{14481}
\DefMacro{results-f-f-rrnn-test-MIN-count-insent}{176430}
\DefMacro{results-f-f-rrnn-test-MIN-count-insent-cs}{45803}
\DefMacro{results-f-f-rrnn-test-MIN-count-insent-noncs}{130627}
\DefMacro{results-f-f-rrnn-test-MIN-count-top-2}{190911}
\DefMacro{results-f-f-rrnn-test-MIN-count-top-3}{190911}
\DefMacro{results-f-f-rrnn-test-MIN-count-top-5}{190911}
\DefMacro{results-f-f-rrnn-test-STDEV-accuracy-G}{0.1\%}
\DefMacro{results-f-f-rrnn-test-STDEV-accuracy-G-G}{0.2\%}
\DefMacro{results-f-f-rrnn-test-STDEV-accuracy-G-L}{0.3\%}
\DefMacro{results-f-f-rrnn-test-STDEV-accuracy-G-V}{0.6\%}
\DefMacro{results-f-f-rrnn-test-STDEV-accuracy-L}{0.1\%}
\DefMacro{results-f-f-rrnn-test-STDEV-accuracy-L-G}{0.3\%}
\DefMacro{results-f-f-rrnn-test-STDEV-accuracy-L-L}{0.1\%}
\DefMacro{results-f-f-rrnn-test-STDEV-accuracy-L-V}{NaN}
\DefMacro{results-f-f-rrnn-test-STDEV-accuracy-V}{0.1\%}
\DefMacro{results-f-f-rrnn-test-STDEV-accuracy-V-G}{0.7\%}
\DefMacro{results-f-f-rrnn-test-STDEV-accuracy-V-L}{NaN}
\DefMacro{results-f-f-rrnn-test-STDEV-accuracy-V-V}{0.0\%}
\DefMacro{results-f-f-rrnn-test-STDEV-accuracy-all}{0.1\%}
\DefMacro{results-f-f-rrnn-test-STDEV-accuracy-betsent}{0.6\%}
\DefMacro{results-f-f-rrnn-test-STDEV-accuracy-insent}{0.0\%}
\DefMacro{results-f-f-rrnn-test-STDEV-accuracy-insent-cs}{0.2\%}
\DefMacro{results-f-f-rrnn-test-STDEV-accuracy-insent-noncs}{0.1\%}
\DefMacro{results-f-f-rrnn-test-STDEV-accuracy-top-2}{0.0\%}
\DefMacro{results-f-f-rrnn-test-STDEV-accuracy-top-3}{0.0\%}
\DefMacro{results-f-f-rrnn-test-STDEV-accuracy-top-5}{0.0\%}
\DefMacro{results-f-f-rrnn-test-STDEV-correct-G}{6617.8\%}
\DefMacro{results-f-f-rrnn-test-STDEV-correct-G-G}{12376.1\%}
\DefMacro{results-f-f-rrnn-test-STDEV-correct-G-L}{5967.1\%}
\DefMacro{results-f-f-rrnn-test-STDEV-correct-G-V}{2948.8\%}
\DefMacro{results-f-f-rrnn-test-STDEV-correct-L}{7387.1\%}
\DefMacro{results-f-f-rrnn-test-STDEV-correct-L-G}{5444.9\%}
\DefMacro{results-f-f-rrnn-test-STDEV-correct-L-L}{2595.3\%}
\DefMacro{results-f-f-rrnn-test-STDEV-correct-L-V}{0.0\%}
\DefMacro{results-f-f-rrnn-test-STDEV-correct-V}{2353.7\%}
\DefMacro{results-f-f-rrnn-test-STDEV-correct-V-G}{3904.1\%}
\DefMacro{results-f-f-rrnn-test-STDEV-correct-V-L}{0.0\%}
\DefMacro{results-f-f-rrnn-test-STDEV-correct-V-V}{665.0\%}
\DefMacro{results-f-f-rrnn-test-STDEV-correct-all}{11512.0\%}
\DefMacro{results-f-f-rrnn-test-STDEV-correct-betsent}{8653.8\%}
\DefMacro{results-f-f-rrnn-test-STDEV-correct-insent}{3196.2\%}
\DefMacro{results-f-f-rrnn-test-STDEV-correct-insent-cs}{11227.1\%}
\DefMacro{results-f-f-rrnn-test-STDEV-correct-insent-noncs}{11172.1\%}
\DefMacro{results-f-f-rrnn-test-STDEV-correct-top-2}{5366.8\%}
\DefMacro{results-f-f-rrnn-test-STDEV-correct-top-3}{4094.2\%}
\DefMacro{results-f-f-rrnn-test-STDEV-correct-top-5}{432.0\%}
\DefMacro{results-f-f-rrnn-test-STDEV-count-G}{0.0\%}
\DefMacro{results-f-f-rrnn-test-STDEV-count-G-G}{0.0\%}
\DefMacro{results-f-f-rrnn-test-STDEV-count-G-L}{0.0\%}
\DefMacro{results-f-f-rrnn-test-STDEV-count-G-V}{0.0\%}
\DefMacro{results-f-f-rrnn-test-STDEV-count-L}{0.0\%}
\DefMacro{results-f-f-rrnn-test-STDEV-count-L-G}{0.0\%}
\DefMacro{results-f-f-rrnn-test-STDEV-count-L-L}{0.0\%}
\DefMacro{results-f-f-rrnn-test-STDEV-count-L-V}{0.0\%}
\DefMacro{results-f-f-rrnn-test-STDEV-count-V}{0.0\%}
\DefMacro{results-f-f-rrnn-test-STDEV-count-V-G}{0.0\%}
\DefMacro{results-f-f-rrnn-test-STDEV-count-V-L}{0.0\%}
\DefMacro{results-f-f-rrnn-test-STDEV-count-V-V}{0.0\%}
\DefMacro{results-f-f-rrnn-test-STDEV-count-all}{0.0\%}
\DefMacro{results-f-f-rrnn-test-STDEV-count-betsent}{0.0\%}
\DefMacro{results-f-f-rrnn-test-STDEV-count-insent}{0.0\%}
\DefMacro{results-f-f-rrnn-test-STDEV-count-insent-cs}{0.0\%}
\DefMacro{results-f-f-rrnn-test-STDEV-count-insent-noncs}{0.0\%}
\DefMacro{results-f-f-rrnn-test-STDEV-count-top-2}{0.0\%}
\DefMacro{results-f-f-rrnn-test-STDEV-count-top-3}{0.0\%}
\DefMacro{results-f-f-rrnn-test-STDEV-count-top-5}{0.0\%}
\DefMacro{results-f-f-rrnn-test-SUM-accuracy-G}{270.3\%}
\DefMacro{results-f-f-rrnn-test-SUM-accuracy-G-G}{277.7\%}
\DefMacro{results-f-f-rrnn-test-SUM-accuracy-G-L}{281.6\%}
\DefMacro{results-f-f-rrnn-test-SUM-accuracy-G-V}{290.0\%}
\DefMacro{results-f-f-rrnn-test-SUM-accuracy-L}{259.4\%}
\DefMacro{results-f-f-rrnn-test-SUM-accuracy-L-G}{240.3\%}
\DefMacro{results-f-f-rrnn-test-SUM-accuracy-L-L}{270.5\%}
\DefMacro{results-f-f-rrnn-test-SUM-accuracy-L-V}{NaN}
\DefMacro{results-f-f-rrnn-test-SUM-accuracy-V}{277.7\%}
\DefMacro{results-f-f-rrnn-test-SUM-accuracy-V-G}{272.4\%}
\DefMacro{results-f-f-rrnn-test-SUM-accuracy-V-L}{NaN}
\DefMacro{results-f-f-rrnn-test-SUM-accuracy-V-V}{288.2\%}
\DefMacro{results-f-f-rrnn-test-SUM-accuracy-all}{267.3\%}
\DefMacro{results-f-f-rrnn-test-SUM-accuracy-betsent}{190.2\%}
\DefMacro{results-f-f-rrnn-test-SUM-accuracy-insent}{273.7\%}
\DefMacro{results-f-f-rrnn-test-SUM-accuracy-insent-cs}{265.6\%}
\DefMacro{results-f-f-rrnn-test-SUM-accuracy-insent-noncs}{276.5\%}
\DefMacro{results-f-f-rrnn-test-SUM-accuracy-top-2}{295.6\%}
\DefMacro{results-f-f-rrnn-test-SUM-accuracy-top-3}{298.3\%}
\DefMacro{results-f-f-rrnn-test-SUM-accuracy-top-5}{299.7\%}
\DefMacro{results-f-f-rrnn-test-SUM-correct-G}{250567}
\DefMacro{results-f-f-rrnn-test-SUM-correct-G-G}{193836}
\DefMacro{results-f-f-rrnn-test-SUM-correct-G-L}{49674}
\DefMacro{results-f-f-rrnn-test-SUM-correct-G-V}{15251}
\DefMacro{results-f-f-rrnn-test-SUM-correct-L}{182417}
\DefMacro{results-f-f-rrnn-test-SUM-correct-L-G}{42387}
\DefMacro{results-f-f-rrnn-test-SUM-correct-L-L}{121268}
\DefMacro{results-f-f-rrnn-test-SUM-correct-L-V}{0}
\DefMacro{results-f-f-rrnn-test-SUM-correct-V}{77379}
\DefMacro{results-f-f-rrnn-test-SUM-correct-V-G}{14344}
\DefMacro{results-f-f-rrnn-test-SUM-correct-V-L}{0}
\DefMacro{results-f-f-rrnn-test-SUM-correct-V-V}{46067}
\DefMacro{results-f-f-rrnn-test-SUM-correct-all}{510363}
\DefMacro{results-f-f-rrnn-test-SUM-correct-betsent}{27536}
\DefMacro{results-f-f-rrnn-test-SUM-correct-insent}{482827}
\DefMacro{results-f-f-rrnn-test-SUM-correct-insent-cs}{121656}
\DefMacro{results-f-f-rrnn-test-SUM-correct-insent-noncs}{361171}
\DefMacro{results-f-f-rrnn-test-SUM-correct-top-2}{564407}
\DefMacro{results-f-f-rrnn-test-SUM-correct-top-3}{569416}
\DefMacro{results-f-f-rrnn-test-SUM-correct-top-5}{572067}
\DefMacro{results-f-f-rrnn-test-SUM-count-G}{278130}
\DefMacro{results-f-f-rrnn-test-SUM-count-G-G}{209415}
\DefMacro{results-f-f-rrnn-test-SUM-count-G-L}{52917}
\DefMacro{results-f-f-rrnn-test-SUM-count-G-V}{15777}
\DefMacro{results-f-f-rrnn-test-SUM-count-L}{210996}
\DefMacro{results-f-f-rrnn-test-SUM-count-L-G}{52920}
\DefMacro{results-f-f-rrnn-test-SUM-count-L-L}{134511}
\DefMacro{results-f-f-rrnn-test-SUM-count-L-V}{0}
\DefMacro{results-f-f-rrnn-test-SUM-count-V}{83607}
\DefMacro{results-f-f-rrnn-test-SUM-count-V-G}{15795}
\DefMacro{results-f-f-rrnn-test-SUM-count-V-L}{0}
\DefMacro{results-f-f-rrnn-test-SUM-count-V-V}{47955}
\DefMacro{results-f-f-rrnn-test-SUM-count-all}{572733}
\DefMacro{results-f-f-rrnn-test-SUM-count-betsent}{43443}
\DefMacro{results-f-f-rrnn-test-SUM-count-insent}{529290}
\DefMacro{results-f-f-rrnn-test-SUM-count-insent-cs}{137409}
\DefMacro{results-f-f-rrnn-test-SUM-count-insent-noncs}{391881}
\DefMacro{results-f-f-rrnn-test-SUM-count-top-2}{572733}
\DefMacro{results-f-f-rrnn-test-SUM-count-top-3}{572733}
\DefMacro{results-f-f-rrnn-test-SUM-count-top-5}{572733}
\DefMacro{results-f-f-ngram-val-AVG-accuracy-G}{93.1\%}
\DefMacro{results-f-f-ngram-val-AVG-accuracy-G-G}{93.0\%}
\DefMacro{results-f-f-ngram-val-AVG-accuracy-G-L}{96.3\%}
\DefMacro{results-f-f-ngram-val-AVG-accuracy-G-V}{98.6\%}
\DefMacro{results-f-f-ngram-val-AVG-accuracy-L}{92.5\%}
\DefMacro{results-f-f-ngram-val-AVG-accuracy-L-G}{93.5\%}
\DefMacro{results-f-f-ngram-val-AVG-accuracy-L-L}{95.8\%}
\DefMacro{results-f-f-ngram-val-AVG-accuracy-L-V}{NaN}
\DefMacro{results-f-f-ngram-val-AVG-accuracy-V}{93.4\%}
\DefMacro{results-f-f-ngram-val-AVG-accuracy-V-G}{92.7\%}
\DefMacro{results-f-f-ngram-val-AVG-accuracy-V-L}{NaN}
\DefMacro{results-f-f-ngram-val-AVG-accuracy-V-V}{97.7\%}
\DefMacro{results-f-f-ngram-val-AVG-accuracy-all}{93.0\%}
\DefMacro{results-f-f-ngram-val-AVG-accuracy-betsent}{72.3\%}
\DefMacro{results-f-f-ngram-val-AVG-accuracy-insent}{94.9\%}
\DefMacro{results-f-f-ngram-val-AVG-accuracy-insent-cs}{95.1\%}
\DefMacro{results-f-f-ngram-val-AVG-accuracy-insent-noncs}{94.8\%}
\DefMacro{results-f-f-ngram-val-AVG-accuracy-top-2}{98.1\%}
\DefMacro{results-f-f-ngram-val-AVG-accuracy-top-3}{98.6\%}
\DefMacro{results-f-f-ngram-val-AVG-accuracy-top-5}{98.8\%}
\DefMacro{results-f-f-ngram-val-AVG-correct-G}{6999900.0\%}
\DefMacro{results-f-f-ngram-val-AVG-correct-G-G}{5266900.0\%}
\DefMacro{results-f-f-ngram-val-AVG-correct-G-L}{1203900.0\%}
\DefMacro{results-f-f-ngram-val-AVG-correct-G-V}{598900.0\%}
\DefMacro{results-f-f-ngram-val-AVG-correct-L}{4788400.0\%}
\DefMacro{results-f-f-ngram-val-AVG-correct-L-G}{1168900.0\%}
\DefMacro{results-f-f-ngram-val-AVG-correct-L-L}{3243400.0\%}
\DefMacro{results-f-f-ngram-val-AVG-correct-L-V}{0.0\%}
\DefMacro{results-f-f-ngram-val-AVG-correct-V}{3580600.0\%}
\DefMacro{results-f-f-ngram-val-AVG-correct-V-G}{564100.0\%}
\DefMacro{results-f-f-ngram-val-AVG-correct-V-L}{0.0\%}
\DefMacro{results-f-f-ngram-val-AVG-correct-V-V}{2298700.0\%}
\DefMacro{results-f-f-ngram-val-AVG-correct-all}{15368900.0\%}
\DefMacro{results-f-f-ngram-val-AVG-correct-betsent}{1024100.0\%}
\DefMacro{results-f-f-ngram-val-AVG-correct-insent}{14344800.0\%}
\DefMacro{results-f-f-ngram-val-AVG-correct-insent-cs}{3535800.0\%}
\DefMacro{results-f-f-ngram-val-AVG-correct-insent-noncs}{10809000.0\%}
\DefMacro{results-f-f-ngram-val-AVG-correct-top-2}{16214000.0\%}
\DefMacro{results-f-f-ngram-val-AVG-correct-top-3}{16298500.0\%}
\DefMacro{results-f-f-ngram-val-AVG-correct-top-5}{16335800.0\%}
\DefMacro{results-f-f-ngram-val-AVG-count-G}{7519200.0\%}
\DefMacro{results-f-f-ngram-val-AVG-count-G-G}{5660400.0\%}
\DefMacro{results-f-f-ngram-val-AVG-count-G-L}{1250500.0\%}
\DefMacro{results-f-f-ngram-val-AVG-count-G-V}{607600.0\%}
\DefMacro{results-f-f-ngram-val-AVG-count-L}{5179400.0\%}
\DefMacro{results-f-f-ngram-val-AVG-count-L-G}{1250600.0\%}
\DefMacro{results-f-f-ngram-val-AVG-count-L-L}{3386800.0\%}
\DefMacro{results-f-f-ngram-val-AVG-count-L-V}{0.0\%}
\DefMacro{results-f-f-ngram-val-AVG-count-V}{3833800.0\%}
\DefMacro{results-f-f-ngram-val-AVG-count-V-G}{608200.0\%}
\DefMacro{results-f-f-ngram-val-AVG-count-V-L}{0.0\%}
\DefMacro{results-f-f-ngram-val-AVG-count-V-V}{2352800.0\%}
\DefMacro{results-f-f-ngram-val-AVG-count-all}{16532400.0\%}
\DefMacro{results-f-f-ngram-val-AVG-count-betsent}{1415500.0\%}
\DefMacro{results-f-f-ngram-val-AVG-count-insent}{15116900.0\%}
\DefMacro{results-f-f-ngram-val-AVG-count-insent-cs}{3716900.0\%}
\DefMacro{results-f-f-ngram-val-AVG-count-insent-noncs}{11400000.0\%}
\DefMacro{results-f-f-ngram-val-AVG-count-top-2}{16532400.0\%}
\DefMacro{results-f-f-ngram-val-AVG-count-top-3}{16532400.0\%}
\DefMacro{results-f-f-ngram-val-AVG-count-top-5}{16532400.0\%}
\DefMacro{results-f-f-ngram-val-MAX-accuracy-G}{93.1\%}
\DefMacro{results-f-f-ngram-val-MAX-accuracy-G-G}{93.0\%}
\DefMacro{results-f-f-ngram-val-MAX-accuracy-G-L}{96.3\%}
\DefMacro{results-f-f-ngram-val-MAX-accuracy-G-V}{98.6\%}
\DefMacro{results-f-f-ngram-val-MAX-accuracy-L}{92.5\%}
\DefMacro{results-f-f-ngram-val-MAX-accuracy-L-G}{93.5\%}
\DefMacro{results-f-f-ngram-val-MAX-accuracy-L-L}{95.8\%}
\DefMacro{results-f-f-ngram-val-MAX-accuracy-L-V}{NaN}
\DefMacro{results-f-f-ngram-val-MAX-accuracy-V}{93.4\%}
\DefMacro{results-f-f-ngram-val-MAX-accuracy-V-G}{92.7\%}
\DefMacro{results-f-f-ngram-val-MAX-accuracy-V-L}{NaN}
\DefMacro{results-f-f-ngram-val-MAX-accuracy-V-V}{97.7\%}
\DefMacro{results-f-f-ngram-val-MAX-accuracy-all}{93.0\%}
\DefMacro{results-f-f-ngram-val-MAX-accuracy-betsent}{72.3\%}
\DefMacro{results-f-f-ngram-val-MAX-accuracy-insent}{94.9\%}
\DefMacro{results-f-f-ngram-val-MAX-accuracy-insent-cs}{95.1\%}
\DefMacro{results-f-f-ngram-val-MAX-accuracy-insent-noncs}{94.8\%}
\DefMacro{results-f-f-ngram-val-MAX-accuracy-top-2}{98.1\%}
\DefMacro{results-f-f-ngram-val-MAX-accuracy-top-3}{98.6\%}
\DefMacro{results-f-f-ngram-val-MAX-accuracy-top-5}{98.8\%}
\DefMacro{results-f-f-ngram-val-MAX-correct-G}{69999}
\DefMacro{results-f-f-ngram-val-MAX-correct-G-G}{52669}
\DefMacro{results-f-f-ngram-val-MAX-correct-G-L}{12039}
\DefMacro{results-f-f-ngram-val-MAX-correct-G-V}{5989}
\DefMacro{results-f-f-ngram-val-MAX-correct-L}{47884}
\DefMacro{results-f-f-ngram-val-MAX-correct-L-G}{11689}
\DefMacro{results-f-f-ngram-val-MAX-correct-L-L}{32434}
\DefMacro{results-f-f-ngram-val-MAX-correct-L-V}{0}
\DefMacro{results-f-f-ngram-val-MAX-correct-V}{35806}
\DefMacro{results-f-f-ngram-val-MAX-correct-V-G}{5641}
\DefMacro{results-f-f-ngram-val-MAX-correct-V-L}{0}
\DefMacro{results-f-f-ngram-val-MAX-correct-V-V}{22987}
\DefMacro{results-f-f-ngram-val-MAX-correct-all}{153689}
\DefMacro{results-f-f-ngram-val-MAX-correct-betsent}{10241}
\DefMacro{results-f-f-ngram-val-MAX-correct-insent}{143448}
\DefMacro{results-f-f-ngram-val-MAX-correct-insent-cs}{35358}
\DefMacro{results-f-f-ngram-val-MAX-correct-insent-noncs}{108090}
\DefMacro{results-f-f-ngram-val-MAX-correct-top-2}{162140}
\DefMacro{results-f-f-ngram-val-MAX-correct-top-3}{162985}
\DefMacro{results-f-f-ngram-val-MAX-correct-top-5}{163358}
\DefMacro{results-f-f-ngram-val-MAX-count-G}{75192}
\DefMacro{results-f-f-ngram-val-MAX-count-G-G}{56604}
\DefMacro{results-f-f-ngram-val-MAX-count-G-L}{12505}
\DefMacro{results-f-f-ngram-val-MAX-count-G-V}{6076}
\DefMacro{results-f-f-ngram-val-MAX-count-L}{51794}
\DefMacro{results-f-f-ngram-val-MAX-count-L-G}{12506}
\DefMacro{results-f-f-ngram-val-MAX-count-L-L}{33868}
\DefMacro{results-f-f-ngram-val-MAX-count-L-V}{0}
\DefMacro{results-f-f-ngram-val-MAX-count-V}{38338}
\DefMacro{results-f-f-ngram-val-MAX-count-V-G}{6082}
\DefMacro{results-f-f-ngram-val-MAX-count-V-L}{0}
\DefMacro{results-f-f-ngram-val-MAX-count-V-V}{23528}
\DefMacro{results-f-f-ngram-val-MAX-count-all}{165324}
\DefMacro{results-f-f-ngram-val-MAX-count-betsent}{14155}
\DefMacro{results-f-f-ngram-val-MAX-count-insent}{151169}
\DefMacro{results-f-f-ngram-val-MAX-count-insent-cs}{37169}
\DefMacro{results-f-f-ngram-val-MAX-count-insent-noncs}{114000}
\DefMacro{results-f-f-ngram-val-MAX-count-top-2}{165324}
\DefMacro{results-f-f-ngram-val-MAX-count-top-3}{165324}
\DefMacro{results-f-f-ngram-val-MAX-count-top-5}{165324}
\DefMacro{results-f-f-ngram-val-MEDIAN-accuracy-G}{93.1\%}
\DefMacro{results-f-f-ngram-val-MEDIAN-accuracy-G-G}{93.0\%}
\DefMacro{results-f-f-ngram-val-MEDIAN-accuracy-G-L}{96.3\%}
\DefMacro{results-f-f-ngram-val-MEDIAN-accuracy-G-V}{98.6\%}
\DefMacro{results-f-f-ngram-val-MEDIAN-accuracy-L}{92.5\%}
\DefMacro{results-f-f-ngram-val-MEDIAN-accuracy-L-G}{93.5\%}
\DefMacro{results-f-f-ngram-val-MEDIAN-accuracy-L-L}{95.8\%}
\DefMacro{results-f-f-ngram-val-MEDIAN-accuracy-L-V}{NaN}
\DefMacro{results-f-f-ngram-val-MEDIAN-accuracy-V}{93.4\%}
\DefMacro{results-f-f-ngram-val-MEDIAN-accuracy-V-G}{92.7\%}
\DefMacro{results-f-f-ngram-val-MEDIAN-accuracy-V-L}{NaN}
\DefMacro{results-f-f-ngram-val-MEDIAN-accuracy-V-V}{97.7\%}
\DefMacro{results-f-f-ngram-val-MEDIAN-accuracy-all}{93.0\%}
\DefMacro{results-f-f-ngram-val-MEDIAN-accuracy-betsent}{72.3\%}
\DefMacro{results-f-f-ngram-val-MEDIAN-accuracy-insent}{94.9\%}
\DefMacro{results-f-f-ngram-val-MEDIAN-accuracy-insent-cs}{95.1\%}
\DefMacro{results-f-f-ngram-val-MEDIAN-accuracy-insent-noncs}{94.8\%}
\DefMacro{results-f-f-ngram-val-MEDIAN-accuracy-top-2}{98.1\%}
\DefMacro{results-f-f-ngram-val-MEDIAN-accuracy-top-3}{98.6\%}
\DefMacro{results-f-f-ngram-val-MEDIAN-accuracy-top-5}{98.8\%}
\DefMacro{results-f-f-ngram-val-MEDIAN-correct-G}{6999900.0\%}
\DefMacro{results-f-f-ngram-val-MEDIAN-correct-G-G}{5266900.0\%}
\DefMacro{results-f-f-ngram-val-MEDIAN-correct-G-L}{1203900.0\%}
\DefMacro{results-f-f-ngram-val-MEDIAN-correct-G-V}{598900.0\%}
\DefMacro{results-f-f-ngram-val-MEDIAN-correct-L}{4788400.0\%}
\DefMacro{results-f-f-ngram-val-MEDIAN-correct-L-G}{1168900.0\%}
\DefMacro{results-f-f-ngram-val-MEDIAN-correct-L-L}{3243400.0\%}
\DefMacro{results-f-f-ngram-val-MEDIAN-correct-L-V}{0.0\%}
\DefMacro{results-f-f-ngram-val-MEDIAN-correct-V}{3580600.0\%}
\DefMacro{results-f-f-ngram-val-MEDIAN-correct-V-G}{564100.0\%}
\DefMacro{results-f-f-ngram-val-MEDIAN-correct-V-L}{0.0\%}
\DefMacro{results-f-f-ngram-val-MEDIAN-correct-V-V}{2298700.0\%}
\DefMacro{results-f-f-ngram-val-MEDIAN-correct-all}{15368900.0\%}
\DefMacro{results-f-f-ngram-val-MEDIAN-correct-betsent}{1024100.0\%}
\DefMacro{results-f-f-ngram-val-MEDIAN-correct-insent}{14344800.0\%}
\DefMacro{results-f-f-ngram-val-MEDIAN-correct-insent-cs}{3535800.0\%}
\DefMacro{results-f-f-ngram-val-MEDIAN-correct-insent-noncs}{10809000.0\%}
\DefMacro{results-f-f-ngram-val-MEDIAN-correct-top-2}{16214000.0\%}
\DefMacro{results-f-f-ngram-val-MEDIAN-correct-top-3}{16298500.0\%}
\DefMacro{results-f-f-ngram-val-MEDIAN-correct-top-5}{16335800.0\%}
\DefMacro{results-f-f-ngram-val-MEDIAN-count-G}{7519200.0\%}
\DefMacro{results-f-f-ngram-val-MEDIAN-count-G-G}{5660400.0\%}
\DefMacro{results-f-f-ngram-val-MEDIAN-count-G-L}{1250500.0\%}
\DefMacro{results-f-f-ngram-val-MEDIAN-count-G-V}{607600.0\%}
\DefMacro{results-f-f-ngram-val-MEDIAN-count-L}{5179400.0\%}
\DefMacro{results-f-f-ngram-val-MEDIAN-count-L-G}{1250600.0\%}
\DefMacro{results-f-f-ngram-val-MEDIAN-count-L-L}{3386800.0\%}
\DefMacro{results-f-f-ngram-val-MEDIAN-count-L-V}{0.0\%}
\DefMacro{results-f-f-ngram-val-MEDIAN-count-V}{3833800.0\%}
\DefMacro{results-f-f-ngram-val-MEDIAN-count-V-G}{608200.0\%}
\DefMacro{results-f-f-ngram-val-MEDIAN-count-V-L}{0.0\%}
\DefMacro{results-f-f-ngram-val-MEDIAN-count-V-V}{2352800.0\%}
\DefMacro{results-f-f-ngram-val-MEDIAN-count-all}{16532400.0\%}
\DefMacro{results-f-f-ngram-val-MEDIAN-count-betsent}{1415500.0\%}
\DefMacro{results-f-f-ngram-val-MEDIAN-count-insent}{15116900.0\%}
\DefMacro{results-f-f-ngram-val-MEDIAN-count-insent-cs}{3716900.0\%}
\DefMacro{results-f-f-ngram-val-MEDIAN-count-insent-noncs}{11400000.0\%}
\DefMacro{results-f-f-ngram-val-MEDIAN-count-top-2}{16532400.0\%}
\DefMacro{results-f-f-ngram-val-MEDIAN-count-top-3}{16532400.0\%}
\DefMacro{results-f-f-ngram-val-MEDIAN-count-top-5}{16532400.0\%}
\DefMacro{results-f-f-ngram-val-MIN-accuracy-G}{93.1\%}
\DefMacro{results-f-f-ngram-val-MIN-accuracy-G-G}{93.0\%}
\DefMacro{results-f-f-ngram-val-MIN-accuracy-G-L}{96.3\%}
\DefMacro{results-f-f-ngram-val-MIN-accuracy-G-V}{98.6\%}
\DefMacro{results-f-f-ngram-val-MIN-accuracy-L}{92.5\%}
\DefMacro{results-f-f-ngram-val-MIN-accuracy-L-G}{93.5\%}
\DefMacro{results-f-f-ngram-val-MIN-accuracy-L-L}{95.8\%}
\DefMacro{results-f-f-ngram-val-MIN-accuracy-L-V}{NaN}
\DefMacro{results-f-f-ngram-val-MIN-accuracy-V}{93.4\%}
\DefMacro{results-f-f-ngram-val-MIN-accuracy-V-G}{92.7\%}
\DefMacro{results-f-f-ngram-val-MIN-accuracy-V-L}{NaN}
\DefMacro{results-f-f-ngram-val-MIN-accuracy-V-V}{97.7\%}
\DefMacro{results-f-f-ngram-val-MIN-accuracy-all}{93.0\%}
\DefMacro{results-f-f-ngram-val-MIN-accuracy-betsent}{72.3\%}
\DefMacro{results-f-f-ngram-val-MIN-accuracy-insent}{94.9\%}
\DefMacro{results-f-f-ngram-val-MIN-accuracy-insent-cs}{95.1\%}
\DefMacro{results-f-f-ngram-val-MIN-accuracy-insent-noncs}{94.8\%}
\DefMacro{results-f-f-ngram-val-MIN-accuracy-top-2}{98.1\%}
\DefMacro{results-f-f-ngram-val-MIN-accuracy-top-3}{98.6\%}
\DefMacro{results-f-f-ngram-val-MIN-accuracy-top-5}{98.8\%}
\DefMacro{results-f-f-ngram-val-MIN-correct-G}{69999}
\DefMacro{results-f-f-ngram-val-MIN-correct-G-G}{52669}
\DefMacro{results-f-f-ngram-val-MIN-correct-G-L}{12039}
\DefMacro{results-f-f-ngram-val-MIN-correct-G-V}{5989}
\DefMacro{results-f-f-ngram-val-MIN-correct-L}{47884}
\DefMacro{results-f-f-ngram-val-MIN-correct-L-G}{11689}
\DefMacro{results-f-f-ngram-val-MIN-correct-L-L}{32434}
\DefMacro{results-f-f-ngram-val-MIN-correct-L-V}{0}
\DefMacro{results-f-f-ngram-val-MIN-correct-V}{35806}
\DefMacro{results-f-f-ngram-val-MIN-correct-V-G}{5641}
\DefMacro{results-f-f-ngram-val-MIN-correct-V-L}{0}
\DefMacro{results-f-f-ngram-val-MIN-correct-V-V}{22987}
\DefMacro{results-f-f-ngram-val-MIN-correct-all}{153689}
\DefMacro{results-f-f-ngram-val-MIN-correct-betsent}{10241}
\DefMacro{results-f-f-ngram-val-MIN-correct-insent}{143448}
\DefMacro{results-f-f-ngram-val-MIN-correct-insent-cs}{35358}
\DefMacro{results-f-f-ngram-val-MIN-correct-insent-noncs}{108090}
\DefMacro{results-f-f-ngram-val-MIN-correct-top-2}{162140}
\DefMacro{results-f-f-ngram-val-MIN-correct-top-3}{162985}
\DefMacro{results-f-f-ngram-val-MIN-correct-top-5}{163358}
\DefMacro{results-f-f-ngram-val-MIN-count-G}{75192}
\DefMacro{results-f-f-ngram-val-MIN-count-G-G}{56604}
\DefMacro{results-f-f-ngram-val-MIN-count-G-L}{12505}
\DefMacro{results-f-f-ngram-val-MIN-count-G-V}{6076}
\DefMacro{results-f-f-ngram-val-MIN-count-L}{51794}
\DefMacro{results-f-f-ngram-val-MIN-count-L-G}{12506}
\DefMacro{results-f-f-ngram-val-MIN-count-L-L}{33868}
\DefMacro{results-f-f-ngram-val-MIN-count-L-V}{0}
\DefMacro{results-f-f-ngram-val-MIN-count-V}{38338}
\DefMacro{results-f-f-ngram-val-MIN-count-V-G}{6082}
\DefMacro{results-f-f-ngram-val-MIN-count-V-L}{0}
\DefMacro{results-f-f-ngram-val-MIN-count-V-V}{23528}
\DefMacro{results-f-f-ngram-val-MIN-count-all}{165324}
\DefMacro{results-f-f-ngram-val-MIN-count-betsent}{14155}
\DefMacro{results-f-f-ngram-val-MIN-count-insent}{151169}
\DefMacro{results-f-f-ngram-val-MIN-count-insent-cs}{37169}
\DefMacro{results-f-f-ngram-val-MIN-count-insent-noncs}{114000}
\DefMacro{results-f-f-ngram-val-MIN-count-top-2}{165324}
\DefMacro{results-f-f-ngram-val-MIN-count-top-3}{165324}
\DefMacro{results-f-f-ngram-val-MIN-count-top-5}{165324}
\DefMacro{results-f-f-ngram-val-STDEV-accuracy-G}{0.0\%}
\DefMacro{results-f-f-ngram-val-STDEV-accuracy-G-G}{0.0\%}
\DefMacro{results-f-f-ngram-val-STDEV-accuracy-G-L}{0.0\%}
\DefMacro{results-f-f-ngram-val-STDEV-accuracy-G-V}{0.0\%}
\DefMacro{results-f-f-ngram-val-STDEV-accuracy-L}{0.0\%}
\DefMacro{results-f-f-ngram-val-STDEV-accuracy-L-G}{0.0\%}
\DefMacro{results-f-f-ngram-val-STDEV-accuracy-L-L}{0.0\%}
\DefMacro{results-f-f-ngram-val-STDEV-accuracy-L-V}{NaN}
\DefMacro{results-f-f-ngram-val-STDEV-accuracy-V}{0.0\%}
\DefMacro{results-f-f-ngram-val-STDEV-accuracy-V-G}{0.0\%}
\DefMacro{results-f-f-ngram-val-STDEV-accuracy-V-L}{NaN}
\DefMacro{results-f-f-ngram-val-STDEV-accuracy-V-V}{0.0\%}
\DefMacro{results-f-f-ngram-val-STDEV-accuracy-all}{0.0\%}
\DefMacro{results-f-f-ngram-val-STDEV-accuracy-betsent}{0.0\%}
\DefMacro{results-f-f-ngram-val-STDEV-accuracy-insent}{0.0\%}
\DefMacro{results-f-f-ngram-val-STDEV-accuracy-insent-cs}{0.0\%}
\DefMacro{results-f-f-ngram-val-STDEV-accuracy-insent-noncs}{0.0\%}
\DefMacro{results-f-f-ngram-val-STDEV-accuracy-top-2}{0.0\%}
\DefMacro{results-f-f-ngram-val-STDEV-accuracy-top-3}{0.0\%}
\DefMacro{results-f-f-ngram-val-STDEV-accuracy-top-5}{0.0\%}
\DefMacro{results-f-f-ngram-val-STDEV-correct-G}{0.0\%}
\DefMacro{results-f-f-ngram-val-STDEV-correct-G-G}{0.0\%}
\DefMacro{results-f-f-ngram-val-STDEV-correct-G-L}{0.0\%}
\DefMacro{results-f-f-ngram-val-STDEV-correct-G-V}{0.0\%}
\DefMacro{results-f-f-ngram-val-STDEV-correct-L}{0.0\%}
\DefMacro{results-f-f-ngram-val-STDEV-correct-L-G}{0.0\%}
\DefMacro{results-f-f-ngram-val-STDEV-correct-L-L}{0.0\%}
\DefMacro{results-f-f-ngram-val-STDEV-correct-L-V}{0.0\%}
\DefMacro{results-f-f-ngram-val-STDEV-correct-V}{0.0\%}
\DefMacro{results-f-f-ngram-val-STDEV-correct-V-G}{0.0\%}
\DefMacro{results-f-f-ngram-val-STDEV-correct-V-L}{0.0\%}
\DefMacro{results-f-f-ngram-val-STDEV-correct-V-V}{0.0\%}
\DefMacro{results-f-f-ngram-val-STDEV-correct-all}{0.0\%}
\DefMacro{results-f-f-ngram-val-STDEV-correct-betsent}{0.0\%}
\DefMacro{results-f-f-ngram-val-STDEV-correct-insent}{0.0\%}
\DefMacro{results-f-f-ngram-val-STDEV-correct-insent-cs}{0.0\%}
\DefMacro{results-f-f-ngram-val-STDEV-correct-insent-noncs}{0.0\%}
\DefMacro{results-f-f-ngram-val-STDEV-correct-top-2}{0.0\%}
\DefMacro{results-f-f-ngram-val-STDEV-correct-top-3}{0.0\%}
\DefMacro{results-f-f-ngram-val-STDEV-correct-top-5}{0.0\%}
\DefMacro{results-f-f-ngram-val-STDEV-count-G}{0.0\%}
\DefMacro{results-f-f-ngram-val-STDEV-count-G-G}{0.0\%}
\DefMacro{results-f-f-ngram-val-STDEV-count-G-L}{0.0\%}
\DefMacro{results-f-f-ngram-val-STDEV-count-G-V}{0.0\%}
\DefMacro{results-f-f-ngram-val-STDEV-count-L}{0.0\%}
\DefMacro{results-f-f-ngram-val-STDEV-count-L-G}{0.0\%}
\DefMacro{results-f-f-ngram-val-STDEV-count-L-L}{0.0\%}
\DefMacro{results-f-f-ngram-val-STDEV-count-L-V}{0.0\%}
\DefMacro{results-f-f-ngram-val-STDEV-count-V}{0.0\%}
\DefMacro{results-f-f-ngram-val-STDEV-count-V-G}{0.0\%}
\DefMacro{results-f-f-ngram-val-STDEV-count-V-L}{0.0\%}
\DefMacro{results-f-f-ngram-val-STDEV-count-V-V}{0.0\%}
\DefMacro{results-f-f-ngram-val-STDEV-count-all}{0.0\%}
\DefMacro{results-f-f-ngram-val-STDEV-count-betsent}{0.0\%}
\DefMacro{results-f-f-ngram-val-STDEV-count-insent}{0.0\%}
\DefMacro{results-f-f-ngram-val-STDEV-count-insent-cs}{0.0\%}
\DefMacro{results-f-f-ngram-val-STDEV-count-insent-noncs}{0.0\%}
\DefMacro{results-f-f-ngram-val-STDEV-count-top-2}{0.0\%}
\DefMacro{results-f-f-ngram-val-STDEV-count-top-3}{0.0\%}
\DefMacro{results-f-f-ngram-val-STDEV-count-top-5}{0.0\%}
\DefMacro{results-f-f-ngram-val-SUM-accuracy-G}{279.3\%}
\DefMacro{results-f-f-ngram-val-SUM-accuracy-G-G}{279.1\%}
\DefMacro{results-f-f-ngram-val-SUM-accuracy-G-L}{288.8\%}
\DefMacro{results-f-f-ngram-val-SUM-accuracy-G-V}{295.7\%}
\DefMacro{results-f-f-ngram-val-SUM-accuracy-L}{277.4\%}
\DefMacro{results-f-f-ngram-val-SUM-accuracy-L-G}{280.4\%}
\DefMacro{results-f-f-ngram-val-SUM-accuracy-L-L}{287.3\%}
\DefMacro{results-f-f-ngram-val-SUM-accuracy-L-V}{NaN}
\DefMacro{results-f-f-ngram-val-SUM-accuracy-V}{280.2\%}
\DefMacro{results-f-f-ngram-val-SUM-accuracy-V-G}{278.2\%}
\DefMacro{results-f-f-ngram-val-SUM-accuracy-V-L}{NaN}
\DefMacro{results-f-f-ngram-val-SUM-accuracy-V-V}{293.1\%}
\DefMacro{results-f-f-ngram-val-SUM-accuracy-all}{278.9\%}
\DefMacro{results-f-f-ngram-val-SUM-accuracy-betsent}{217.0\%}
\DefMacro{results-f-f-ngram-val-SUM-accuracy-insent}{284.7\%}
\DefMacro{results-f-f-ngram-val-SUM-accuracy-insent-cs}{285.4\%}
\DefMacro{results-f-f-ngram-val-SUM-accuracy-insent-noncs}{284.4\%}
\DefMacro{results-f-f-ngram-val-SUM-accuracy-top-2}{294.2\%}
\DefMacro{results-f-f-ngram-val-SUM-accuracy-top-3}{295.8\%}
\DefMacro{results-f-f-ngram-val-SUM-accuracy-top-5}{296.4\%}
\DefMacro{results-f-f-ngram-val-SUM-correct-G}{209997}
\DefMacro{results-f-f-ngram-val-SUM-correct-G-G}{158007}
\DefMacro{results-f-f-ngram-val-SUM-correct-G-L}{36117}
\DefMacro{results-f-f-ngram-val-SUM-correct-G-V}{17967}
\DefMacro{results-f-f-ngram-val-SUM-correct-L}{143652}
\DefMacro{results-f-f-ngram-val-SUM-correct-L-G}{35067}
\DefMacro{results-f-f-ngram-val-SUM-correct-L-L}{97302}
\DefMacro{results-f-f-ngram-val-SUM-correct-L-V}{0}
\DefMacro{results-f-f-ngram-val-SUM-correct-V}{107418}
\DefMacro{results-f-f-ngram-val-SUM-correct-V-G}{16923}
\DefMacro{results-f-f-ngram-val-SUM-correct-V-L}{0}
\DefMacro{results-f-f-ngram-val-SUM-correct-V-V}{68961}
\DefMacro{results-f-f-ngram-val-SUM-correct-all}{461067}
\DefMacro{results-f-f-ngram-val-SUM-correct-betsent}{30723}
\DefMacro{results-f-f-ngram-val-SUM-correct-insent}{430344}
\DefMacro{results-f-f-ngram-val-SUM-correct-insent-cs}{106074}
\DefMacro{results-f-f-ngram-val-SUM-correct-insent-noncs}{324270}
\DefMacro{results-f-f-ngram-val-SUM-correct-top-2}{486420}
\DefMacro{results-f-f-ngram-val-SUM-correct-top-3}{488955}
\DefMacro{results-f-f-ngram-val-SUM-correct-top-5}{490074}
\DefMacro{results-f-f-ngram-val-SUM-count-G}{225576}
\DefMacro{results-f-f-ngram-val-SUM-count-G-G}{169812}
\DefMacro{results-f-f-ngram-val-SUM-count-G-L}{37515}
\DefMacro{results-f-f-ngram-val-SUM-count-G-V}{18228}
\DefMacro{results-f-f-ngram-val-SUM-count-L}{155382}
\DefMacro{results-f-f-ngram-val-SUM-count-L-G}{37518}
\DefMacro{results-f-f-ngram-val-SUM-count-L-L}{101604}
\DefMacro{results-f-f-ngram-val-SUM-count-L-V}{0}
\DefMacro{results-f-f-ngram-val-SUM-count-V}{115014}
\DefMacro{results-f-f-ngram-val-SUM-count-V-G}{18246}
\DefMacro{results-f-f-ngram-val-SUM-count-V-L}{0}
\DefMacro{results-f-f-ngram-val-SUM-count-V-V}{70584}
\DefMacro{results-f-f-ngram-val-SUM-count-all}{495972}
\DefMacro{results-f-f-ngram-val-SUM-count-betsent}{42465}
\DefMacro{results-f-f-ngram-val-SUM-count-insent}{453507}
\DefMacro{results-f-f-ngram-val-SUM-count-insent-cs}{111507}
\DefMacro{results-f-f-ngram-val-SUM-count-insent-noncs}{342000}
\DefMacro{results-f-f-ngram-val-SUM-count-top-2}{495972}
\DefMacro{results-f-f-ngram-val-SUM-count-top-3}{495972}
\DefMacro{results-f-f-ngram-val-SUM-count-top-5}{495972}
\DefMacro{results-f-f-ngram-test-AVG-accuracy-G}{94.2\%}
\DefMacro{results-f-f-ngram-test-AVG-accuracy-G-G}{94.3\%}
\DefMacro{results-f-f-ngram-test-AVG-accuracy-G-L}{97.5\%}
\DefMacro{results-f-f-ngram-test-AVG-accuracy-G-V}{97.5\%}
\DefMacro{results-f-f-ngram-test-AVG-accuracy-L}{92.3\%}
\DefMacro{results-f-f-ngram-test-AVG-accuracy-L-G}{95.0\%}
\DefMacro{results-f-f-ngram-test-AVG-accuracy-L-L}{96.2\%}
\DefMacro{results-f-f-ngram-test-AVG-accuracy-L-V}{NaN}
\DefMacro{results-f-f-ngram-test-AVG-accuracy-V}{93.9\%}
\DefMacro{results-f-f-ngram-test-AVG-accuracy-V-G}{90.1\%}
\DefMacro{results-f-f-ngram-test-AVG-accuracy-V-L}{NaN}
\DefMacro{results-f-f-ngram-test-AVG-accuracy-V-V}{97.7\%}
\DefMacro{results-f-f-ngram-test-AVG-accuracy-all}{93.4\%}
\DefMacro{results-f-f-ngram-test-AVG-accuracy-betsent}{69.0\%}
\DefMacro{results-f-f-ngram-test-AVG-accuracy-insent}{95.4\%}
\DefMacro{results-f-f-ngram-test-AVG-accuracy-insent-cs}{95.7\%}
\DefMacro{results-f-f-ngram-test-AVG-accuracy-insent-noncs}{95.4\%}
\DefMacro{results-f-f-ngram-test-AVG-accuracy-top-2}{98.4\%}
\DefMacro{results-f-f-ngram-test-AVG-accuracy-top-3}{98.9\%}
\DefMacro{results-f-f-ngram-test-AVG-accuracy-top-5}{99.2\%}
\DefMacro{results-f-f-ngram-test-AVG-correct-G}{8732200.0\%}
\DefMacro{results-f-f-ngram-test-AVG-correct-G-G}{6581100.0\%}
\DefMacro{results-f-f-ngram-test-AVG-correct-G-L}{1719500.0\%}
\DefMacro{results-f-f-ngram-test-AVG-correct-G-V}{512700.0\%}
\DefMacro{results-f-f-ngram-test-AVG-correct-L}{6489100.0\%}
\DefMacro{results-f-f-ngram-test-AVG-correct-L-G}{1676600.0\%}
\DefMacro{results-f-f-ngram-test-AVG-correct-L-L}{4313500.0\%}
\DefMacro{results-f-f-ngram-test-AVG-correct-L-V}{0.0\%}
\DefMacro{results-f-f-ngram-test-AVG-correct-V}{2617600.0\%}
\DefMacro{results-f-f-ngram-test-AVG-correct-V-G}{474500.0\%}
\DefMacro{results-f-f-ngram-test-AVG-correct-V-L}{0.0\%}
\DefMacro{results-f-f-ngram-test-AVG-correct-V-V}{1562200.0\%}
\DefMacro{results-f-f-ngram-test-AVG-correct-all}{17838900.0\%}
\DefMacro{results-f-f-ngram-test-AVG-correct-betsent}{998800.0\%}
\DefMacro{results-f-f-ngram-test-AVG-correct-insent}{16840100.0\%}
\DefMacro{results-f-f-ngram-test-AVG-correct-insent-cs}{4383300.0\%}
\DefMacro{results-f-f-ngram-test-AVG-correct-insent-noncs}{12456800.0\%}
\DefMacro{results-f-f-ngram-test-AVG-correct-top-2}{18780500.0\%}
\DefMacro{results-f-f-ngram-test-AVG-correct-top-3}{18877900.0\%}
\DefMacro{results-f-f-ngram-test-AVG-correct-top-5}{18934800.0\%}
\DefMacro{results-f-f-ngram-test-AVG-count-G}{9271000.0\%}
\DefMacro{results-f-f-ngram-test-AVG-count-G-G}{6980500.0\%}
\DefMacro{results-f-f-ngram-test-AVG-count-G-L}{1763900.0\%}
\DefMacro{results-f-f-ngram-test-AVG-count-G-V}{525900.0\%}
\DefMacro{results-f-f-ngram-test-AVG-count-L}{7033200.0\%}
\DefMacro{results-f-f-ngram-test-AVG-count-L-G}{1764000.0\%}
\DefMacro{results-f-f-ngram-test-AVG-count-L-L}{4483700.0\%}
\DefMacro{results-f-f-ngram-test-AVG-count-L-V}{0.0\%}
\DefMacro{results-f-f-ngram-test-AVG-count-V}{2786900.0\%}
\DefMacro{results-f-f-ngram-test-AVG-count-V-G}{526500.0\%}
\DefMacro{results-f-f-ngram-test-AVG-count-V-L}{0.0\%}
\DefMacro{results-f-f-ngram-test-AVG-count-V-V}{1598500.0\%}
\DefMacro{results-f-f-ngram-test-AVG-count-all}{19091100.0\%}
\DefMacro{results-f-f-ngram-test-AVG-count-betsent}{1448100.0\%}
\DefMacro{results-f-f-ngram-test-AVG-count-insent}{17643000.0\%}
\DefMacro{results-f-f-ngram-test-AVG-count-insent-cs}{4580300.0\%}
\DefMacro{results-f-f-ngram-test-AVG-count-insent-noncs}{13062700.0\%}
\DefMacro{results-f-f-ngram-test-AVG-count-top-2}{19091100.0\%}
\DefMacro{results-f-f-ngram-test-AVG-count-top-3}{19091100.0\%}
\DefMacro{results-f-f-ngram-test-AVG-count-top-5}{19091100.0\%}
\DefMacro{results-f-f-ngram-test-MAX-accuracy-G}{94.2\%}
\DefMacro{results-f-f-ngram-test-MAX-accuracy-G-G}{94.3\%}
\DefMacro{results-f-f-ngram-test-MAX-accuracy-G-L}{97.5\%}
\DefMacro{results-f-f-ngram-test-MAX-accuracy-G-V}{97.5\%}
\DefMacro{results-f-f-ngram-test-MAX-accuracy-L}{92.3\%}
\DefMacro{results-f-f-ngram-test-MAX-accuracy-L-G}{95.0\%}
\DefMacro{results-f-f-ngram-test-MAX-accuracy-L-L}{96.2\%}
\DefMacro{results-f-f-ngram-test-MAX-accuracy-L-V}{NaN}
\DefMacro{results-f-f-ngram-test-MAX-accuracy-V}{93.9\%}
\DefMacro{results-f-f-ngram-test-MAX-accuracy-V-G}{90.1\%}
\DefMacro{results-f-f-ngram-test-MAX-accuracy-V-L}{NaN}
\DefMacro{results-f-f-ngram-test-MAX-accuracy-V-V}{97.7\%}
\DefMacro{results-f-f-ngram-test-MAX-accuracy-all}{93.4\%}
\DefMacro{results-f-f-ngram-test-MAX-accuracy-betsent}{69.0\%}
\DefMacro{results-f-f-ngram-test-MAX-accuracy-insent}{95.4\%}
\DefMacro{results-f-f-ngram-test-MAX-accuracy-insent-cs}{95.7\%}
\DefMacro{results-f-f-ngram-test-MAX-accuracy-insent-noncs}{95.4\%}
\DefMacro{results-f-f-ngram-test-MAX-accuracy-top-2}{98.4\%}
\DefMacro{results-f-f-ngram-test-MAX-accuracy-top-3}{98.9\%}
\DefMacro{results-f-f-ngram-test-MAX-accuracy-top-5}{99.2\%}
\DefMacro{results-f-f-ngram-test-MAX-correct-G}{87322}
\DefMacro{results-f-f-ngram-test-MAX-correct-G-G}{65811}
\DefMacro{results-f-f-ngram-test-MAX-correct-G-L}{17195}
\DefMacro{results-f-f-ngram-test-MAX-correct-G-V}{5127}
\DefMacro{results-f-f-ngram-test-MAX-correct-L}{64891}
\DefMacro{results-f-f-ngram-test-MAX-correct-L-G}{16766}
\DefMacro{results-f-f-ngram-test-MAX-correct-L-L}{43135}
\DefMacro{results-f-f-ngram-test-MAX-correct-L-V}{0}
\DefMacro{results-f-f-ngram-test-MAX-correct-V}{26176}
\DefMacro{results-f-f-ngram-test-MAX-correct-V-G}{4745}
\DefMacro{results-f-f-ngram-test-MAX-correct-V-L}{0}
\DefMacro{results-f-f-ngram-test-MAX-correct-V-V}{15622}
\DefMacro{results-f-f-ngram-test-MAX-correct-all}{178389}
\DefMacro{results-f-f-ngram-test-MAX-correct-betsent}{9988}
\DefMacro{results-f-f-ngram-test-MAX-correct-insent}{168401}
\DefMacro{results-f-f-ngram-test-MAX-correct-insent-cs}{43833}
\DefMacro{results-f-f-ngram-test-MAX-correct-insent-noncs}{124568}
\DefMacro{results-f-f-ngram-test-MAX-correct-top-2}{187805}
\DefMacro{results-f-f-ngram-test-MAX-correct-top-3}{188779}
\DefMacro{results-f-f-ngram-test-MAX-correct-top-5}{189348}
\DefMacro{results-f-f-ngram-test-MAX-count-G}{92710}
\DefMacro{results-f-f-ngram-test-MAX-count-G-G}{69805}
\DefMacro{results-f-f-ngram-test-MAX-count-G-L}{17639}
\DefMacro{results-f-f-ngram-test-MAX-count-G-V}{5259}
\DefMacro{results-f-f-ngram-test-MAX-count-L}{70332}
\DefMacro{results-f-f-ngram-test-MAX-count-L-G}{17640}
\DefMacro{results-f-f-ngram-test-MAX-count-L-L}{44837}
\DefMacro{results-f-f-ngram-test-MAX-count-L-V}{0}
\DefMacro{results-f-f-ngram-test-MAX-count-V}{27869}
\DefMacro{results-f-f-ngram-test-MAX-count-V-G}{5265}
\DefMacro{results-f-f-ngram-test-MAX-count-V-L}{0}
\DefMacro{results-f-f-ngram-test-MAX-count-V-V}{15985}
\DefMacro{results-f-f-ngram-test-MAX-count-all}{190911}
\DefMacro{results-f-f-ngram-test-MAX-count-betsent}{14481}
\DefMacro{results-f-f-ngram-test-MAX-count-insent}{176430}
\DefMacro{results-f-f-ngram-test-MAX-count-insent-cs}{45803}
\DefMacro{results-f-f-ngram-test-MAX-count-insent-noncs}{130627}
\DefMacro{results-f-f-ngram-test-MAX-count-top-2}{190911}
\DefMacro{results-f-f-ngram-test-MAX-count-top-3}{190911}
\DefMacro{results-f-f-ngram-test-MAX-count-top-5}{190911}
\DefMacro{results-f-f-ngram-test-MEDIAN-accuracy-G}{94.2\%}
\DefMacro{results-f-f-ngram-test-MEDIAN-accuracy-G-G}{94.3\%}
\DefMacro{results-f-f-ngram-test-MEDIAN-accuracy-G-L}{97.5\%}
\DefMacro{results-f-f-ngram-test-MEDIAN-accuracy-G-V}{97.5\%}
\DefMacro{results-f-f-ngram-test-MEDIAN-accuracy-L}{92.3\%}
\DefMacro{results-f-f-ngram-test-MEDIAN-accuracy-L-G}{95.0\%}
\DefMacro{results-f-f-ngram-test-MEDIAN-accuracy-L-L}{96.2\%}
\DefMacro{results-f-f-ngram-test-MEDIAN-accuracy-L-V}{NaN}
\DefMacro{results-f-f-ngram-test-MEDIAN-accuracy-V}{93.9\%}
\DefMacro{results-f-f-ngram-test-MEDIAN-accuracy-V-G}{90.1\%}
\DefMacro{results-f-f-ngram-test-MEDIAN-accuracy-V-L}{NaN}
\DefMacro{results-f-f-ngram-test-MEDIAN-accuracy-V-V}{97.7\%}
\DefMacro{results-f-f-ngram-test-MEDIAN-accuracy-all}{93.4\%}
\DefMacro{results-f-f-ngram-test-MEDIAN-accuracy-betsent}{69.0\%}
\DefMacro{results-f-f-ngram-test-MEDIAN-accuracy-insent}{95.4\%}
\DefMacro{results-f-f-ngram-test-MEDIAN-accuracy-insent-cs}{95.7\%}
\DefMacro{results-f-f-ngram-test-MEDIAN-accuracy-insent-noncs}{95.4\%}
\DefMacro{results-f-f-ngram-test-MEDIAN-accuracy-top-2}{98.4\%}
\DefMacro{results-f-f-ngram-test-MEDIAN-accuracy-top-3}{98.9\%}
\DefMacro{results-f-f-ngram-test-MEDIAN-accuracy-top-5}{99.2\%}
\DefMacro{results-f-f-ngram-test-MEDIAN-correct-G}{8732200.0\%}
\DefMacro{results-f-f-ngram-test-MEDIAN-correct-G-G}{6581100.0\%}
\DefMacro{results-f-f-ngram-test-MEDIAN-correct-G-L}{1719500.0\%}
\DefMacro{results-f-f-ngram-test-MEDIAN-correct-G-V}{512700.0\%}
\DefMacro{results-f-f-ngram-test-MEDIAN-correct-L}{6489100.0\%}
\DefMacro{results-f-f-ngram-test-MEDIAN-correct-L-G}{1676600.0\%}
\DefMacro{results-f-f-ngram-test-MEDIAN-correct-L-L}{4313500.0\%}
\DefMacro{results-f-f-ngram-test-MEDIAN-correct-L-V}{0.0\%}
\DefMacro{results-f-f-ngram-test-MEDIAN-correct-V}{2617600.0\%}
\DefMacro{results-f-f-ngram-test-MEDIAN-correct-V-G}{474500.0\%}
\DefMacro{results-f-f-ngram-test-MEDIAN-correct-V-L}{0.0\%}
\DefMacro{results-f-f-ngram-test-MEDIAN-correct-V-V}{1562200.0\%}
\DefMacro{results-f-f-ngram-test-MEDIAN-correct-all}{17838900.0\%}
\DefMacro{results-f-f-ngram-test-MEDIAN-correct-betsent}{998800.0\%}
\DefMacro{results-f-f-ngram-test-MEDIAN-correct-insent}{16840100.0\%}
\DefMacro{results-f-f-ngram-test-MEDIAN-correct-insent-cs}{4383300.0\%}
\DefMacro{results-f-f-ngram-test-MEDIAN-correct-insent-noncs}{12456800.0\%}
\DefMacro{results-f-f-ngram-test-MEDIAN-correct-top-2}{18780500.0\%}
\DefMacro{results-f-f-ngram-test-MEDIAN-correct-top-3}{18877900.0\%}
\DefMacro{results-f-f-ngram-test-MEDIAN-correct-top-5}{18934800.0\%}
\DefMacro{results-f-f-ngram-test-MEDIAN-count-G}{9271000.0\%}
\DefMacro{results-f-f-ngram-test-MEDIAN-count-G-G}{6980500.0\%}
\DefMacro{results-f-f-ngram-test-MEDIAN-count-G-L}{1763900.0\%}
\DefMacro{results-f-f-ngram-test-MEDIAN-count-G-V}{525900.0\%}
\DefMacro{results-f-f-ngram-test-MEDIAN-count-L}{7033200.0\%}
\DefMacro{results-f-f-ngram-test-MEDIAN-count-L-G}{1764000.0\%}
\DefMacro{results-f-f-ngram-test-MEDIAN-count-L-L}{4483700.0\%}
\DefMacro{results-f-f-ngram-test-MEDIAN-count-L-V}{0.0\%}
\DefMacro{results-f-f-ngram-test-MEDIAN-count-V}{2786900.0\%}
\DefMacro{results-f-f-ngram-test-MEDIAN-count-V-G}{526500.0\%}
\DefMacro{results-f-f-ngram-test-MEDIAN-count-V-L}{0.0\%}
\DefMacro{results-f-f-ngram-test-MEDIAN-count-V-V}{1598500.0\%}
\DefMacro{results-f-f-ngram-test-MEDIAN-count-all}{19091100.0\%}
\DefMacro{results-f-f-ngram-test-MEDIAN-count-betsent}{1448100.0\%}
\DefMacro{results-f-f-ngram-test-MEDIAN-count-insent}{17643000.0\%}
\DefMacro{results-f-f-ngram-test-MEDIAN-count-insent-cs}{4580300.0\%}
\DefMacro{results-f-f-ngram-test-MEDIAN-count-insent-noncs}{13062700.0\%}
\DefMacro{results-f-f-ngram-test-MEDIAN-count-top-2}{19091100.0\%}
\DefMacro{results-f-f-ngram-test-MEDIAN-count-top-3}{19091100.0\%}
\DefMacro{results-f-f-ngram-test-MEDIAN-count-top-5}{19091100.0\%}
\DefMacro{results-f-f-ngram-test-MIN-accuracy-G}{94.2\%}
\DefMacro{results-f-f-ngram-test-MIN-accuracy-G-G}{94.3\%}
\DefMacro{results-f-f-ngram-test-MIN-accuracy-G-L}{97.5\%}
\DefMacro{results-f-f-ngram-test-MIN-accuracy-G-V}{97.5\%}
\DefMacro{results-f-f-ngram-test-MIN-accuracy-L}{92.3\%}
\DefMacro{results-f-f-ngram-test-MIN-accuracy-L-G}{95.0\%}
\DefMacro{results-f-f-ngram-test-MIN-accuracy-L-L}{96.2\%}
\DefMacro{results-f-f-ngram-test-MIN-accuracy-L-V}{NaN}
\DefMacro{results-f-f-ngram-test-MIN-accuracy-V}{93.9\%}
\DefMacro{results-f-f-ngram-test-MIN-accuracy-V-G}{90.1\%}
\DefMacro{results-f-f-ngram-test-MIN-accuracy-V-L}{NaN}
\DefMacro{results-f-f-ngram-test-MIN-accuracy-V-V}{97.7\%}
\DefMacro{results-f-f-ngram-test-MIN-accuracy-all}{93.4\%}
\DefMacro{results-f-f-ngram-test-MIN-accuracy-betsent}{69.0\%}
\DefMacro{results-f-f-ngram-test-MIN-accuracy-insent}{95.4\%}
\DefMacro{results-f-f-ngram-test-MIN-accuracy-insent-cs}{95.7\%}
\DefMacro{results-f-f-ngram-test-MIN-accuracy-insent-noncs}{95.4\%}
\DefMacro{results-f-f-ngram-test-MIN-accuracy-top-2}{98.4\%}
\DefMacro{results-f-f-ngram-test-MIN-accuracy-top-3}{98.9\%}
\DefMacro{results-f-f-ngram-test-MIN-accuracy-top-5}{99.2\%}
\DefMacro{results-f-f-ngram-test-MIN-correct-G}{87322}
\DefMacro{results-f-f-ngram-test-MIN-correct-G-G}{65811}
\DefMacro{results-f-f-ngram-test-MIN-correct-G-L}{17195}
\DefMacro{results-f-f-ngram-test-MIN-correct-G-V}{5127}
\DefMacro{results-f-f-ngram-test-MIN-correct-L}{64891}
\DefMacro{results-f-f-ngram-test-MIN-correct-L-G}{16766}
\DefMacro{results-f-f-ngram-test-MIN-correct-L-L}{43135}
\DefMacro{results-f-f-ngram-test-MIN-correct-L-V}{0}
\DefMacro{results-f-f-ngram-test-MIN-correct-V}{26176}
\DefMacro{results-f-f-ngram-test-MIN-correct-V-G}{4745}
\DefMacro{results-f-f-ngram-test-MIN-correct-V-L}{0}
\DefMacro{results-f-f-ngram-test-MIN-correct-V-V}{15622}
\DefMacro{results-f-f-ngram-test-MIN-correct-all}{178389}
\DefMacro{results-f-f-ngram-test-MIN-correct-betsent}{9988}
\DefMacro{results-f-f-ngram-test-MIN-correct-insent}{168401}
\DefMacro{results-f-f-ngram-test-MIN-correct-insent-cs}{43833}
\DefMacro{results-f-f-ngram-test-MIN-correct-insent-noncs}{124568}
\DefMacro{results-f-f-ngram-test-MIN-correct-top-2}{187805}
\DefMacro{results-f-f-ngram-test-MIN-correct-top-3}{188779}
\DefMacro{results-f-f-ngram-test-MIN-correct-top-5}{189348}
\DefMacro{results-f-f-ngram-test-MIN-count-G}{92710}
\DefMacro{results-f-f-ngram-test-MIN-count-G-G}{69805}
\DefMacro{results-f-f-ngram-test-MIN-count-G-L}{17639}
\DefMacro{results-f-f-ngram-test-MIN-count-G-V}{5259}
\DefMacro{results-f-f-ngram-test-MIN-count-L}{70332}
\DefMacro{results-f-f-ngram-test-MIN-count-L-G}{17640}
\DefMacro{results-f-f-ngram-test-MIN-count-L-L}{44837}
\DefMacro{results-f-f-ngram-test-MIN-count-L-V}{0}
\DefMacro{results-f-f-ngram-test-MIN-count-V}{27869}
\DefMacro{results-f-f-ngram-test-MIN-count-V-G}{5265}
\DefMacro{results-f-f-ngram-test-MIN-count-V-L}{0}
\DefMacro{results-f-f-ngram-test-MIN-count-V-V}{15985}
\DefMacro{results-f-f-ngram-test-MIN-count-all}{190911}
\DefMacro{results-f-f-ngram-test-MIN-count-betsent}{14481}
\DefMacro{results-f-f-ngram-test-MIN-count-insent}{176430}
\DefMacro{results-f-f-ngram-test-MIN-count-insent-cs}{45803}
\DefMacro{results-f-f-ngram-test-MIN-count-insent-noncs}{130627}
\DefMacro{results-f-f-ngram-test-MIN-count-top-2}{190911}
\DefMacro{results-f-f-ngram-test-MIN-count-top-3}{190911}
\DefMacro{results-f-f-ngram-test-MIN-count-top-5}{190911}
\DefMacro{results-f-f-ngram-test-STDEV-accuracy-G}{0.0\%}
\DefMacro{results-f-f-ngram-test-STDEV-accuracy-G-G}{0.0\%}
\DefMacro{results-f-f-ngram-test-STDEV-accuracy-G-L}{0.0\%}
\DefMacro{results-f-f-ngram-test-STDEV-accuracy-G-V}{0.0\%}
\DefMacro{results-f-f-ngram-test-STDEV-accuracy-L}{0.0\%}
\DefMacro{results-f-f-ngram-test-STDEV-accuracy-L-G}{0.0\%}
\DefMacro{results-f-f-ngram-test-STDEV-accuracy-L-L}{0.0\%}
\DefMacro{results-f-f-ngram-test-STDEV-accuracy-L-V}{NaN}
\DefMacro{results-f-f-ngram-test-STDEV-accuracy-V}{0.0\%}
\DefMacro{results-f-f-ngram-test-STDEV-accuracy-V-G}{0.0\%}
\DefMacro{results-f-f-ngram-test-STDEV-accuracy-V-L}{NaN}
\DefMacro{results-f-f-ngram-test-STDEV-accuracy-V-V}{0.0\%}
\DefMacro{results-f-f-ngram-test-STDEV-accuracy-all}{0.0\%}
\DefMacro{results-f-f-ngram-test-STDEV-accuracy-betsent}{0.0\%}
\DefMacro{results-f-f-ngram-test-STDEV-accuracy-insent}{0.0\%}
\DefMacro{results-f-f-ngram-test-STDEV-accuracy-insent-cs}{0.0\%}
\DefMacro{results-f-f-ngram-test-STDEV-accuracy-insent-noncs}{0.0\%}
\DefMacro{results-f-f-ngram-test-STDEV-accuracy-top-2}{0.0\%}
\DefMacro{results-f-f-ngram-test-STDEV-accuracy-top-3}{0.0\%}
\DefMacro{results-f-f-ngram-test-STDEV-accuracy-top-5}{0.0\%}
\DefMacro{results-f-f-ngram-test-STDEV-correct-G}{0.0\%}
\DefMacro{results-f-f-ngram-test-STDEV-correct-G-G}{0.0\%}
\DefMacro{results-f-f-ngram-test-STDEV-correct-G-L}{0.0\%}
\DefMacro{results-f-f-ngram-test-STDEV-correct-G-V}{0.0\%}
\DefMacro{results-f-f-ngram-test-STDEV-correct-L}{0.0\%}
\DefMacro{results-f-f-ngram-test-STDEV-correct-L-G}{0.0\%}
\DefMacro{results-f-f-ngram-test-STDEV-correct-L-L}{0.0\%}
\DefMacro{results-f-f-ngram-test-STDEV-correct-L-V}{0.0\%}
\DefMacro{results-f-f-ngram-test-STDEV-correct-V}{0.0\%}
\DefMacro{results-f-f-ngram-test-STDEV-correct-V-G}{0.0\%}
\DefMacro{results-f-f-ngram-test-STDEV-correct-V-L}{0.0\%}
\DefMacro{results-f-f-ngram-test-STDEV-correct-V-V}{0.0\%}
\DefMacro{results-f-f-ngram-test-STDEV-correct-all}{0.0\%}
\DefMacro{results-f-f-ngram-test-STDEV-correct-betsent}{0.0\%}
\DefMacro{results-f-f-ngram-test-STDEV-correct-insent}{0.0\%}
\DefMacro{results-f-f-ngram-test-STDEV-correct-insent-cs}{0.0\%}
\DefMacro{results-f-f-ngram-test-STDEV-correct-insent-noncs}{0.0\%}
\DefMacro{results-f-f-ngram-test-STDEV-correct-top-2}{0.0\%}
\DefMacro{results-f-f-ngram-test-STDEV-correct-top-3}{0.0\%}
\DefMacro{results-f-f-ngram-test-STDEV-correct-top-5}{0.0\%}
\DefMacro{results-f-f-ngram-test-STDEV-count-G}{0.0\%}
\DefMacro{results-f-f-ngram-test-STDEV-count-G-G}{0.0\%}
\DefMacro{results-f-f-ngram-test-STDEV-count-G-L}{0.0\%}
\DefMacro{results-f-f-ngram-test-STDEV-count-G-V}{0.0\%}
\DefMacro{results-f-f-ngram-test-STDEV-count-L}{0.0\%}
\DefMacro{results-f-f-ngram-test-STDEV-count-L-G}{0.0\%}
\DefMacro{results-f-f-ngram-test-STDEV-count-L-L}{0.0\%}
\DefMacro{results-f-f-ngram-test-STDEV-count-L-V}{0.0\%}
\DefMacro{results-f-f-ngram-test-STDEV-count-V}{0.0\%}
\DefMacro{results-f-f-ngram-test-STDEV-count-V-G}{0.0\%}
\DefMacro{results-f-f-ngram-test-STDEV-count-V-L}{0.0\%}
\DefMacro{results-f-f-ngram-test-STDEV-count-V-V}{0.0\%}
\DefMacro{results-f-f-ngram-test-STDEV-count-all}{0.0\%}
\DefMacro{results-f-f-ngram-test-STDEV-count-betsent}{0.0\%}
\DefMacro{results-f-f-ngram-test-STDEV-count-insent}{0.0\%}
\DefMacro{results-f-f-ngram-test-STDEV-count-insent-cs}{0.0\%}
\DefMacro{results-f-f-ngram-test-STDEV-count-insent-noncs}{0.0\%}
\DefMacro{results-f-f-ngram-test-STDEV-count-top-2}{0.0\%}
\DefMacro{results-f-f-ngram-test-STDEV-count-top-3}{0.0\%}
\DefMacro{results-f-f-ngram-test-STDEV-count-top-5}{0.0\%}
\DefMacro{results-f-f-ngram-test-SUM-accuracy-G}{282.6\%}
\DefMacro{results-f-f-ngram-test-SUM-accuracy-G-G}{282.8\%}
\DefMacro{results-f-f-ngram-test-SUM-accuracy-G-L}{292.4\%}
\DefMacro{results-f-f-ngram-test-SUM-accuracy-G-V}{292.5\%}
\DefMacro{results-f-f-ngram-test-SUM-accuracy-L}{276.8\%}
\DefMacro{results-f-f-ngram-test-SUM-accuracy-L-G}{285.1\%}
\DefMacro{results-f-f-ngram-test-SUM-accuracy-L-L}{288.6\%}
\DefMacro{results-f-f-ngram-test-SUM-accuracy-L-V}{NaN}
\DefMacro{results-f-f-ngram-test-SUM-accuracy-V}{281.8\%}
\DefMacro{results-f-f-ngram-test-SUM-accuracy-V-G}{270.4\%}
\DefMacro{results-f-f-ngram-test-SUM-accuracy-V-L}{NaN}
\DefMacro{results-f-f-ngram-test-SUM-accuracy-V-V}{293.2\%}
\DefMacro{results-f-f-ngram-test-SUM-accuracy-all}{280.3\%}
\DefMacro{results-f-f-ngram-test-SUM-accuracy-betsent}{206.9\%}
\DefMacro{results-f-f-ngram-test-SUM-accuracy-insent}{286.3\%}
\DefMacro{results-f-f-ngram-test-SUM-accuracy-insent-cs}{287.1\%}
\DefMacro{results-f-f-ngram-test-SUM-accuracy-insent-noncs}{286.1\%}
\DefMacro{results-f-f-ngram-test-SUM-accuracy-top-2}{295.1\%}
\DefMacro{results-f-f-ngram-test-SUM-accuracy-top-3}{296.6\%}
\DefMacro{results-f-f-ngram-test-SUM-accuracy-top-5}{297.5\%}
\DefMacro{results-f-f-ngram-test-SUM-correct-G}{261966}
\DefMacro{results-f-f-ngram-test-SUM-correct-G-G}{197433}
\DefMacro{results-f-f-ngram-test-SUM-correct-G-L}{51585}
\DefMacro{results-f-f-ngram-test-SUM-correct-G-V}{15381}
\DefMacro{results-f-f-ngram-test-SUM-correct-L}{194673}
\DefMacro{results-f-f-ngram-test-SUM-correct-L-G}{50298}
\DefMacro{results-f-f-ngram-test-SUM-correct-L-L}{129405}
\DefMacro{results-f-f-ngram-test-SUM-correct-L-V}{0}
\DefMacro{results-f-f-ngram-test-SUM-correct-V}{78528}
\DefMacro{results-f-f-ngram-test-SUM-correct-V-G}{14235}
\DefMacro{results-f-f-ngram-test-SUM-correct-V-L}{0}
\DefMacro{results-f-f-ngram-test-SUM-correct-V-V}{46866}
\DefMacro{results-f-f-ngram-test-SUM-correct-all}{535167}
\DefMacro{results-f-f-ngram-test-SUM-correct-betsent}{29964}
\DefMacro{results-f-f-ngram-test-SUM-correct-insent}{505203}
\DefMacro{results-f-f-ngram-test-SUM-correct-insent-cs}{131499}
\DefMacro{results-f-f-ngram-test-SUM-correct-insent-noncs}{373704}
\DefMacro{results-f-f-ngram-test-SUM-correct-top-2}{563415}
\DefMacro{results-f-f-ngram-test-SUM-correct-top-3}{566337}
\DefMacro{results-f-f-ngram-test-SUM-correct-top-5}{568044}
\DefMacro{results-f-f-ngram-test-SUM-count-G}{278130}
\DefMacro{results-f-f-ngram-test-SUM-count-G-G}{209415}
\DefMacro{results-f-f-ngram-test-SUM-count-G-L}{52917}
\DefMacro{results-f-f-ngram-test-SUM-count-G-V}{15777}
\DefMacro{results-f-f-ngram-test-SUM-count-L}{210996}
\DefMacro{results-f-f-ngram-test-SUM-count-L-G}{52920}
\DefMacro{results-f-f-ngram-test-SUM-count-L-L}{134511}
\DefMacro{results-f-f-ngram-test-SUM-count-L-V}{0}
\DefMacro{results-f-f-ngram-test-SUM-count-V}{83607}
\DefMacro{results-f-f-ngram-test-SUM-count-V-G}{15795}
\DefMacro{results-f-f-ngram-test-SUM-count-V-L}{0}
\DefMacro{results-f-f-ngram-test-SUM-count-V-V}{47955}
\DefMacro{results-f-f-ngram-test-SUM-count-all}{572733}
\DefMacro{results-f-f-ngram-test-SUM-count-betsent}{43443}
\DefMacro{results-f-f-ngram-test-SUM-count-insent}{529290}
\DefMacro{results-f-f-ngram-test-SUM-count-insent-cs}{137409}
\DefMacro{results-f-f-ngram-test-SUM-count-insent-noncs}{391881}
\DefMacro{results-f-f-ngram-test-SUM-count-top-2}{572733}
\DefMacro{results-f-f-ngram-test-SUM-count-top-3}{572733}
\DefMacro{results-f-f-ngram-test-SUM-count-top-5}{572733}

To implement learning and suggesting of spacing in Coq files based on our models, we modified the SerAPI library~\cite{Gallego2016} to serialize tokens in Coq files, organized at the sentence level. We then serialized all sentences in our MathComp corpus, and extracted information on token kind and spacing using the source file location information included in each token. \begin{wraptable}{r}{6cm}

\begin{small}
\begin{center}
\caption{\TableCaptionResultsF}
\begin{tabular}{l r r}
\toprule
\TableHeadModel
& \UseMacro{table-head-results-f-acc-all}
& \UseMacro{table-head-results-f-acc-top-3}
 \\
\midrule
\UseMacro{f-brnn}
& 
\textbf{\UseMacro{results-f-f-brnn-test-AVG-accuracy-all}}
& 
\textbf{\UseMacro{results-f-f-brnn-test-AVG-accuracy-top-3}}
\\
\UseMacro{f-ngram}
& 
\UseMacro{results-f-f-ngram-test-AVG-accuracy-all}
& 
\UseMacro{results-f-f-ngram-test-AVG-accuracy-top-3}
\\
\bottomrule
\end{tabular}
\end{center}
\end{small}
\vspace{\UseMacro{vspace-results-f}}

 \end{wraptable} Finally, we implemented our language models using the PyTorch machine learning framework. To evaluate the models using our implementation, we divided corpus files into training, validation, and testing sets, and calculated the \topone and \topthree accuracy of space prediction on the testing set after training.  According to the results, which can be seen in Table~\ref{tbl:results-f}, both the \UseMacro{f-ngram-text} and the \UseMacro{f-brnn-text} model are able to learn and suggest formatting conventions with high accuracy. However, the more sophisticated \UseMacro{f-brnn-text} model performs significantly better than the \UseMacro{f-ngram-text} model.

\subsection*{Challenges and Future Directions}

Despite the high accuracy achieved by our preliminary implementation even when using the baseline \UseMacro{f-ngram-text} model, we believe our spacing prediction (based only on raw token streams) needs significant tuning for practical use. For example, newlines before \CoqIn{Qed} sentences often get mispredicted, and unlike for name suggestions~\cite{Nie2020}, it is usually inconvenient to inspect more than the \topone suggestion for spacing. Moreover, for MathComp, we were able to construct, with help from maintainers, a sufficiently large corpus with strict adherence to conventions; for other projects, it may be more challenging, e.g., due to project size or lack of consensus on conventions. We ask the Coq community for input on the following challenges and directions:

\MyPara{Measuring successful predictions} Certain formatting errors, such as improper mid-sentence line breaks, are usually considered worse than others. Can we collaboratively define Coq-specific measures of formatting prediction success and use them to improve the models?

\MyPara{Finding conventions and corpus code} Which files in which projects are idiomatically formatted? What are the main coding styles used in the Coq community? With agreement on these questions, style conformity for files and whole projects can be precisely measured.

\MyPara{Manually improving generated suggestions} How do we best represent and apply rule-based conventions to do reranking of suggestions generated by a trained language model? How should we weigh manually specified conventions against learned ones?

\MyPara{Refactoring of code to adhere to conventions} Our preliminary implementation only modifies spacing, but code may require refactoring to properly address convention requirements, most simply, introducing or changing bullets and specific notations.

\MyPara{Integrating suggestions into the development process} How do we best provide our tools for suggesting conventions to the community? For example, displaying formatting suggestions during code reviews of pull requests on GitHub may work well for large projects, but small projects may have different workflows and thus benefit more from integration with IDEs.

{\footnotesize \label{sec:bib} \bibliographystyle{abbrv} \bibliography{bib} }

\begin{thebibliography}{1}

\bibitem{AllamanisETAL18Survey}
M.~Allamanis, E.~T. Barr, P.~Devanbu, and C.~Sutton.
\newblock A survey of machine learning for big code and naturalness.
\newblock {\em ACM Computing Surveys}, 51(4):81, 2018.

\bibitem{Gallego2016}
E.~J. Gallego~Arias.
\newblock {SerAPI}: Machine-friendly, data-centric serialization for {Coq}.
\newblock Technical report, MINES ParisTech, 2016.
\newblock \url{https://hal-mines-paristech.archives-ouvertes.fr/hal-01384408}.

\bibitem{Nie2020}
P.~Nie, K.~Palmskog, J.~J. Li, and M.~Gligoric.
\newblock Deep generation of {Coq} lemma names using elaborated terms.
\newblock In {\em IJCAR}, 2020.
\newblock To appear. Extended version at
  \url{https://arxiv.org/abs/2004.07761}.

\bibitem{PetersETAL17Semi-supervised}
M.~E. Peters, W.~Ammar, C.~Bhagavatula, and R.~Power.
\newblock Semi-supervised sequence tagging with bidirectional language models.
\newblock In {\em ACL}, pages 1756--1765, 2017.

\end{thebibliography}

\end{document}